\documentclass[aps,prd,twocolumn,showpacs,groupedaddress,amsmath,amssymb,superscriptaddress,nofootinbib,floatfix,noeprint]{aa} 

\usepackage[dvipsnames]{xcolor}

\newcommand\changed[1]{{\color{blue}#1}}
\usepackage{graphicx}
\usepackage{dcolumn}
\usepackage{multirow} 
\usepackage{bm}
\usepackage{hyperref}

\usepackage{color}
\usepackage[utf8]{inputenc}
\usepackage[bottom]{footmisc}
\usepackage[T1]{fontenc}
\usepackage{mathptmx}
\usepackage{url}
\usepackage{relsize}
\usepackage{orcidlink}
\usepackage[switch]{lineno}    

\newcommand\unit[1]{{\rm #1}}
\newcommand\E[1]{{\left\langle #1 \right \rangle}}

\newcommand\mc{{\cal M}_c}

\newcommand\skipme[1]{}

\def\RIT{Center for Computational Relativity and Gravitation, Rochester Institute of Technology, Rochester, New York 14623, USA}
\def\NCSU{Department of Physics, North Carolina State University, Raleigh, North Carolina 27695, USA}
\def\Kenyon{Department of Physics, Kenyon College, Gambier, Ohio 43022, USA}

\begin{document}
	
\title{Astrophysical assumptions and equation of state framework have larger impact on equation of state inference than individual neutron star observations}

\author{
Atul Kedia\orcidlink{0000-0002-3023-0371}\inst{1,2} \and 
Richard O'Shaughnessy\orcidlink{0000-0001-5832-8517}\inst{1} \and 
Leslie Wade\orcidlink{0000-0002-8135-9351}\inst{3} \and 
Anjali Yelikar\orcidlink{0000-0002-8065-1174}\inst{1}}
\institute{\RIT \and \NCSU \and \Kenyon }


\abstract
{The wide range of nuclear densities achieved in neutron stars makes them probes of dense nuclear behavior in the form of the nuclear
equation of state (EoS). Studying neutron stars both in isolation, with X-ray measurements and pulse profiling, and in dynamic events, such as neutron star mergers, have provided insight into these high nuclear densities.  
Though nominally congruent, here we highlight impact of implicit assumptions embedded in joint analysis
of these messengers and their systematic impact on EoS inference.
We show that astrophysical assumptions and EoS framework can have a larger effect on inferred EoS than individual
contemporary neutron star observations.
Performing a proof-of-concept demonstration using the chronologically first few observational constraints, after the application of  5 to 6 observational constraints,  additional
observations  provided diminishing returns and
modified the inferred EoS by shifting the radius of a 1.4 $\unit{M_\odot}$ NS by $\sim$ 0.1 km. By contrast, astrophysical
priors, specifically the spin and mass ratio motivated by astrophysical source population uncertainties, and EoS
framework tend to impact EoS inference much more substantially by shifting the 1.4 $\unit{M_\odot}$ NS radius by $\sim$
0.5 km and by modifying shape of inferred mass-radius relationship.
The inferred EoS depends strongly on the adopted choice of spectral parameterizations: when we employ a framework which explicitly enforces causality, we find   a strong phase-transition-like feature at $\sim 10^{14.5}$ g cm$^{-3}$.}


\titlerunning{Astrophysical assumptions and equation of state framework have impact on the inferred EoS}
\maketitle

\section{Introduction}

Neutron star observations, made over the last decades have provided a critical perspective into matter at high
densities, complementing rapid theoretical and experimental progress 
\citep{LIGO-GW170817-bns,LIGO-GW170817-SourceProperties,LIGO-GW170817-EOS, 2020GReGr..52..109C, Landry:2020vaw,2023PhRvD.107d3034M,%
Reed:2021nqk,Essick:2021kjb,%
Agathos:2015uaa, Lackey:2014fwa, Alvarez-Castillo:2016oln, Margalit:2017dij, Annala:2017llu, Most:2018hfd,%
Radice:2017lry, Rezzolla:2017aly, Riley:2018ekf, Tews:2018iwm, Greif:2018njt, Kiuchi:2019lls, Landry:2020vaw,%
Landry:2018prd, Shibata:2019ctb, 2020NatAs...4..625C, 2020PhRvR...2c3514P, 2020ApJ...893...61Z, Essick:2020flb,%
2020ApJ...893L..21R, Dietrich:2020efo, 2020EPJST.229.3663C, 2020PhRvC.102d5807L, 2021PhRvC.103c5802X,%
2021PhRvL.126f1101A}. 
High-precision pulsar observations have provided evidence for a wide range of neutron star masses, constraining the
neutron star maximum mass from below \citep{2016ARA&A..54..401O}.  X-ray
measurements of neutron star radii have provided additional constraints  \citep{2021PhRvL.126f1101A,2016ApJ...831..184B,
  Steiner:2017vmg, 2023JCAP...02..016F}.  The NICER mission has enabled unique neutron star periodic profile monitoring, enabling simultaneous
mass and radius constraints on a handful of neutron stars including J0740, J0437, and J0614
\citep{Miller:2019cac,Riley:2019yda, Raaijmakers:2019qny,Bilous:2019knh,2019ApJ...887L..27G, Bogdanov:2019ixe, Bogdanov:2019qjb, 2021ApJ...914L..15B, Miller:2021qha, Riley:2021pdl,%
2021ApJ...918L..29R, 2020ApJ...894L...8C, Raaijmakers:2019dks,Legred:2021, Pang:2021jta,%
2024ApJ...976...58S,2025arXiv250614883M,2024ApJ...971L..20C}
Observations of merging neutron stars have also provided several signatures which potentially limit the nuclear equation of state (EoS). Gravitational wave measurements of GW170817 limit the effective inspiral tidal deformability \citep{LIGO-GW170817-EOS, Margalit:2017dij, De:2018uhw, Most:2018hfd, 2018ApJ...852L..29R, 2017ApJ...850L..34B,2020PhRvR...2c3514P}. 
Simultaneous electromagnetic observations also provide suggestive hints into neutron star compactness, deduced from
the presence and energetics of gamma-ray  \citep{2018ApJ...852L..25R} and kilonova \citep{Dietrich:2020efo,
Coughlin:2018miv, Coughlin:2018fis} emission 
\citep{LIGO-GW170817-EOS, Margalit:2017dij, Annala:2017llu, Most:2018hfd, Raithel:2018ncd, 2018PhRvD..97b1501R,
Shibata:2019ctb, Most:2018eaw, 2021ApJ...908L..28N}. 
While the origin of quasi-periodic oscillations remain unclear, if these can be attributed to neutron stars, then additional constraints can be formulated \citep{2023A&A...676A..65S}.

Initial interpretations of the first GW and two NICER measurements suggested a consensus towards a comparable NS radius
  (see, e.g., \cite{2021PhRvL.126f1101A} and references therein), as deduced by many groups employing similar  Bayesian
methods from various independent observational sources and techniques
(see \cite{2019PASA...36...10T} for a review of Bayesian statistics for astrophysical observations).
Even after modest revisions with updated modeling systematics, the original consensus remained intact; see, e.g.,
\cite{2024ApJ...971L..19R,2025PhLB..86539501L,2025PhRvX..15b1014K} and references therein.
However, at the time of the the most recent observation of J0614 \citep{2025arXiv250614883M} , the full range of the
  handful of NICER
  observations are individually consistent with a broad range of NS radii, particularly once measurement systematis are
  included  \citep{Riley:2019yda,2019ApJ...887L..27G,2024ApJ...961...62V}. 
Similarly, intial multimessenger inferences of GW and the kilonova associated with AT2017gfo suggested the light curves
provided modest additional information further supporting the initial consensus \citep{Dietrich:2020efo}.
Since then, however,  several studies have pointed out that
the amount and nature of material ejected from a merger like GW170817 is currently inconsistent with this consensus and our
present understanding of kilonova emission
\citep{2021ApJ...906...98N,2020ARNPS..7013120R,2021ApJ...910..116K,PhysRevResearch.5.013168,Ristic22,2025arXiv250312320R}.
In sum, preliminary conclusions about the nuclear EOS based early multimessenger evidence are narrower than what now
seems appropriate given the latest observations and understanding of instrumental and modeling systematics.


In this work, motivated by the apparent tension between early estimates and present observations, we revisit the observations  used to establish the initial consensus, to seek out 
underappreciated or unknown   systematic factors in the data analysis strategy which can significantly impact the
inferred EOS.  Using an independent
Bayesian framework to combine multiple independent observations,
we demonstrate  these inferences rely on rarely discussed  and of necessity
strong assumptions about the neutron star populations pertinent to each class of observation.  We argue these often inconsistently-applied prior factors have a
substantial effect and may cause analyses to inadvertently rule out a priori consistent EoS models.  Conversely, we
demonstrate that these measurements are largely consistent, with modest tension with each other and
our priors even for our most extreme scenarios. 


Unlike many recent studies which rely on increasingly complex models for the nuclear EoS, to preserve our emphasis on the
observational ingredients and priors we employ an extremely simple low-dimensional model for the nuclear EoS:
%
two  conventional spectral representations \citep{Lindblom:2010bb, Lindblom:2018rfr, PhysRevD.105.063031}
based on approximating a functional versus (log) pressure with a Taylor series.
These low-dimensional smooth approximations are known to have limited flexibility compared with more expressive infinite-dimensional nonparametric models like Gaussian processes \citep{Landry:2018prd, Landry:2020vaw, Landry:2021hvl}. 
However, these models are more than sufficient to express contemporary observational uncertainties, helping us indicate
avenues where  systematics may play an important role.
Other studies have thoroughly explored more flexible approaches including multiple phase transitions; see, e..g,
\citep{PhysRevD.110.123009} and references therein.

This paper is organized as follows. In Section \ref{sec:method} we review
our specific implementation of hierarchical Bayesian
inference for the nuclear equation of state.  Specifically, in Sections   \ref{sec:sub:m_max}, \ref{sec:sub:NS_obs}, \ref{sec:sub:GWanalysis}
 we will highlight how the likelihood has factors which depend explicitly on astrophysical assumptions about the source
 population.  We will then discuss a range of possible source populations, particularly for the exceptional source
 GW170817.
In this section we also review
pertinent observations of neutron stars;   our specific procedure and priors for each class of measurement,
and   our phenomenological parameterizations of the nuclear
equation of state.
For galactic neutron stars we employ radio observations of heavy neutron stars PSR-J0348
\citep{Antoniadis:2013pzd} and PSR-J1641 \citep{2018ApJS..235...37A, 2023ApJ...951L...9A}, NICER X-ray pulse profiling for
PSR-J0030, PSR-J0740, and X-ray spectra and distance analyzed for HESS J1731 \citep{2022NatAs...6.1444D} which retains the
possibility of consisting of strange quark matter, however for this study we assume it to be nucleonic.  For
gravitational wave sources, we employ only GW170817.
In Section \ref{sec:results} we provide inferences for the nuclear EoS using galactic pulsars and nuclear symmetry
energy; update that inference using information from
GW170817, and then reassess those results by incorporating more indirect interpretations of observations, including limits on the NS maximum mass.
In Section \ref{sec:discussion} we elaborate on novel, unfamiliar, or unexpected features of our analysis and
  their broader implications, focusing on how our choces of NS population models and representation of the EOS posterior
  differ from other work.
Finally, we present our concluding remarks in Sec. \ref{sec:conclusion}.

\section{Methods}
\label{sec:method}

In this section, we carefully review hierarchical Bayesian inference for the nuclear EOS, to indicate how assumptions about the astrophysical population of NS responsible for different phenomena, particularly the
  unique binary multimessenger source GW170817,  must be incorporated
  into our calculations.  To assess the relative significance of the binary NS population model, we compare its impact
  to two other well-studied factors, as enumerated in Table \ref{table:observations}: (a) the inclusion or
  omission of several other astrophysical constraints,  and (b) the use of an alternative EOS parameterization and
  prior.  (In our Appendicies, we futher investigate additional even more subdominant factors, such as the impact of
 systematic uncertainty in the GW signal model.)

\subsection{Bayesian framework}
At root, Bayesian inference expresses a posterior as a likelihood times a prior (modulo a normalization).
  Because independent observations of any event have independent statistical uncertainties and are associated with disctinct
  astrophysical phenomena, we combine independent observations by   multiplying their likelihoods together.
Thus, if ${\cal Y}$ characterizes EOS parameters having a prior 
$p({\cal Y})$, then the likelihood used in this work will have a form 
\begin{align}
{\cal L}_{\rm net}({\cal Y}) p({\cal Y}) = 
\left[\prod_n Z_n({\cal Y}) \right ]
\left[\prod_k {\cal L}_k \right] 
\left[\prod_g {\cal E}_g\right ] 
 p({\cal Y}) 
\label{eq:L_net}
\end{align}
where the factors in this expression are briefly defined below and elaborated in subsequent sections.
In this expression,  $n$ indexes the neutron star mass radius observations and $Z_k$ is computed according to
Eq. (\ref{eq:marginal_integral}); $k$ indexes the pulsar mass measurements providing a useful constraint on the maximum
neutron star mass, with ${\cal L}_k$ computed using one of the factors in Eq. (\ref{eq:L_mmax}); and $g$ indexes the
gravitational wave observations, with the GW likelihood computed via Eq. (\ref{eq:Evidence_gw}).
For brevity, we may use a subscript or argument $\alpha$ to refer to an expression conditioned on a specific
  EOS realization ${\cal Y}_\alpha$.
In this expression, almost all factors represent marginal likelihoods and depend on assumptions about the isolated or binary NS population responsible for those
  events, as described in the subsections below which describe each factor in detail.

In the equation presented above, for simplicity we have not incorporated two factors superfluous to our study but essential to
real analyses of the unknown neutron star populations using a large census of NS observations: survey selection biases and models for the  unknown neutron star
population.  For example, unbiased estimates of the nuclear equation of state require self-consistent estimates of the
binary neutron star population \citep{2020arXiv200101747W}.  In turn, self-consistent population estimates must correctly account for
parameter-dependent selection effects effects \citep{2019MNRAS.486.1086M,2019PhRvD.100d3012W}.
To simplify our discussion and focus on the systematic effects that different astrophysical assumptions about
  the binary NS population can introduce, we perform a proof-of-concept calculation using only a handful of
  well-motivated population assumptions, rather than a broad parameterized model family.

\begin{table*}
\caption{The astrophysical and experimental constraints applied and sections and figures in which the impact of respective constraints are discussed.}
\scalebox{0.87}{\begin{tabular}{@{\extracolsep{1pt}} c c c c c c c c c c c c c@{}}
    \hline \hline
    \multirow{2}{*}{Section} & \multirow{2}{*}{Figure} & \multirow{2}{*}{EoS} & \multicolumn{9}{c}{Constraint} & \multirow{2}{*}{Notes} \\
    \cline{4-12}  &  &  & GW170817 & J0348 & J1614 & J0740 & J0030 & PREX & J1731 & J0437 & $M_{max}$ \\ \hline
    \ref{sec:results:galactic_pulsars} & \ref{fig:PSR_only} & $\Gamma$ & - & \checkmark & \checkmark & \checkmark & \checkmark & \checkmark & - & - & - & Base NS inference\\
    \ref{sec:results:galactic_pulsars} &  \ref{fig:causal} & $\Upsilon$ & - & \checkmark & \checkmark & \checkmark & \checkmark & \checkmark & - & - & - & EoS framework impact\\
    \ref{sec:sub:sub:GWanalysis_alone} &  \ref{fig:GW_parameter_posterior}, \ref{fig:GW_parameter_posterior:Aggressive}, \ref{fig:170817:R1p4_nominal} & - & \checkmark & - & - & - & - & - & - & - & - & NS prior impact\\
    \ref{sec:sub:sub:GWanalysis_joint} &  \ref{fig:joint_gw_psr} & $\Gamma$ & \checkmark & \checkmark & \checkmark & \checkmark & \checkmark & \checkmark & - & - & - & GW+PSR+PREX joint inference\\
    \ref{sec:sub:sub:GWanalysis_joint} &  \ref{fig:joint_gw_psr:ExtremeQ} & $\Gamma$ & \checkmark & \checkmark & \checkmark & \checkmark & \checkmark & \checkmark & - & - & - & NS prior impact in joint inference\\
    \ref{sec:results:mmax_limit} &  \ref{fig:Mmax_posteriors} & $\Gamma$ & - & \checkmark & \checkmark & \checkmark & \checkmark & \checkmark & - & - & \checkmark & Upper limit on $M_{max}$\\
    \ref{sec:results:hess_j1731} &  \ref{fig:add_HESS} & $\Gamma$ & - & \checkmark & \checkmark & \checkmark & \checkmark & \checkmark & \checkmark & - & - & Impact of J1731\\
    App. \ref{ap:170817} &  \ref{fig:ap:MR:ResumS} & $\Gamma$ & \checkmark & \checkmark & \checkmark & \checkmark & \checkmark & \checkmark & - & - & - & Impact of waveform models\\
    App. \ref{sec:sub:revised_0437} &  \ref{fig:j0437} & $\Gamma$ & - & \checkmark & \checkmark & \checkmark & \checkmark & \checkmark & - & \checkmark & - & Impact of J0437\\
    \hline \hline
\end{tabular}}
\label{table:observations}
\end{table*}

  \subsection{NS with known extremal masses}
  \label{sec:sub:m_max}
Most neutron star mass measurements provide information only about the
empirical pulsar star mass distribution, rather than any direct insight into the nuclear equation of state.  However, a
few very massive pulsars' existence provides useful insight into  the neutron star maximum mass, bounding it  from below:
J0740 \citep{2021ApJ...915L..12F,Cromartie:2019kug}, with pulsar mass consistent with a Gaussian with mean and standard
deviation $2.14 ~\unit{M_\odot}$ and $0.1 ~\unit{M_\odot}$;
J0348+0432 \citep{2013Sci...340..448A}, with pulsar mass $2.01~\unit{M_\odot} \pm 0.04 ~\unit{M_\odot}$; and 
J1614 \citep{2010Natur.467.1081D}, with pulsar mass $1.908~\unit{M_\odot} \pm 0.016 ~\unit{M_\odot}$.
We follow \citep{Dietrich:2020efo} to express the likelihood of an EoS consistent with these observations as the product
of standard normal distributions:
\begin{align}
{\cal L}_\alpha = \prod_k  \Phi\left( \frac{M_{max,\alpha} - M_{*,k}}{\sigma_{*,k}} \right)
\label{eq:L_mmax}
\end{align}
where $\Phi(z)$ is the cumulative distribution of the standard normal distribution for each maximum mass measurement and are shown in Fig. \ref{fig:demo_gaussian} as horizontal bands.
Using these expressions and as illustrated concretely with plots of the pertinent single-event likelihoods in Section
  \ref{sec:results:galactic_pulsars}, these NS observations strongly suggest a familiar conclusion: that the maximum NS mass is larger than
$2.1 M_\odot$.  We adopt this empirical limit to define a broad NS population.

\subsection{Neutron star population model 1: galactic pulsars}
\label{sec:sub:NS}
Deducing the equation of state from observations of  neutron star properties always requires some model for the neutron
star population: empirical observables limit but do not uniquely determine the neutron star mass, spin, radius, and tidal
deformability. In this work, unless otherwise noted and particularly for galactic pulsars, we will assume the isolated (and binary) neutron star population
is uniformly distributed in gravitational mass $m$ between fixed, EoS-independent limits
$m\in[m_{min},m_{max}]$.  We specifically adopt $m_{min}=0.4 ~\unit{M_\odot}$ and $m_{max}=2.1 ~\unit{M_\odot}$, with the lower and upper mass limits chosen because of the collection of NSs we will use to constrain the nuclear EoS. Similarly, unless otherwise noted, we will assume the (aligned component) of neutron star spin
is distributed isotropically in a sphere and uniformly in magnitude, with a maximum dimensionless value $a=S/m^2 <
0.05$.  This fiducial population closely resembles the prior assumptions used when inferring the nature of low-mass gravitational
wave sources.
We emphasize that our prior represents our assumptions about the \emph{astrophysical distribution} of neutron stars \emph{as realized in
  nature}, which need not reach the maximum mass allowed by the nuclear equation of state.

Aside from its EOS-independent upper limit, this fiducial NS mass distribution has one other unconventional
  choice: its lower limit.  Looking ahead to analyses including HESS J1731, this choice allows us to incorporate the
  most observationally likely parameters, avoiding baking in strong prior assumptions  into our analysis about the
  plausibility of subsolar-mass NS. For analyses   omitting the HESS source, the lower limit has a negligible impact, only impacting our overall normalization.

\subsection{Neutron star mass-radius constraints}
\label{sec:sub:NS_obs}
NICER observations of periodic X-ray emission on millisecond pulsars provides information about their masses and radii. Though their data product is reported as a weighted posterior distribution, we find these results can be well approximated by a simple Gaussian distribution, as illustrated in Figure \ref{fig:demo_gaussian} for the pulsar PSR-J0030 (red). In Appendix \ref{ap:nicer_gaussian} we describe the best-fitting gaussian to each NICER observations alongwith HESS-J1731, and the specific method we used to
deduce its parameters. In terms of this Gaussian likelihood $L(M,R)$, we can evaluate the marginal likelihood for each equation of state $\alpha$ according to \citep{gwastro-PENR-RIFT,LIGO-GW170817-EOSrank, 2019PASA...36...10T}
\begin{align}
    Z(\alpha) &= \int \bigg[\int \mathcal{L}(M,R) \delta(R-R_\alpha(s)) \\
    & ~~~~~~~~~~~~~~~ \delta(M-M_\alpha(s)) \mbox{d}M \mbox{d}R \bigg]p(s) \mbox{d}s   \nonumber \\
    & = \int \mathcal{L}(M(s),R_\alpha(M(s))) ~~ p(s) \frac{\mbox{d}M}{\mbox{d}M_\alpha/\mbox{d}s}  \nonumber \\
    & = {\cal F} \int \mathcal{L}(M(s),R_\alpha(M(s))) ~~ \mbox{d} M 
    \label{eq:marginal_integral}
\end{align}
where $s$ is a fiducial parameter characterizing our dimensionless prior along each one-parameter family of each EoS; 
$p(s)$ is a prior along that one-parameter family; and  in the last line we've chosen $p(s)$ to be compatible with our
uniform prior over the neutron star mass, with ${\cal F}$ the appropriate normalization constant in mass space: ${\cal
  F}=1/(m_{max}-m_{min})$.  
For simplicity, in this work we adopt the gravitational mass $M$ as our fiducial prior (i.e., so the denominator
$(dM_\alpha/ds)/p(s)$ is constant), while recognizing the utility of other global choices like the pseudo-enthalpy.
This choice of prior along each family strongly downweights contributions to the evidence near turning points  in the
mass-radius plane at the high and low mass ends of the curve (i.e., $dM/dR \simeq 0$). That said, our choice of prior
enables us to use the same functional form for all equations of state, independent of their maximum mass.
%


\begin{figure}
    \centering
    \includegraphics[trim={0.5cm 0cm 2cm 1.5cm},clip,width=0.48\textwidth]{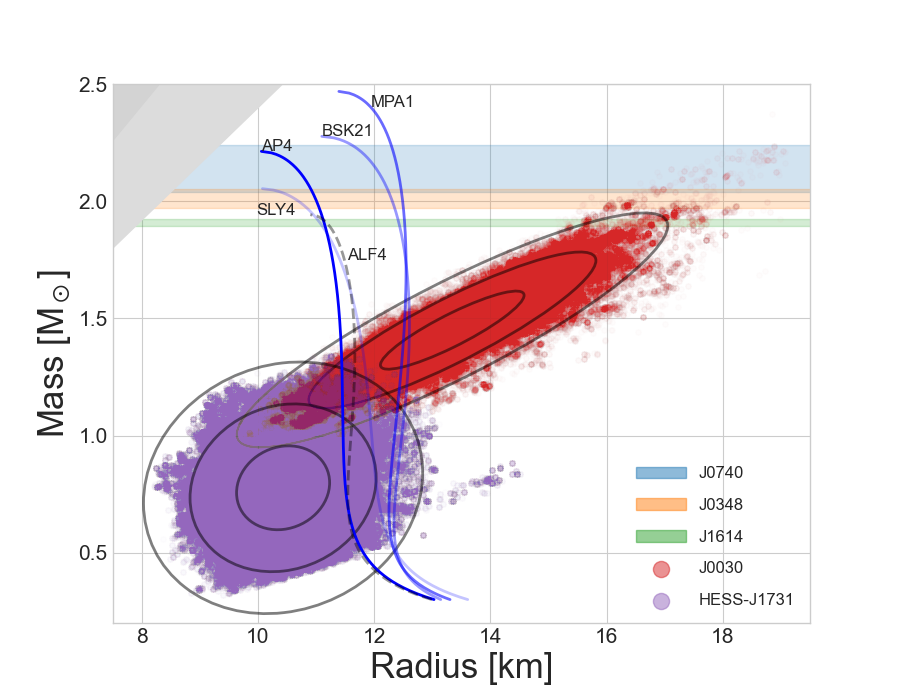}
    \caption{\textbf{Methodology} Mass-Radius likelihoods evaluated for sample cold nuclear EoSs. The pulsar constraints used for the likelihoods are shown in the background in various colors. Darker colors of the M-R curves indicate higher $\log$ likelihood for each EoS.
    The EoS with the lowest likelihood is represented with a dashed curve.
    The red and purple patches are the posterior samples from NICER pulse profile of PSR-J0030 and the HESS-J1731 respectively.
    The horizontal color bands are mass constraints of J0740, J0348, J1614.}
    \label{fig:demo_gaussian}
\end{figure}


As a demo, Figure \ref{fig:demo_gaussian} shows the Mass-Radius curves for sample nuclear EoSs, AP4, MPA1, BSK21, SLY4, and ALF4, plotted over the NICER constraints posteriors by their likelihoods.
The EoSs are selected to demonstrate how inference is conducted in this work via the multiple different channels informing about consistency with the particular measurement.
The darker shaded M-R curves of the various EoSs (MPA1, and AP4 in this example) indicates greater relative likelihood of those EoSs in comparison to the lighter colored EoSs (SLY4 and BSK21). Note ALF4, which is hardly visible, is so due to it not supporting the maximum mass observations. In our likelihood evaluations for grids of EoSs we employ this method for applying Mass-Radius constraints.

\subsection{Neutron star population models 2: Merging GW sources}
\label{sec:sub:prior_binary_gw_pop}

Most compact object formation models suggest that the mass and spin distribution of NS in merging binaries will differ
  substantially from the mass and spin distribution of isolated or accreting NS.   Addtionally, possibly-strong
  selection biases on binary, GW, and especially multimessenger sources can further modify the appropriate population
  model needed for EOS inference.  For this reason,  we employ five distinct mass
  and spin   distributions for binary NS sources.  Unless otherwise noted and following convention for GW sources, we
  assume a maximum dimensionless spin of $|\chi_i|<0.05$ for each NS, with $\chi_{i,z}$ distributed consistent with a
  uniform distribution of spin magnitudes and an  isotropic distribution of spin directions according to
  $p_\chi(\chi_{i,z}|\chi_{\rm max})$.
  Our three primary mass and spin distributions can then be enumerated as follows
  \begin{itemize}
\item  (a) our fiducual base model, in which each NS is drawn independently from a uniform
  mass distribution [$p_0(m_1,m_2, \chi_{1,z}\chi_{2,z})= 2 {\cal F}^2 I(m_1)I(m_2)\Theta(m_1-m_2)p_\chi(\chi_{1,z})p_\chi(\chi_{2,z})$ where $I$ is
    nonzero for $m\in [m_{\rm min}, m_{\rm max}]$ and where the Heavyside step function $\Theta$ ensures that $m_1>m_2$];
\item   (b) a distribution which requires highly unequal masses and allows for large spin
  [$p_{as}(m_1,m_2,\chi_{1,z},\chi_{2,z}) \propto   p_0$ subject to   $m_2/m_1 \le     0.6$ and using $\chi_{\rm
      max}=0.5$];
  (At GW170817's source-frame chirp mass $\mc = 1.186 ~\unit{M_\odot} $, this constraint
  requires more extreme choices for source-frame primary and secondary mass:  $m_2 < 1.062~\unit{M_\odot}$ and $m_1>1.77
  ~\unit{M_\odot}$.)
\item  and (c) a distribution which both favors symmetric masses and requires preferentially aligned spins [i.e., in mass
  adoping $p_0$ but requiring $\chi_{i,z}>0$ and $m_2/m_1 > 0.9$], with the specific choice of mass ratio cutoff motivated by
   the range expected for GW-detected BNS sources derived from the galactic population \citep{2020ApJ...900L..41A}.
  \end{itemize}

Operationally, all of these astrophysical prior choices can be implemented as  alternative
options for  the upper and lower
bounds adopted for $q,\chi_{i,z}$,  when analyzing GW170817 with a nonprecessing GW model.
Several sources of external information provide good reason to adopt alternative informed priors.
In the discussion below, we elaborate on our motivation behind these three choices.

The third distribution (c) is strongly motivated by
galactic
observations and conventional field formation scenarios.
These scenarios
strongly favor nearly equal masses
\citep{2020ApJ...900L..41A,2020A&A...639A.123K,Farrow:2019xnc,2018MNRAS.481.4009V,2020PASA...37...38V,2020ApJ...900L..41A,2020A&A...639A.123K}.
and also usually rule out large or preferentially negatively-oriented spins \citep{2000ApJ...541..319K}.
(Of course, the limited statistics available from the GW census are consistent with a much broader
mass ratio distribution  \citep{Landry:2021hvl,LIGO-O3-O3bpop} and provide minimal constraints on plausible NS spin
orientations; additionally, non-merging galactic BNS binaries have a much broader mass ratio distribution
\citep{2023MNRAS.521.4669Y}.)

The mass distribution adopted for  (b) is motivated by the unexpectedly large dynamical ejecta inferred for AT2017gfo.%
  As most predictions for this ejecta are
suppressed for $q\simeq 1$ \citep{Henkel:2022naw}, others have suggested favoring asymmetric masses where the
dynamical ejecta isn't suppressed \citep{2018ApJ...866...60P,2020Natur.583..211F}.  For example, \citet{2025arXiv250312320R} argue that, if contemporary ejecta models for both the mass and velocity are applied
  directly, a self-consistent interpretation of AT2017gfo either requires implausible NS radii or extremely asymmetric
  binary mergers.

As will be demonstrated later, due to strong spin-mass ratio correlations, this constraint also requires
  substantial NS spin. To avoid adopting overconstraining prior assumptions, for this analysis we will substantially
  relax the maximum spin allowed.  To better understand the impact of relaxing the maximum spin, we will also present a
  handful of results using the fiducial model but with a larger $\chi_{\rm max}$.    This relaxed spin prior can also be
well-justified astrophysically on similar grounds:others have demonstrated that a
strongly-spinning secondary neutron
star ($m_2$) will produce substantially enhanced ejecta, as its outer layers are less bound \citep{2024MNRAS.530.2336R,2022MNRAS.513.3646P}.
Strongly spinning NS mergers can occur for example if the secondary does not have time to spin down, due to fortuitous
NS kicks \citep{2010ApJ...716..615O}  or Kozai-Lidov effects in hierarchical triples  \citep{2023arXiv230210350B}.  (Such a rapid formation scenario isn't
  consistent with the large geometric offset for AT2017gfo but could occur in general; as noted in
  \citep{2023arXiv230210350B,2024arXiv240203696C} this rapid formation scenario would reconcile NS merger formation with r-process-enriched but metal-poor stars.)
While the specific form adopted for the prior (b) is somewhat ad hoc, particularly given the systematics associated with contemporary ejecta calculations, an exhaustive quantitative
investigation needed to develop and calibrate another approximation merits its own dedicated study.  



\subsection{Gravitational wave observations}
\label{sec:sub:GWanalysis}
Gravitational wave observations of GW170817 constrain the binary masses $m_i$ and tidal deformabilities $\Lambda_i$ for
each component of the binary
\citep{LIGO-GW170817-bns,LIGO-GW170817-SourceProperties,LIGO-GW170817-EOS,PhysRevResearch.2.043039}.  
Specifically, the presence of matter at leading order impacts the binary's inspiral, through each components' dimensionless tidal deformability parameter  $\Lambda = (2/3) k_2 (c^2 R / G M)^5$~\citep{Flanagan:2007ix}, where $k_2$ is the $l = 2$ Love number, $M$ is the mass, and $R$ is the radius.
The leading order contribution to the phase evolution of an GW inspiral is given by the weighted combination of
$\Lambda$ terms
\begin{align}
    \tilde{\Lambda} = \frac{16}{13} \frac{(m_1 + 12 m_2) m_1^4 \Lambda_1 + (m_2 + 12 m_1) m_2^4 \Lambda_2}{(m_1 + m_2)^5}
\end{align}
The impact of tidal effects are included in many conventional approximate phenomenological
estimates of outgoing radiation from merging compact binaries. In this work, we will focus on analyses with a
contemporary state-of-the-art model, IMRPhenomPv2\_NRTidalv2 \citep{2019PhRvD..99b4029D,2019PhRvD.100d4003D}, which
incorporates precession physics but omits higher-order modes.  We note  other
contemporary models like NRHybSur3dq8Tidal \citep{2020PhRvD.102b4031B} and TEOBResumS \citep{2018PhRvD..98j4052N} include
these higher-order modes and tidal effects.
In Appendix \ref{ap:170817} we present a single proof-of-concept  inference using one of these alternative
  waveform models, to demonstrate its effect is subdominant to the issues emphasized in this work.

When performing phenomenolgical inference for binary intrinsic parameters
  $X=(\mc,\eta,\mathbf{\chi}_1,\mathbf{\chi}_2,\Lambda_1,\Lambda_2)$, the RIFT parameter inference engine
  \citep{gwastro-PENR-RIFT,2023PhRvD.107b4040W} iteratively assesses the gravitational wave marginal likelihood $L(X)$ on many
  candidate points $X$, producing estimates of the posterior $\propto L(X) p_{gw}(X)$ for $X$ using (fiducial) population priors for these intrinsic
  parameters $p_{gw}(X)$ and an
  interpolated estimate $\hat{L}(X)$ of the true likelihood $L(X)$ obtained by interpolation. 
For our application,  the RIFT marginal likelihood data and interpolation provide a framework to compute an
  evidence for any proposed EOS \citep{gwastro-PENR-RIFT,LIGO-GW170817-EOSrank,2024arXiv240715753V}. Specifically, using the marginal likelihood $L(X)$ obtained by interpolating the \texttt{RIFT} marginal likelihoods and the EoS-derived
relationship connecting  $\Lambda(m|{\cal Y})$ to the neutron star gravitational mass $m$ and equation of state model
${\cal Y}$,  we compute a marginal evidence
\begin{align}
{\cal E} = \int L(X(x,{\cal Y}))p(x) dx
\label{eq:Evidence_gw}
\end{align}
where $x=(m_1,m_2,\mathbf{\chi}_1,\mathbf{\chi}_2)$ are binary parameters omitting tides and $X(x,{\cal Y})$ supplements these
parameters with the EoS-derived tidal parameters (i.e.,
$X=(\mc,\eta,\mathbf{\chi}_1,\mathbf{\chi}_2,\Lambda(m_1|{\cal Y}),\Lambda(m_2|{\cal Y}))$).  
For simplicity, we also use \texttt{RIFT} to perform this integral \citep{gwastro-PENR-RIFT,LIGO-GW170817-EOSrank,gwastro-RIFT_FinerNet}, using the \texttt{RIFT} marginal likelihood estimator $\hat{L}(X)$ for $L(X)$.  These EoS evidence infrastructure has been validated in previous work \citep{LIGO-GW170817-EOSrank}.
We do not employ any customary approximations when evaluating this integral numerically (e.g., that the
chirp mass is well known).

While other objects putatively containing matter have been identified via their GW signature (i.e., GW190425 \citep{2020ApJ...892L...3A}, GW200105,
and  GW200115 \citep{LIGO-O3-NSBH}), their GW signature also
does not meaningfully contain the nature of nuclear matter.  For example, the two objects in NSBH binaries have putative
matter effects that minimally impact $\tilde{\Lambda}$.  Conversely, in GW190425, the large NS masses imply very small
tidal deformabilities, but that low-amplitude GW measurement only weakly constraints $\tilde{\Lambda}$.  For this
reason, we omit further discussion of these three objects.

\subsubsection*{Phenomenology of $\Lambda,~R$}
To provide a preliminary estimate of the impact of the NS population priors on EoS inference derived from
GW170817, we follow  Zhao and Lattimer 2018 \citep{Zhao:2018nyf}, and adopt simple approximate relationships between tidal
deformability and fiducial NS radius $R_{1.4}$.  In brief, they argue that to a good approximation $\Lambda\propto
(R/M)^{6}$ (Eq. (13) in \cite{Zhao:2018nyf}); that therefore  $\tilde{\Lambda}$ is to a good approximation proportional to $(R_{1.4}/\mc)^6$, with a
nearly mass-ratio-independent constant (Eq. (15) in \cite{Zhao:2018nyf}); and therefore for a binary with well-measured chirp mass, the
radius of a fiducial NS can be estimated as (evaluated for GW170817)
\begin{align}
\label{eq:ZL}
R_{1.4} \simeq 13.4~\unit{km}(\tilde{\Lambda}/800)^{1/6} \; .
\end{align}

\subsection{Nuclear equation of state phenomenology}
Following previous investigations, when parameterizing a nuclear EOS $\alpha$, we characterise it  with a low-dimensional
phenomenological model.
Rather than  employ a sophisticated but high-dimensional phenomenological approach based on
microphysics-motivated calculations with many tunable parameters such as relativistic mean field theory \citep{1989RPPh...52..439R,2003RvMP...75..121B,2022PhRvC.106e5804A,2023PhRvC.108b5809Z,2023ApJ...943..163Z,2018PhRvC..97b5805M,2024MNRAS.529.4650H,2023PhRvD.107j3018M}, for simplicity
and lacking observationally-driven necessity, we adopt two distinct analytic representations.

\begin{figure}
\includegraphics[trim={0.5cm 0cm 1.5cm 1.5cm},clip,width=\columnwidth]{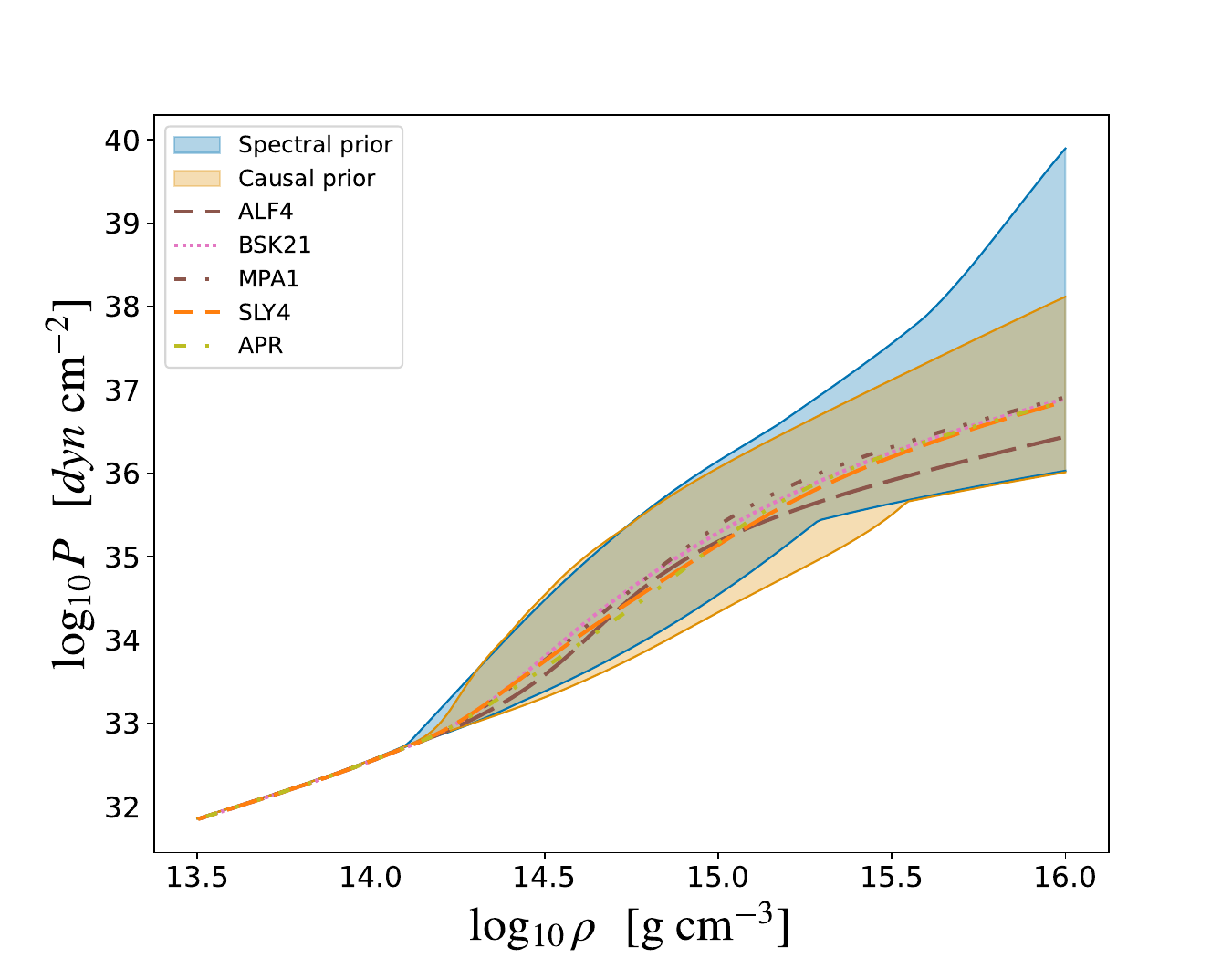} \\
\includegraphics[trim={1cm 0cm 1.5cm 1.5cm},clip,width=\columnwidth]{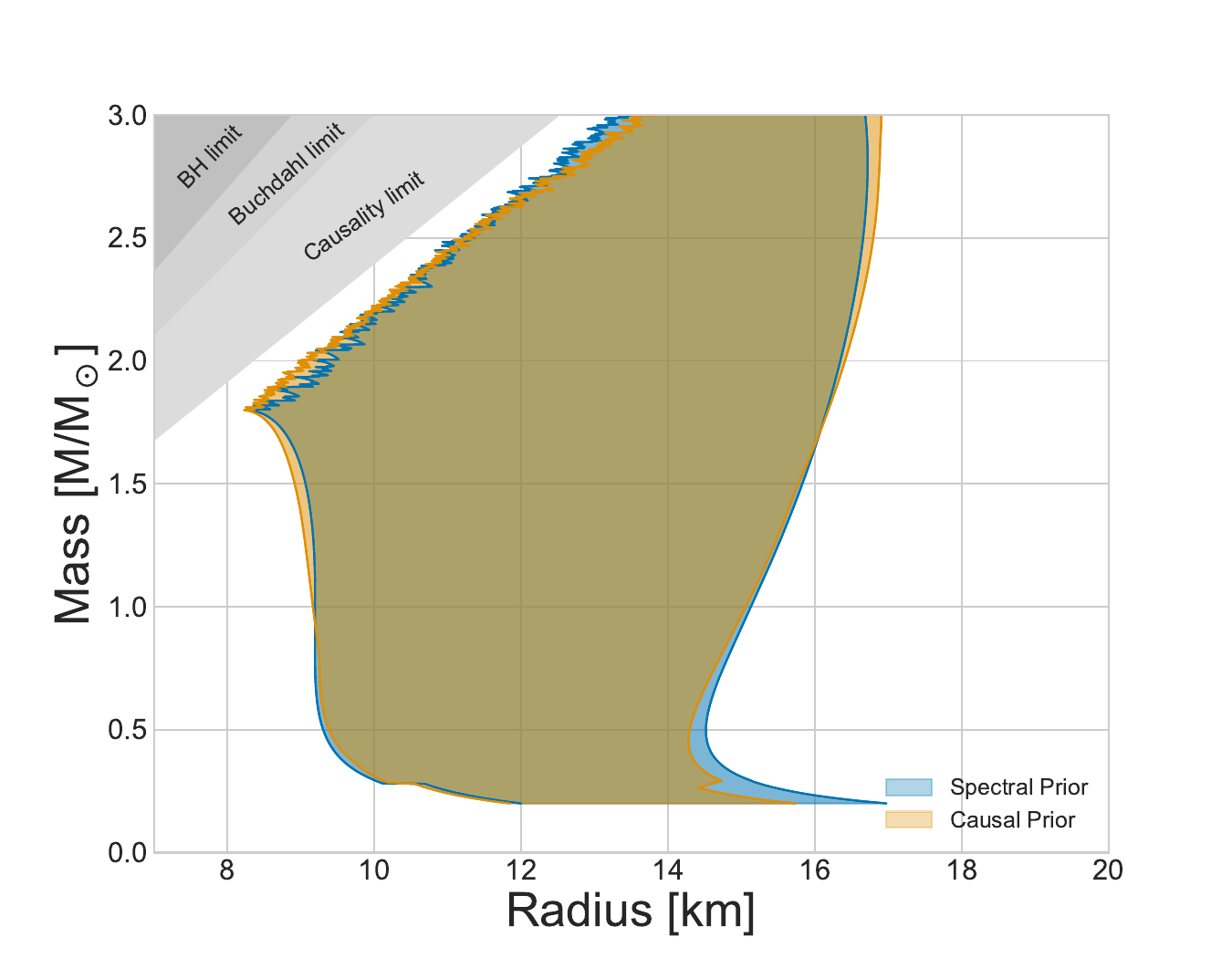}
\centering
\caption{\textbf{Priors} Pressure $p$ versus energy density $\rho$ for cold nuclear matter prior utilized. The
  shaded region shows the 100\% credible interval implied by our fiducial prior on the $\Gamma$-spectral model (blue) and $\Upsilon$-spectral model (orange).
  Also shown are several fiducial equations of state in dashed and dotted curves, for reference.
  \\
  Top panel: Pressure-Density priors for the causal and regular spectral representations along with five fiducial
  EoSs. The interval shown encompasses all draws from our prior (here, shown for $10^4$ samples). \\
   Bottom panel: Mass-Radius prior range for the EoSs in the above range.
  The grey shaded regions indicate various constraints on single-stable NS; `BH limit' is where the Schwarzchild radius is larger than the NS radius, 
  `Buchdahl limit' is the range where an NS would need infinite pressure \citep{PhysRev.116.1027}, and `Causality limit' is obtained by finding the maximally compact (M/R) solution for Mass-Radius in the limiting physical range of pressure-density \citep{2012ARNPS..62..485L}.
  }
\label{fig:p_rho_m_r_prior}

\end{figure}

In this work we principally employ a specific and widely-used spectral parameterization \citep{Lindblom:2010bb}, referred
to as the $\Gamma$-spectral parameterization from here onwards, in which the natural log of the zero-temperature adiabatic index $\Gamma = d\ln p/d\ln n$ is expressed as a power series in $\log p$:
\begin{align}
    \ln \Gamma = \sum_k \gamma_k (\ln p/p_o)^k
\end{align}
For this model, we adopt a uniform prior in its hyperparameters, with limits consistent with previous studies \citep{Carney:2018sdv,LIGO-GW170817-SourceProperties}.
Though we do not explicitly enforce other customary constraints (e.g., on the adiabatic index), our TOV solver requires
EoS that are nearly causal, so our calculations de facto reject highly unphysical samples as implausible.
As noted in Appendix B of\citet{2020arXiv200101747W}, this EoS model family and prior has a narrow, strongly correlated subspace of physical realizations: only a fraction $5\times 10^{-5}$ of the prior volume is physical. For this reason, adaptive or well-tuned sampling will be an essential to obtain  robust results with this framework.
To illustrate the impact of model and prior systematics, we also employ another spectral formulation,  referred to as $\Upsilon$-spectral from here
onwards (though denoted the ``improved causal spectral'' formulation in \citet{PhysRevD.105.063031}), designed to ensure smooth transition from the fixed low density piece to the fit higher density part and to ensure causality all throughout. In particular we use the pressure based formulation described in \citep{PhysRevD.105.063031}, where the spectral expansion is constructed for the sound speed, $v(p)$, dependent parameter $\Upsilon (p)$ defined as
\begin{align}
\Upsilon(p) = \frac{c^2 - v^2(p)}{v^2(p)} ~~~ \,
\end{align}
for which the spectral expansion is 
\begin{align}
\Upsilon (p) = \exp \Bigg[{\sum_{k=0}^\infty} \lambda_k \Bigg(\log\bigg(\frac{p}{p_0}\bigg)\Bigg)^k \Bigg] \; .
\label{eq:causal_decomp}
\end{align}
This decomposition is very similar to the previous causal-$\Gamma$-spectral representation \citep{Lindblom:2018rfr} with the only difference being that the zeroth coefficient, $\lambda_0$, is predetermined. Here $\lambda_0$ is fixed such that sound speed continuity is ensured at the matching point, $p_0$, between the low density and high density pieces
\begin{align}
\lambda_0 = \log\Bigg(\frac{c^2 - v_0^2}{v_0^2}\Bigg)
\end{align}
We restrict our analyses to the first four independent parameter in this decomposition, $k=1,2,3,4$, in order to capture systematic features of the EoSs while keeping the exploration manageable.
For simplicity,  we choose a uniform prior for these four  spectral parameters, over $\lambda_1 \in [-3, 3.2]$, $\lambda_2 \in [-4, 2]$,
$\lambda_3 \in [-0.2, 0.8]$, and $\lambda_4 \in [-0.05, 0.02]$,
where these ranges of each parameter are motivated by covering existing EOS as well as the
$\Gamma$-spectral posterior
for GW170817 \citep{2020arXiv200101747W}.
While this prior range nominally includes configurations with a NS maximum mass $M_{\rm max}$ smaller than the
  largest NS mass in our fiducial population 
  ($m_{\rm max}=2.1   M_\odot$), in practice when performing joint inference we always require consistency with the known massive NS and thus reject entirely any NS with
  maximum mass smaller than $m_{\rm max}$.  [For the purposes of illustrating trends in single-constraint marginal likelihoods versus
    parameters, we at times relax this requirement and incorporate EOS realizations and marginal likelihoods which do
    not conform to this constraint.]
Figure \ref{fig:p_rho_m_r_prior} shows our implied prior on $p(\rho)$ and nonrotating neutron star gravitational mass and radius.
We use these two parametric approaches to qualitatively assess the kinds of systematic uncertainty exhibited among some
of the commonly used parameterizations, such as piecewise polytropic \citep{Read:2009iy}, dynamical-piecewise
polytropic \citep{PhysRevD.106.103027, Carney:2018sdv}, and  similar spectral methods
\citep{Lindblom:2010bb,Lindblom:2018rfr, PhysRevD.105.063031} which adopt different basis functions or parameters \citep{2023PhRvD.107l3017L}.

We follow previous work that employs Gaussian quadrature to evaluate the energy-density at given pressure on the \texttt{lalsuite} implementation \citep{LIGO-GW170817-SourceProperties,Carney:2018sdv}. 
We adopt a fixed low-density equation of state with the first four terms in the expansion determining the EoS. This parameterization has been previously constrained by comparison of its input GW170817 \citep{LIGO-GW170817-SourceProperties} and NICER measurements \citep{Miller:2019cac, Miller:2021qha}.
For all EoSs, we solve the Tolman-Oppenheimer-Volkoff (TOV) equation \citep{Oppenheimer:1939ne, Tolman:1939jz} for nonrotating neutron star structure in hydrostatic equilibrium using  \texttt{RePrimAnd}
\citep{2024arXiv240411346K,wolfgang_kastaun_2023_7700296} to find one-parameter families of gravitational mass $M$, radius
$R$, and tidal deformability $\Lambda$.
The output for an equation of state $\alpha$ is characterized by a one-parameter family $R_\alpha(s),M_{\alpha}(s)$,
where $s$ is a fiducial parameter characterizing our dimensionless prior.
For simplicity, in this work we adopt the pseudo-enthalpy h, and only investigate the lowest stable branch of solutions.

\subsection{Additional sources of external information}
\label{sec:sub:ext}

In the preceding subsections, we have highlighted the essential components for any joint analysis for the nuclear EoS from these astrophysical
observations, each with   relatively well-understood observational and theoretical systematics.  
However, other groups have proposed employing other sources of information about the nuclear equation of state to
further sharpen their analysis.   Below, we summarize three particularly tractable approaches to incorporate specific
sources of external information.

\begin{figure}
\includegraphics[trim={0.5cm 0cm 0.5cm 1.5cm},clip,width=1.1\columnwidth]{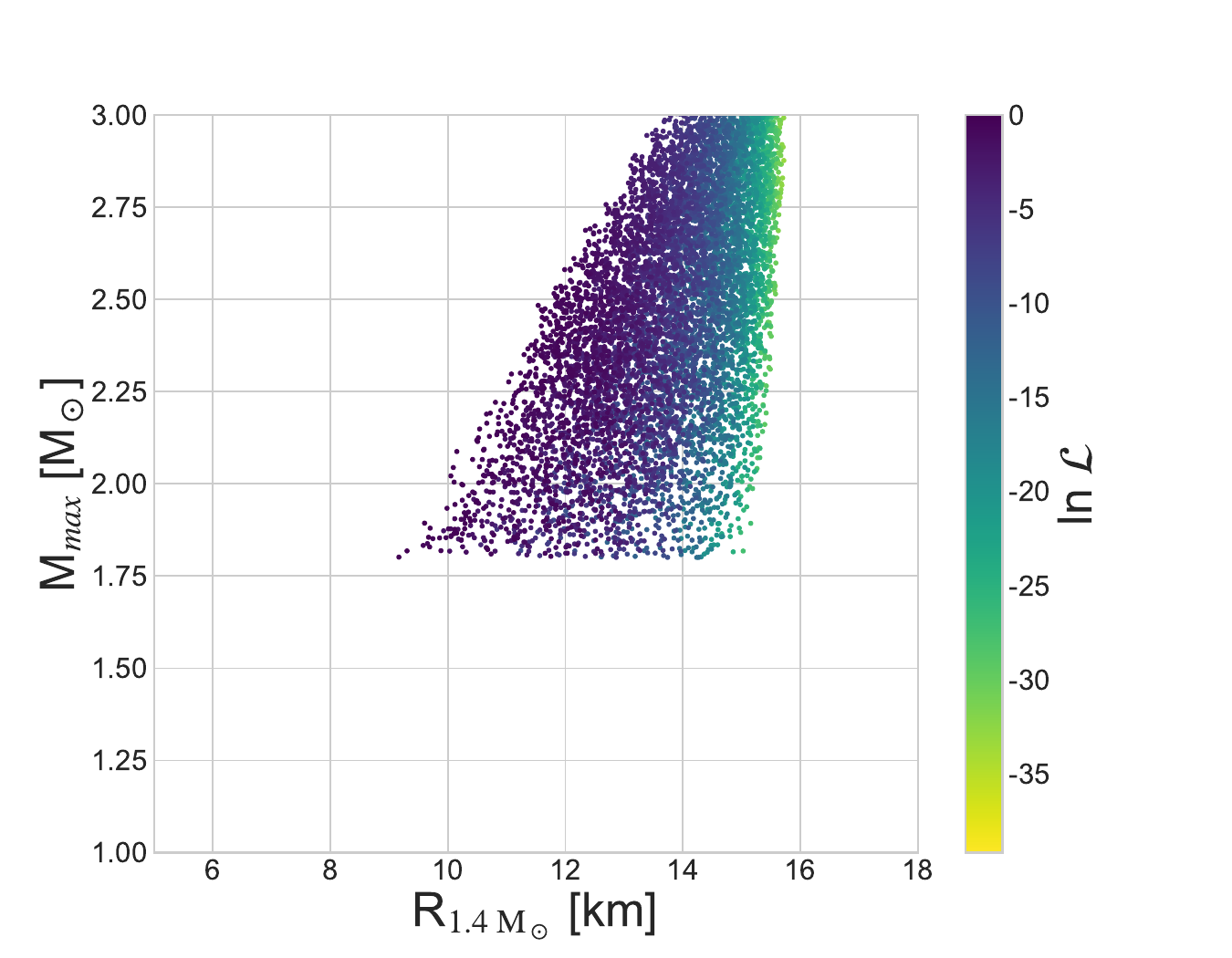}
\caption{\textbf{Marginal likelihood for symmetry energy}: scatterplot of the symmetry energy likelihood evaluated for
  many $\Gamma$-spectral EoS realizations, versus the corresponding $R_{1.4}, M_{max}$ evaluated for the same EoS
  realizations. Roughly speaking and within the context of this model family, this symmetry energy constraint favors
  smaller radii at each fixed $M_{max}$.
\label{fig:lnL:symmetry}
}
\end{figure}

The nuclear EoS must be consistent with properties of nuclei  \citep{2012ARNPS..62..485L,2021ARNPS..71..433L}.  Following
\citep{Miller:2021qha}, one strategy for incorporating information about nuclear-density matter is to require the baryon
density $n$ and energy density $\rho c^2$ to be consistent with constraints on the  nuclear symmetry energy $S\simeq
\rho/n - m_n c^2$, assumed to be known to be normally constrained with a mean of $32~\unit{MeV}$ and a standard
deviation of $2~\unit{MeV}$ at saturation density, assumed to be 2.7 $\times 10^{14}~\rm{g~cm^{-3}}$ \citep{PhysRevD.110.123009, PhysRevC.86.015803, 2019EPJA...55..117L}.
Figure \ref{fig:lnL:symmetry} shows the marginal likelihood for different EoS realizations, plotted versus two
EoS properties $R_{1.4}$ and $M_{\rm max}$.  The symmetry energy constraint favors smaller radii at each fixed $M_{max}$.

Based on observations of AT2017gfo, several groups have argued that the observed electromagnetic energetics cannot be
consistent with prompt collapse of the remnant to a black hole, using that qualitative constraint to  infer an
\emph{upper} bound to the NS maximum mass; see, e.g., 
\citep{2017ApJ...850L..34B,2017LRR....20....3M,2018ApJ...852L..25R,2022PhRvD.105j3022K,2018ApJ...852L..25R,2023MNRAS.519.2615E} and references therein.  Roughly
speaking, these arguments suggest $M_{max}<2.3 ~\unit{M_\odot}$ \citep{2018ApJ...852L..25R}, extremely close to the mass inferred for J0952-0607 \citep{2023MNRAS.519.2615E}.

 As noted in Section \ref{sec:sub:NS}, the observation of AT2017gfo favors large amounts of ejecta.
While these large ejecta masses can be accommodated with exceptional NS spins or mass ratios, these outcomes can also be
produced with large neutron stars.  Using this rationale, some groups have argued that the NS tidal deformability should
be bounded below by approximately $\tilde{\Lambda}>400$ \citep{2018ApJ...852L..29R}.  
This proposed constraint on $\tilde{\Lambda}$ has comparable qualitative impact on the information obtained
from GW170817 as our constraint that $q>0.9$ and $\chi_{i,z}>0$, as will be seen in Section \ref{sec:sub:GWanalysis_results} Figure \ref{fig:GW_parameter_posterior}.
For brevity, we omit further discussion of this  proposal.

A preliminary analysis with the inclusion of a recently measured galactic pulsar J0437 \citep{2024ApJ...971L..20C} is discussed it in Appendix \ref{sec:sub:revised_0437}.


\subsection{Hyperparameter posterior pipeline: \texttt{HyperPipe}}
\label{sec:sub:hyperpipe}
To generate the posterior distribution from ${\cal L}_{\rm net}$ and $p({\cal Y})$ we introduce a
  generalization of the \texttt{RIFT} iterative
inference strategy: \texttt{HyperPipe}.
Starting with some proposed parameters  ${\cal Y}_\alpha$, we evaluate ${\cal L}_{\rm net}({\cal
  Y}_\alpha) \equiv {\cal L}_{net,\alpha}$; build an approximation $\hat{\cal L}({\cal Y})$; and use that approximation
with an adaptive Monte Carlo integration program to generate both an estimated posterior distribution for ${\cal Y}$ as
well as a new set of proposed parameters ${\cal Y}_{\alpha'}$ meriting further investigation.  
This pipeline is able to adaptively explore the space of selected hyperparameters to resolve distinct features in the posterior distribution. The estimate is initiated in a small region of the hyperparameter space (in this case a subsection of the space of ${\cal Y}=\{(\gamma_1,\gamma_2\ldots)\}$). 
We refer to the proposed parameter selection, ${\cal Y}_\alpha$, as the \textit{initial grid} of parameters and the resultant physical properties as \textit{initial samples}.
Following the first iteration of evaluating likelihoods a posterior is estimated by performing Monte Carlo integration. Following this new parameters are sampled in the regions of the posterior with lower resolution in order to not exclude important finer details in addition to samples placed in exploratory regions. Following likelihood evaluations for these new points an updated posterior is evaluated by using prior and marginal likelihood information from samples of the previous iteration(s) and the current iteration, hence being better informed. Following this the process is repeated a few times where new sample points are generated and the posterior updated by Monte Carlo integration until the posterior does not deviate between iteration to iteration, resulting in a stable posterior estimate.

For this, we have extended the open-source \texttt{RIFT} inference package to include a general-purpose implementation of
this algorithm.  Our implementation not only provides a physics-neutral implementation of the \texttt{RIFT} algorithm, but also
can interface with  generic codes which compute  marginal evidence factors appearing in multi-event and multi-messenger inference like
Eq. (\ref{eq:L_net}).  
Moreover, though not used here as the pertinent factors are constant in this study, our implementation can include the necessary population-averaged integrals needed to
account for a flexible
population model and survey selection effects associated each class of measurement (i.e., the  full inhomogeneous
Poisson likelihood \citep{2019PhRvD.100d3012W} or the rate-marginalized likelihood
\citep{2019MNRAS.486.1086M,2019RNAAS...3...66F}).  
Similarly, we can also easily account for multinomial leave-few-out averaged results; for example, to marginalize over
including the HESS result or not, the factor ${\cal E}$ for the HESS result can be replaced by ${\cal E} p + (1-p)$ and
then marginalized over confidence $p$ in that analysis.
The algorithm and implementation for \texttt{HyperPipe} completely follows the original RIFT implementation \citep{gwastro-PENR-RIFT,2023PhRvD.107b4040W,gwastro-RIFT_FinerNet} and
even   re-uses most of its components, differing only in its ability to handle more generic non-GW model parameters
${\cal Y}$ and externally-supplied, arbitrary marginal likelihood calculations associated with factors in the net marginal likelihood as in Eq. (\ref{eq:L_net}).


\begin{figure}
\includegraphics[trim={1cm 0cm 1.5cm 1.5cm},clip,width=\columnwidth]{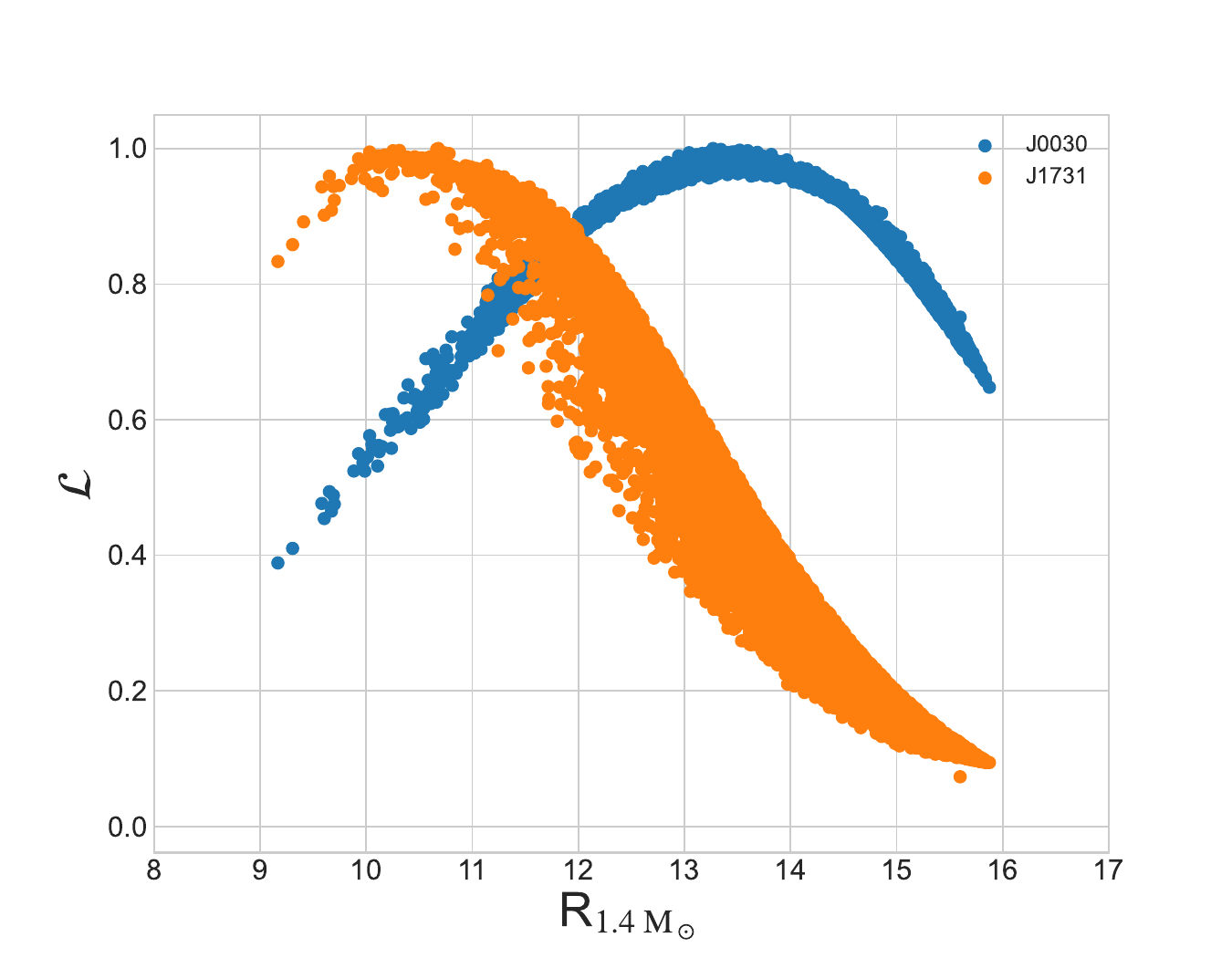}
\includegraphics[trim={1cm 0cm 1.5cm 1.5cm},clip,width=\columnwidth]{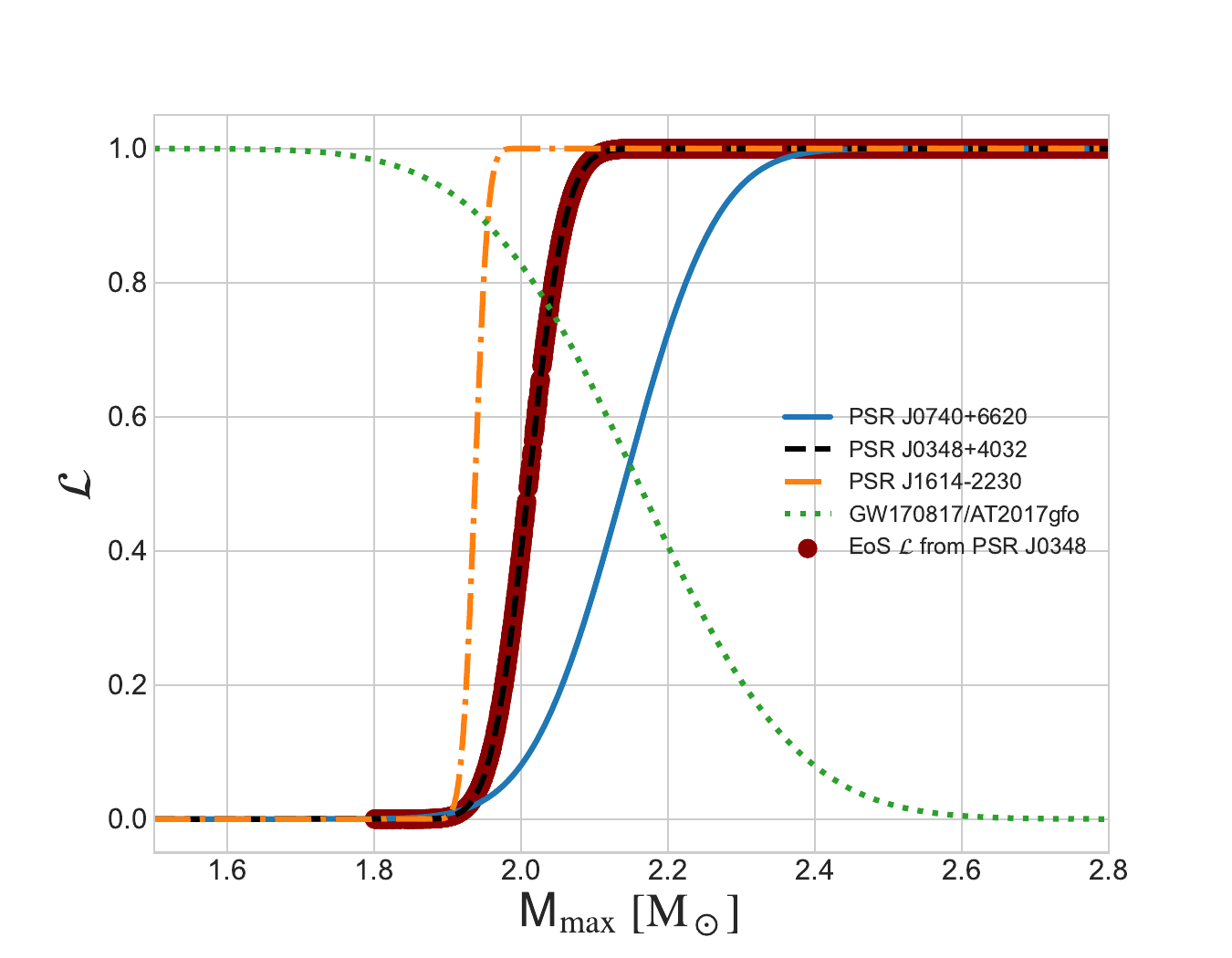}
\caption{\label{fig:lnL_1d}\emph{Top panel}: Likelihood ${\cal Z}$ versus $R_{1.4}$  deduced by comparing
  realizations of the $\Gamma$
  spectral EoS family to the NICER mass-radius constraints of J0030 (blue) and HESS J1731 (orange) 
\emph{Bottom panel}: Likelihood $ {\cal L}$ versus $M_{max}$.  The blue solid, black dashed, and orange dash-dotted lines show
Eq. (\ref{eq:L_mmax}), evaluated for each known pulsar mass.  The dark-red points show a direct evaluation using  realizations of the $\Gamma$
  spectral EoS family to the pulsar J0348, for expediency using \texttt{lalsuite} rather than
  \texttt{RePrimAnd} to evaluate $M_{max}$ for each EoS.  The dotted green line shows the hypothesized lower bound on the maximum
  mass implied by prompt collapse to a black hole; see Section \ref{sec:sub:ext} for discussion.
}
\end{figure}

\subsection{Diagnostics}
As the posterior distribution is directly proportional to the likelihood ${\cal L}_{\rm net}$, we employ several
representations of $\ln {\cal L}_{\rm net}$ and its component factors, to characterize the constraining power of each
individual observation.  
As a demonstration, the top panel in Figure \ref{fig:lnL_1d} shows our marginal log-likelihood $\ln {\cal Z}$
evaluated using EoS drawn from the full set of marginal likelihood evaluations produced during a hierarchical inference
of the EoS, rendered  versus $R_{1.4}$, the radius of a fiducial
neutron star.   In this calculation, the list of equations of state $\alpha$ are drawn from the broad analyses performed
in this work.   Even more transparently than the two-dimensional approach adopted in Figure \ref{fig:demo_gaussian},
this one-dimensional representation allows us to visually illustrate how well each individual measurement constrains one
feature of the equation of state.
Similarly, the bottom panel in Figure  \ref{fig:lnL_1d} shows our marginal log-likelihood $\ln {\cal L}$ versus $M_{\rm
  max}$.

These diagnostics allow us to quickly propagate and interpret the effect of different prior choices about the neutron
star population.  For example, adopting an alternative neutron star mass distribution that is uniform out to $M_{\rm
  max}(\alpha)$ instead of a fixed value (such as in this study) shifts log likelihood for each individual likelihood contribution in $\ln {\cal Z}$ by a factor $-\ln \frac{M_{\rm max}-m_{min}}{2.1~\unit{M_\odot}-m_{min}}$.
Because this systematic factor appears along with every mass-radius measurement, its effects stack as more measurements accumulate,
increasingly penalizing configurations with exceptional  $M_{\rm max}$ relative to our approach.    For expected choices
for $M_{\rm max}$, however, this factor is small and should have minimal impact on our posteriors.

\section{Results}
\label{sec:results}

In this section we describe inferences about the EoS.
In Section \ref{sec:results:galactic_pulsars} we first demonstrate our method using only galactic pulsars and
the nuclear symmetry energy.  We show how our HyperPipe implementation, like RIFT, provides interpretable net marginal
likelihoods, allowing us to visually assess the posterior and our coverage similar to typical RIFT analyses.  Using the concrete example of the two spectral EoS parameterizations described earlier
($\Gamma$ and $\Upsilon$), we show that different EoS model families can produce qualitatively different results.
In Section \ref{sec:sub:GWanalysis_results}, we describe how we incorporate information from GW170817 into EoS
inference using different
assumptions about the pertinent binary NS population.   Specfically, in Section \ref{sec:sub:sub:GWanalysis_alone} we
  describe EoS-agnostic RIFT source parameter inference of GW170817 conditioned on several different alternatives.
  population.  We demonstrate that even before explicitly conditoning on an EoS model family, these population
  assumptions have substantial impact on the range of $\tilde{\Lambda}$ with significant support in each scenario.
  Then, in Section \ref{sec:sub:sub:GWanalysis_joint} we present EoS inference conditioned on our fiducial assumptions
  as well as with these alternative binary NS population assumptions.  We find the choice of NS population assumptions
  will significantly change the inferred radius of a typical NS.   In Section \ref{sec:results:mmax_limit}, for context
  and a sense of scale,  we review for
  comparison the extent to which additional ad hoc constraints on the NS maximum mass can further inform the EoS.
  Similarly and last, in Section \ref{sec:results:hess_j1731} we discuss the marginal impact that the HESS J1731 source
  has on our overall inferences, compared to alternative binary NS assumptions for GW170817.

\subsection{Inferences using galactic pulsars}
\label{sec:results:galactic_pulsars}

\begin{figure*}
    \centering
    \includegraphics[width=\columnwidth]{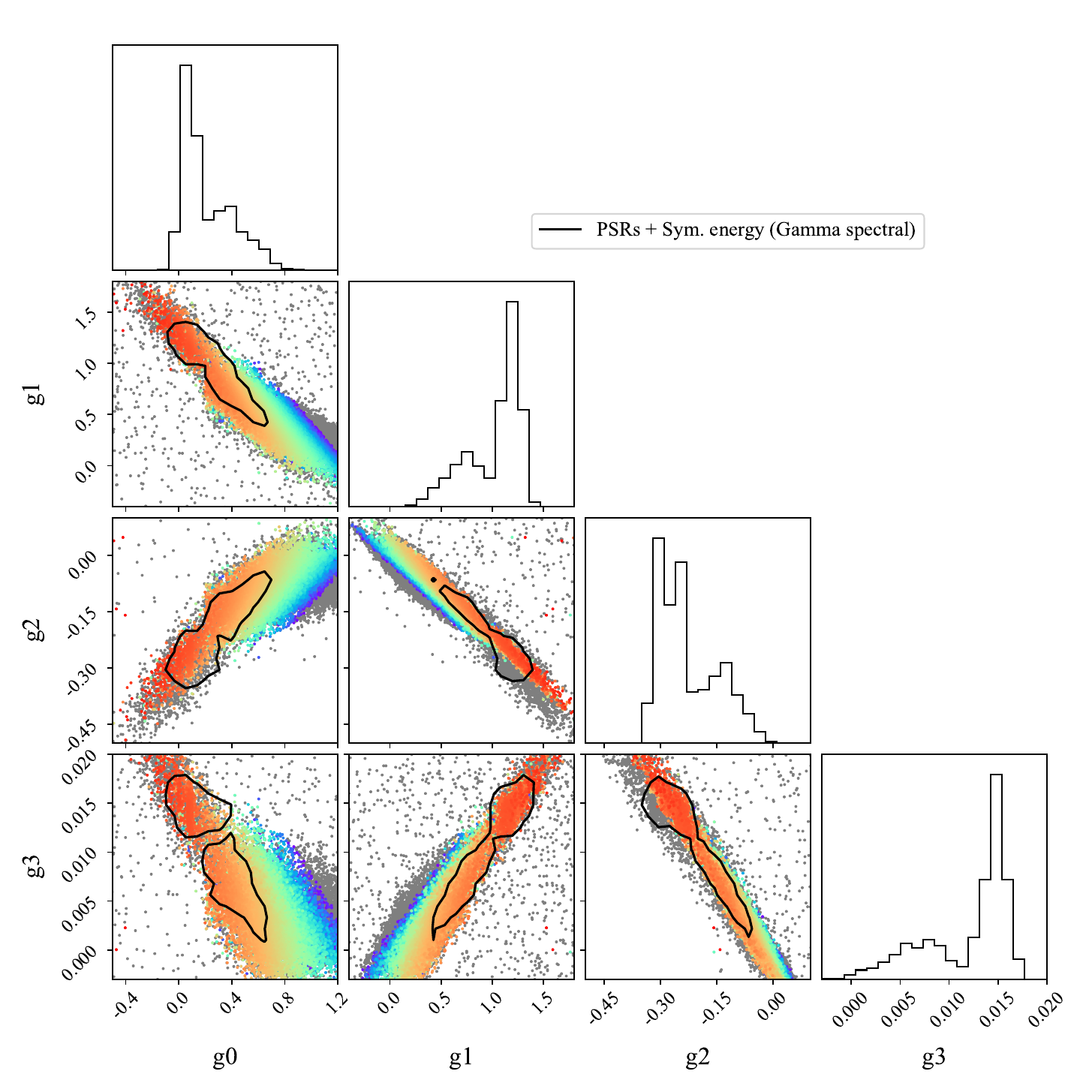}
    \includegraphics[trim={0.5cm 0cm 2cm 2cm},clip,width=\columnwidth]{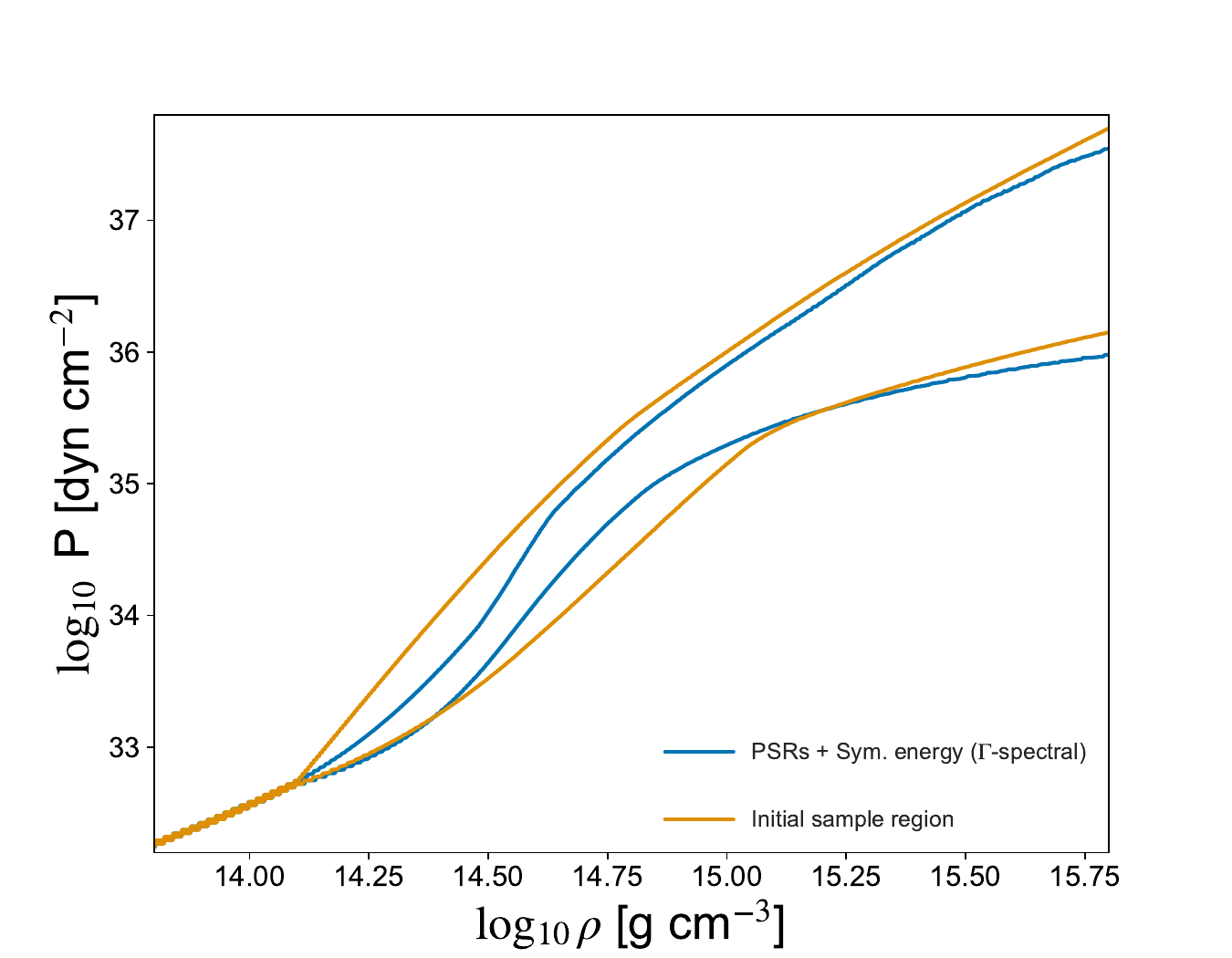}\\
    \includegraphics[trim={0cm 0cm 0cm 1.5cm},clip,width=\columnwidth]{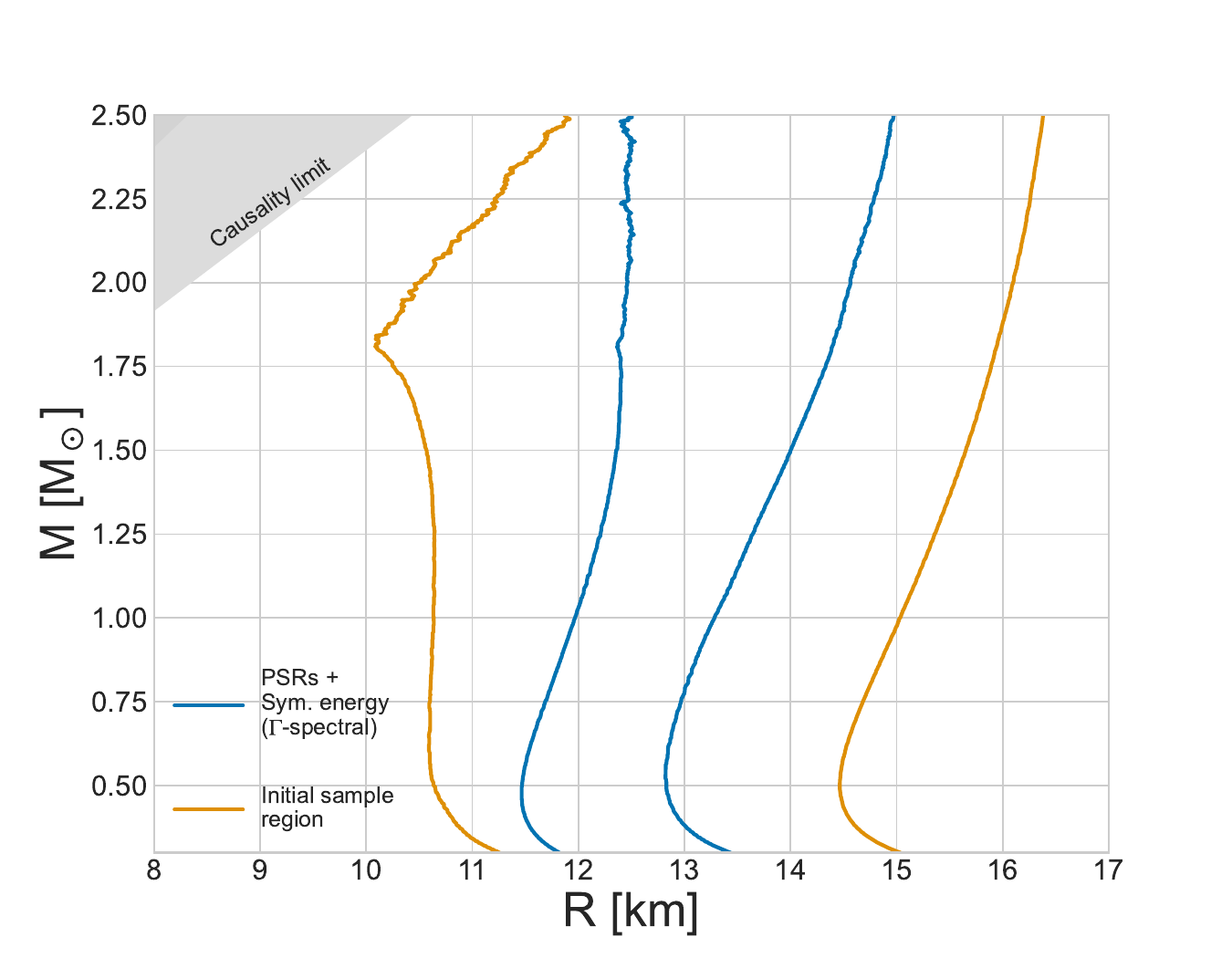}
    \caption{\textbf{EoS inference from galactic pulsars and symmetry energy, $\Gamma$-spectral family}: Marginal posterior
      distribution for the combined inference of observations PSRs J0030, J0348, J0740, J1614, and applying the PREX symmetry
      energy constraint. \emph{Top-left panel}: Distribution of $\Gamma$-spectral hyperparameters.  Color scale shows the marginal likelihood,
      with red indicating the largest values and the solid contour shows the 90\% credible interval, while
      one-dimensional histograms show a one-dimensional marginal distribution. \emph{Top-right panel}: 90\% credible
      interval of pressure versus density, evaluated at each density.  \emph{Bottom panel}: 90\% credible interval for
      radius evaluated at each NS mass, rendered as mass versus radius. In the figures in the top-right and bottom the 98\% interval of the region covered by the initial samples supplied to \texttt{HyperPipe} is also shown.  }
    \label{fig:PSR_only}
\end{figure*}

\begin{figure*}
    \centering
   \includegraphics[trim={0.5cm 0cm 2cm 1.5cm},clip,width=\columnwidth]{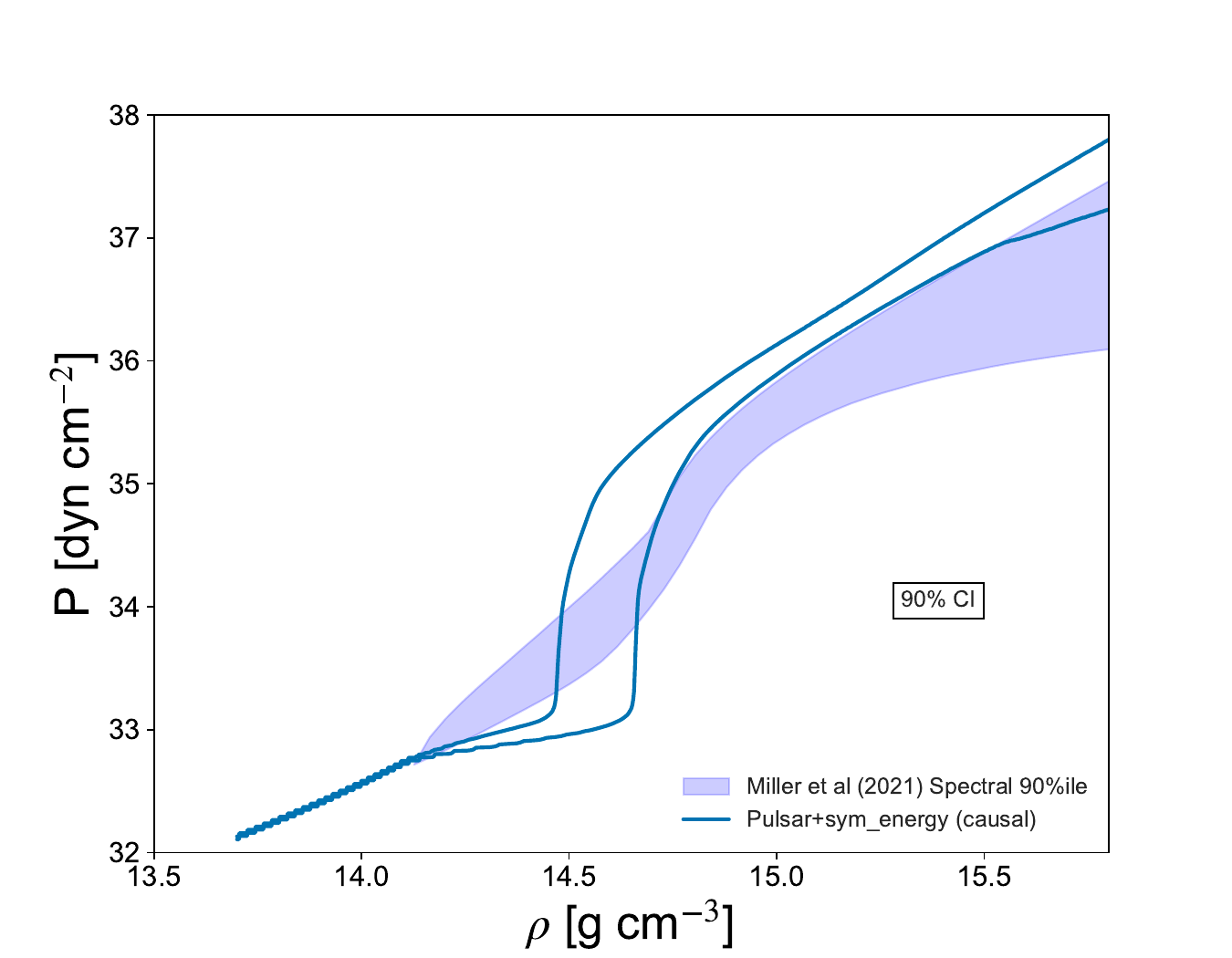} 
    \includegraphics[trim={0.5cm 0cm 2cm 1.5cm},clip,width=\columnwidth]{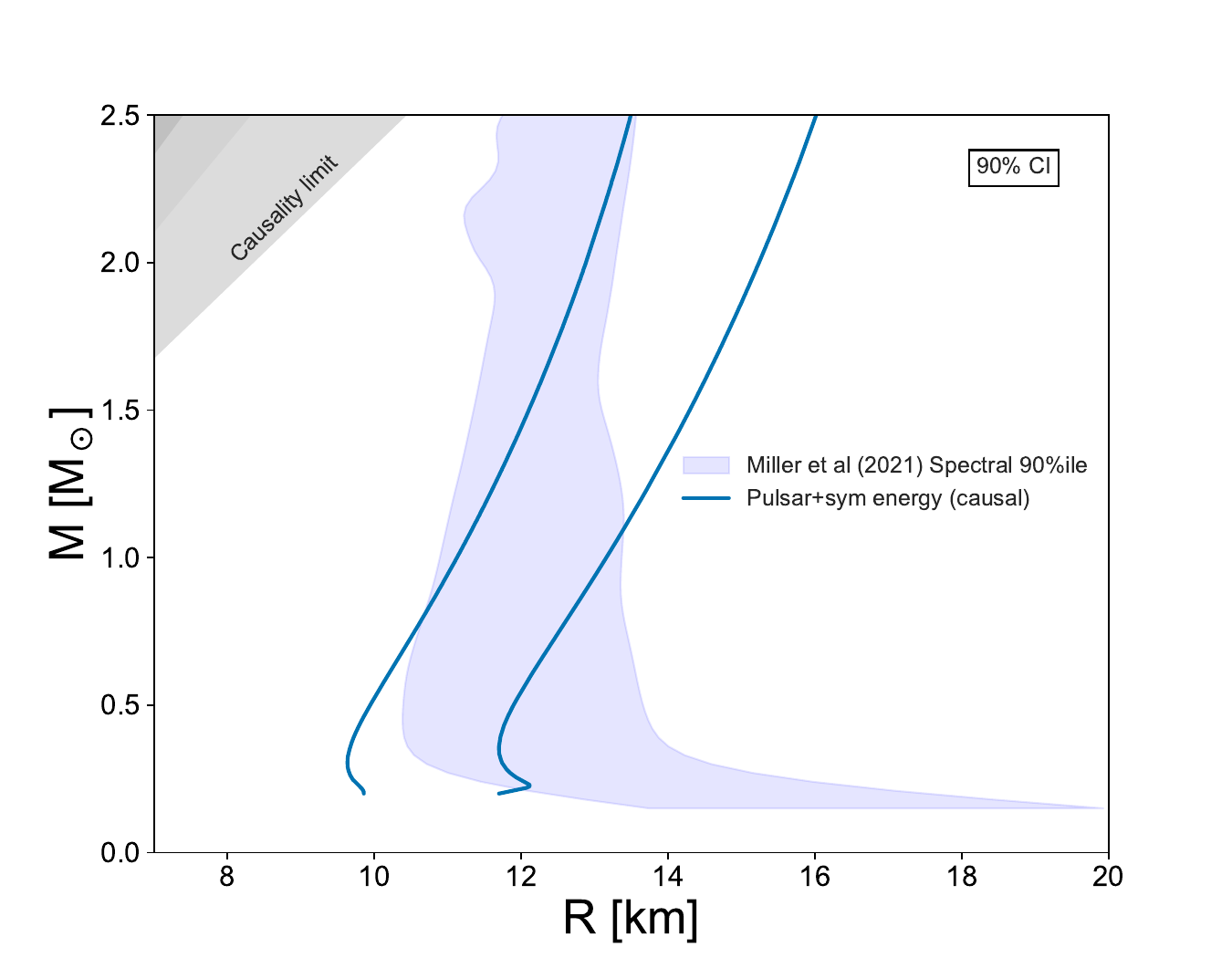}
    \caption{
\noindent \textbf{Impact of parameterization on EoS inference}: Like Figure \ref{fig:PSR_only}, but adopting the
$\Upsilon$-spectral EoS, and constraining for only the PSRs J0030, J0348, J0740, J1614 and PREX symmetry energy. We also
provide a comparison analysis which adopted the same PSRs, and symmetry energy, but includes GW constraint and uses $\Gamma$-spectral parameterization for these constraints for reference \citep{Miller:2021qha}.
    }
    \label{fig:causal}
\end{figure*}

We first reanalyze previously-studied events with our approach, initially using the original spectral parameterizations and priors for the assumed nuclear equation of state.  
To illustrate how our analysis proceeds using comparable constraints as in \citep{Miller:2021qha} (i.e., galactic neutron
star observations and the nuclear symmetry energy), the colored scatterplot points in
top left panel of Figure \ref{fig:PSR_only} show the marginal
likelihoods versus the four model hyperparameters ($\Gamma$-spectral parameters in this case).  In this example, we are employing all the
galactic pulsar mass and radius measurements mentioned in Section \ref{sec:sub:NS_obs} (i.e., mass-radius information for PSR-J0030, and mass measurements for
J0740, J0348, J1614),  as well as the lower bounds on maximum neutron star mass from observed massive pulsars,
as indicated in Figure \ref{fig:demo_gaussian}.  As described in Section \ref{sec:sub:hyperpipe}, these likelihood evaluations have been identified  by \hyperref[sec:sub:hyperpipe]{\texttt{HyperPipe}} by
conducting an iterative exploration of the model space, in the process recovering the posterior distribution, whose 90\% credible
intervals and one-dimensional marginal distributions are also shown in the same panel of
 Figure \ref{fig:PSR_only}.
 As expected given how broad and weak our present observational
constraints on neutron star radius as illustrated in Figure \ref{fig:demo_gaussian}, we find our marginal likelihoods
are relatively uniform over a wide range of the parameter space, insofar as EoS compatible with those parameters exist.
(Most of our nominally uniform prior volume encompasses incompatible EoS.)
%
Next, the curves in the top-right (bottom) panels express the 90\% credible interval of our inferred
equation of state in $p(\rho)$ ($M(R)$).   
Once again, we see that our posterior 90\% credible pointwise interval only differs modestly from the full range suggested by our initial
assumptions, being narrower at all Mass-Radius estimates. The initial sample space is only indicative of the EoSs taken into consideration at first, as \hyperref[sec:sub:hyperpipe]{\texttt{HyperPipe}} selects new points to explore, the posterior is free to go beyond this initial sample limits to attain the best fit.
Overall, because our prior assumptions are very well adapted to contemporary theoretical preferences, the new NICER 
observations add relatively little new additional information over and beyond our original preferences, with one important
exception highlighted below.

To investigate the impact of the EoS parameterization (and prior), we also inferred the EoS with the $\Upsilon$-spectral
family. 
Figure \ref{fig:causal} shows a comparable analysis derived using the $\Upsilon$- spectral parameterization
(e.g. Eq. \ref{eq:causal_decomp}), illustrating the inferred relationships between pressure and density, and mass and radius. In this figure we construct $p(\rho)$ posterior distributions
  using one-dimensional credible intervals at each density, while $M(R)$ is constructed
similarly versus $M$. In sharp contrast to our fiducial analysis with the $\Gamma$-spectral parameterization, the
  $\Upsilon$-spectral parameterization converges to extreme EoS at low and high density.  Close to our matching density
  with the fixed crust EoS, this $\Upsilon$-spectral posterior converges to sharp density transitions, consistent with
  $c_s \simeq 0$. At the densities $\gtrsim
 10^{14.5 }$ g cm$^{-3}$, however, $c_s \to 1$.
 In other words,  this alternative parameterization (with broader priors) best fits galactic pulsar observations with
 strong low-density phase transitions and
  unexpected high-density behavior, so the $M(R)$ curve aligns more closely with astrophysical results, notably the
  NICER observation. While the EoS $p(\rho)$ and $c_s(\rho)$ have unexpected behavior, this model
  family of course produces mass-radius behavior consistent with astrophysical constraints, which are most informative
  in radius over a narrow mass range.  

Considering only the observational marginal likelihood $\ln {\cal L}_{\rm net}$ and disregarding EoS priors, our inferences favor the $\Upsilon$-spectral
family relative to the $\Gamma$-spectral EoS.  Specifically comparing the largest marginal likelihood $\ln {\cal L}_{\rm
  net}$ identified in each analysis, we find the $\Upsilon$-spectral recovers a slightly-larger peak marginal
likelihood, with
\begin{equation}
\frac{{\cal L}_\Upsilon, {\rm max}}{{\cal L}_\Gamma, {\rm max}}\sim 2.39 ~~.
\end{equation}
Moreover, the best-fitting EoS identified by the
$\Upsilon$-spectral family has qualitative differences from the EoS explored with the $\Gamma$-spectral EoS: a regime of EoSs with an early phase-transition followed by a sharp upturn such that $c_s \to 1$.
We attribute the differences between the two parameterizations' results to their underlying form: given our fiducial
hyperparameter priors, the $\Gamma$-spectral
framework is not sufficiently flexible to produce  EoSs with such sharp jumps.
%
Our study corroborates many previous investigations on how different EoS parameterizations,
priors, and phase transitions implementations can remain consistent with observations yet correspond to very different underlying
conclusions about nuclear physics.

\subsection{Adding information from GW sources}
\label{sec:sub:GWanalysis_results}

In order to jointly infer the nuclear EOS simultaneously using PSR and GW observations, we need the gravitational wave likelihood $L(X)$ in Eq. (\ref{eq:Evidence_gw}), to compute the appropriate marginal evidence ${\cal E}$ for each EOS.   In this section, we first describe our reanalysis of GW170817, with which we construct the necessary marginal likelihood. 
We then demonstrate, using this phenomenological approach alone and without recourse to detailed EOS modeling, that the posterior for $\tilde{\Lambda}$ depends strongly on the assumed BNS population prior adopted, specifically using three concrete examples motivated in Section \ref{sec:sub:GWanalysis}.
We discuss the correlations and features of the GW posterior that cause these prior changes to have the large effect on the $\tilde{\Lambda}$ posterior, and reflect on whether they are robust to other modeling systematics.  Finally, using the likelihood provided by this analysis, we perform full-end-to-end EOS inference with the GW and PSR likelihoods together, for the different BNS population prior models discussed above.  As anticipated by our simple order-of-magnitude argument using only the posterior for $\tilde{\Lambda}$, we find the BNS population priors have a substantial effect on conclusions about the nuclear EOS.

\subsubsection{Reanalysis of GW170817 alone}
\label{sec:sub:sub:GWanalysis_alone}
As a benchmark analysis including tidal effects, we analyze open GW data for GW170817 available  from GWOSC \citep{ligo-O1O2-opendata
}, using the same PSDs provided with GWTC-1  \citep{LIGO-O2-Catalog,LIGO-O2-Catalog-PSDRelease}, over a
frequency range from $23\unit{Hz}$ to $1700\unit{Hz}$ \citep{TheLIGOScientific:2014jea, TheVirgo:2014hva}.
By default, we adopt a low-spin prior $|\chi_i|<0.05$, with known sky location and source
luminosity distance derived from the electromagnetically-identified host galaxy.  We do not impose any constraints on
the tidal deformability, allowing both $\Lambda_i$ to take on arbitrary values.   Other prior assumptions are customary
(e.g., uniform in detector-frame component masses).   
We include prior information about the alignment between the binary's angular momentum direction and the line of sight,
inferred from late-time radio afterglow observation \citep{2019NatAs...3..940H,2018Natur.554..207M}.  For simplicity and
without loss of generality we assume both component spins are parallel to the orbital angular momentum direction,
primarily to facilitate external constraints on the binary's orbital inclination.  
Parameter inference is performed with the most recent \texttt{RIFT} compact binary parameter inference engine \citep{gwastro-PENR-RIFT,gwastro-PENR-RIFT-GPU,PhysRevD.107.024040}. 

\begin{figure}
\includegraphics[trim={0cm 0cm 0cm 1cm},clip,width=\columnwidth]{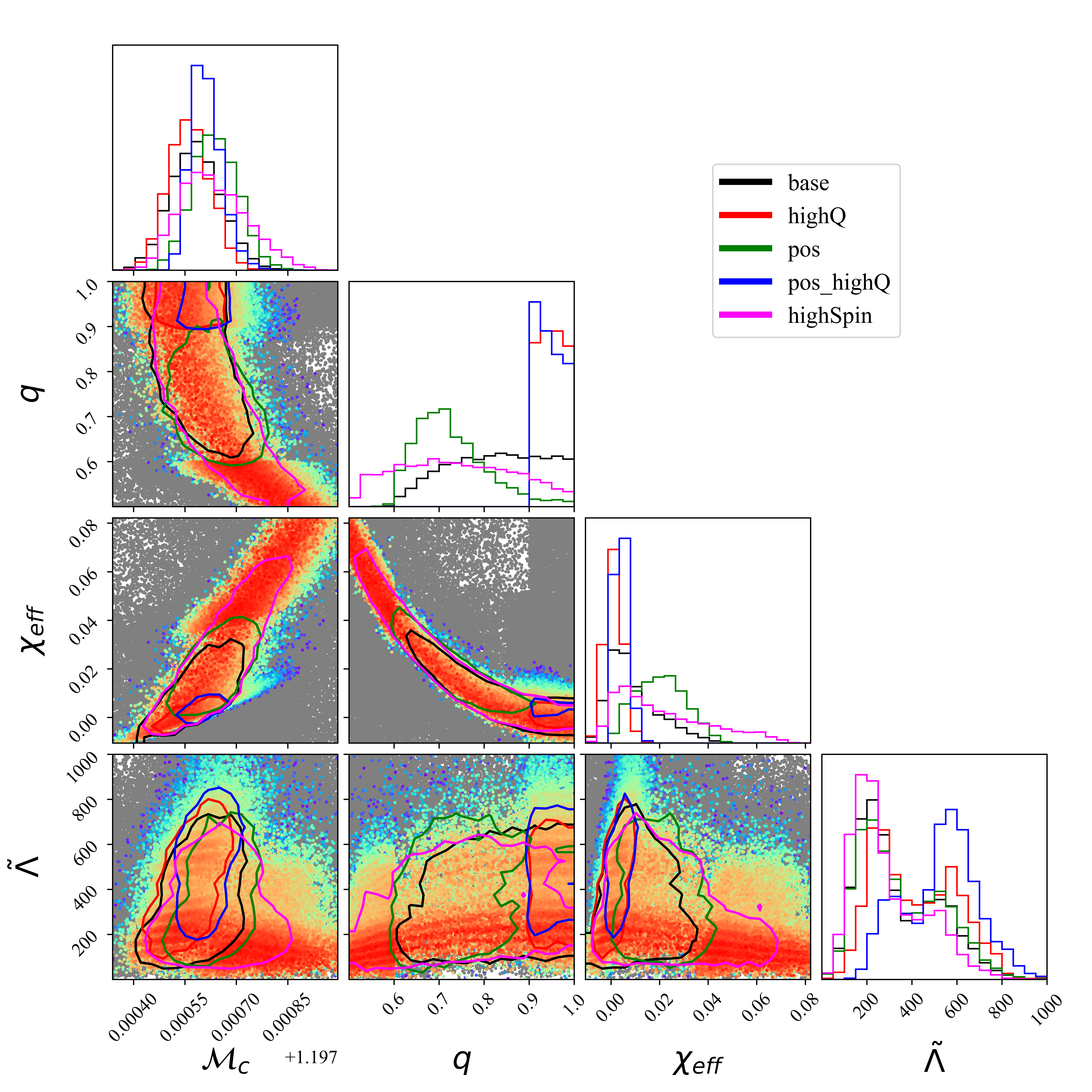}
\caption{\textbf{GW inference} Benchmark inference of properties of GW170817 using \texttt{RIFT} and IMRPhenomPv2\_NRTidalv2.  Color scale shows the
  inferred marginal likelihood, increasing from blue to red over a range of 15 in log-likelihood (with gray indicating
  values more than $15$ below the peak); 2d contours show the 90\% credible intervals; and one-dimensional distributions show the
 one-dimensional marginal posterior distributions.  Parameters shown are the detector-frame chirp mass (denoted here by ${\cal M}_c$), mass ratio $q=m_2/m_1$, aligned effective inspiral spin $\chi_{\rm eff}$, and tidal deformability $\tilde{\Lambda}$.   Results for our fiducial analysis (in black) shown here are very similar to published results available
 through GWOSC
    \citep{LIGO-GW170817-bns,LIGO-GW170817-SourceProperties}.  Colors indicate results derived
    using fiducial priors (black); $q>0.8$ (red);    $\chi_{i,z}>0$ (green); and both (blue). The magenta
    curve shows results derived with a spin prior extending up to $|\chi_i|=0.4$.
}
\label{fig:GW_parameter_posterior}
\end{figure}

Figure \ref{fig:GW_parameter_posterior} shows our results for the joint posterior
in mass, aligned effective inspiral spin $\chi_{\rm eff}$,
and tidal deformability $\tilde{\Lambda}$, showing both the posterior (contours) and marginal likelihood (points), obtained using different prior assumptions about the BNS population.
All marginal likelihoods accumulated during multiple independent analyses of GW170817, performed with different intrinsic priors but with the same data handling and priors over extrinsic  parameters are shown.
Some of these prior assumptions by construction limit the range of spins and $q$ allowed, so sharp edges in posteriors associated with the prior constraints are expected and natural.
For
the fiducial analysis (black), the
marginal distributions of mass, mass ratio, spin, and $\tilde{\Lambda}$ are consistent with previously published
interpretations using comparable models, settings, and data \citep{LIGO-GW170817-bns,LIGO-GW170817-SourceProperties}.
For comparison and to highlight the impact of astrophysically plausible prior assumptions on source parameter inferences, this figure also shows analyses in which we require
either 
$q>0.9$,   $\chi_{i,z}>0$, or  both.  As the likelihood contains strong correlated features, these 
constraints impact the posterior distribution of tidal deformability $\tilde{\Lambda}$. As seen in the $\tilde{\Lambda}-\chi_{\rm eff}$ 2-d posterior, when applying only the $q>0.9$ condition the strong correlation between $\tilde{\Lambda}$ and $\chi_{\rm eff}$ requires negative $\chi_{\rm eff}$ for small $\tilde{\Lambda}$.  As a result, by further requiring that $\chi_i>0$, we eliminate the possibility that $\tilde{\Lambda}$ can be small (blue).
I.e., when applied individually,
these two constraints have modest impact.  However, when imposed simultaneously, even these modest constraints
produce qualitative changes in the  inferred tidal deformability.

\begin{figure}
\includegraphics[trim={0cm 0cm 0cm 1cm},clip,width=\columnwidth]{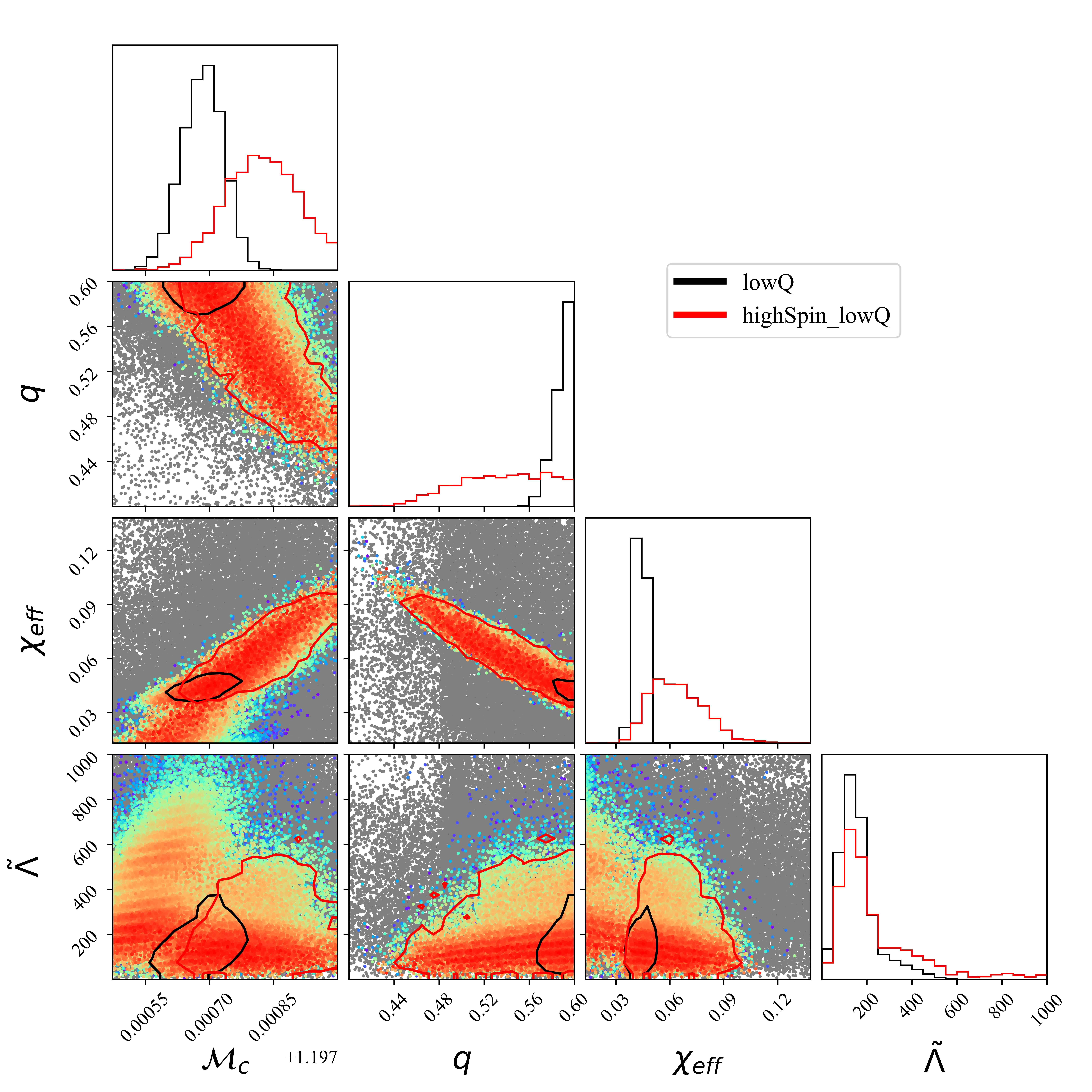}
\caption{\textbf{GW inference assuming high mass ratio} A reanalysis of GW170817 using \texttt{RIFT} and IMRPhenomPv2\_NRTidalv2
  while requiring $m_2/m_1<0.6$   As in Figure \ref{fig:GW_parameter_posterior}, the color scale shows the marginal
  likelihood and lines show one or two-dimensional marginal distributions. The solid black line adopts the fiducial spin
  prior $|\chi_i|<0.05$; the red line adopts a prior of similar form but allows much larger component spins
  $|\chi_i|<0.4$.   Unlike the results presented in Figure \ref{fig:GW_parameter_posterior}, these investigations use a uniform prior on $\chi_{i,z}$,
  rather than the ``z prior'', to ensure comprehensive coverage.   Due to strong correlations between spin and mass ratio, the fiducial spin prior's arbitrary upper limit strongly constrains
  the extent of the mass ratio distribution.
}
\label{fig:GW_parameter_posterior:Aggressive}
\end{figure}

Figures \ref{fig:GW_parameter_posterior} and \ref{fig:GW_parameter_posterior:Aggressive} also illustrate inferences derived using even more extreme prior assumptions
about the pertinent neutron star spin  and mass ratio.  For example, for the magenta curve provided in Figure
\ref{fig:GW_parameter_posterior}, the analysis merely allowed for larger possible values of the NS maximum spin.
Relaxing the upper limit on $\chi_i$  allows a lower mass for $m_2$ to remain consistent with the data
\citep{LIGO-GW170817-SourceProperties}, while increasing $\chi_i$ only incrementally.   Though this single changed assumption allows for more extreme masses, in
isolation this one relaxed assumption doesn't change our qualitative interpretation; notably, the distribution of
$\tilde{\Lambda}$ is nearly unchanged. By contrast, Figure
\ref{fig:GW_parameter_posterior:Aggressive} shows inferences derived assuming $m_2/m_1 < 0.6$, both adopting fiducial
spin magnitudes (black) and larger NS spins (red).  
Notably, the likelihood includes comparable values for a wide range of mass ratios, including masses $m_2<1~\unit{M_\odot}$.
Contrasting the likelihood (points) to the two posterior marginals (black and red) shows how strongly the arbitrary
upper limit in our fiducial \emph{spin prior} limits the mass ratio distribution.  
Both extreme-$q$ analyses nonetheless produce comparable $\tilde{\Lambda}$ posterior distributions.  
Due to  correlations between mass ratio and $\tilde{\Lambda}$, the constraint $q<0.6$ causes the posterior distribution of
$\tilde{\Lambda}$ to systematically decrease.
%
In Appendix \ref{ap:170817}, we show that the median $\tilde{\Lambda}$ steadily decreases versus an imposed
  maximum mass ratio $q$.

\begin{figure}
\includegraphics[width=\columnwidth]{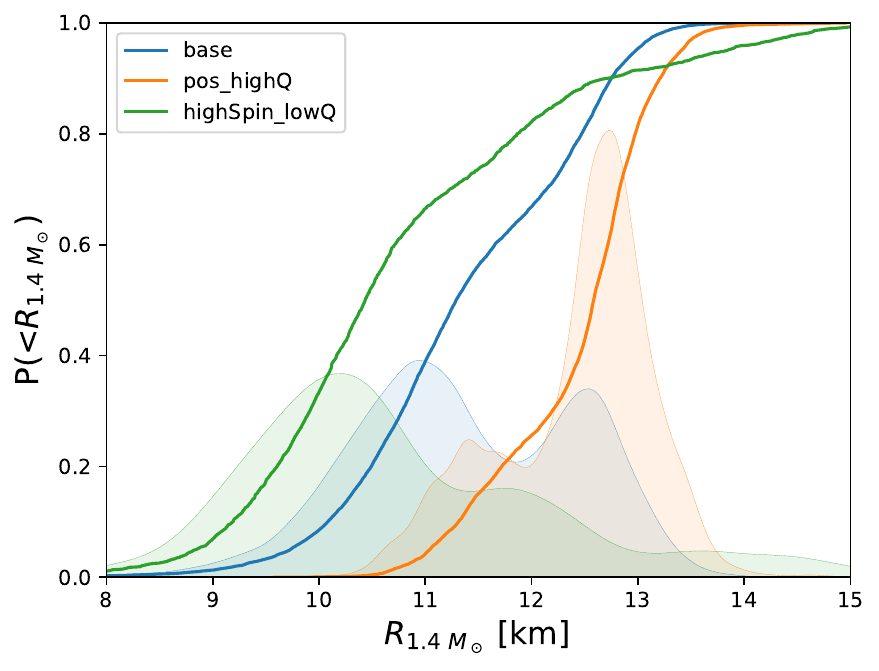}
    \caption{\textbf{GW-derived NS radius} Nominal distribution of $R_{1.4}$ derived from
  Eq. (\ref{eq:ZL}) and three GW-only inferences of GW170817 presented in Figures \ref{fig:GW_parameter_posterior},
  \ref{fig:GW_parameter_posterior:Aggressive}: our fiducial analysis (blue); the comparable mass positive spin analysis (orange); and the
high mass ratio analysis (green). Cumulative distribution is shown in solid lines and probability distribution is shown in lighter shade.
}
\label{fig:170817:R1p4_nominal}
\end{figure}

These GW-only reanalyses suggest that astrophysical priors change $\tilde{\Lambda}$ by roughly a factor of 2, increasing the
median $\tilde{\Lambda}$  (if we require comparable mass and positive aligned spins) or decreasing it (if we require
$q<0.6$) respectively.  However, because of the  exponent relating $R_{1.4}$ and $\tilde{\Lambda}$, see Eq. (\ref{eq:ZL}), these
factors of 2 propagate to a roughly 10\% change in the inferred $R_{1.4}$, comparable to the statistical measurement
errors usually presented for many of the  individual measurements described earlier in Section \ref{sec:sub:NS_obs}.
Figure \ref{fig:170817:R1p4_nominal} shows the cumulative distribution of $R_{1.4}$ derived from this expression and 
three GW-only analyses: our fiducial analysis (blue); the comparable mass positive spin analysis (orange); and the
high mass ratio analysis (green). Note, the bimodality in the probability distributions of the fiducial and the comparable-mass positive spin analyses arise due to bimodality in $\tilde{\Lambda}$ seen in Fig. \ref{fig:GW_parameter_posterior}.

We emphasize this simple discussion presents radius constraints directly in terms of conventional nonphysical GW
priors on $\Lambda_i$, treating each as independent and uniformly distributed.    These statistical errors are not
informed by any EoS model families, nor even by the requirement that both NS have the same EoS. In addition, the jointly-uniform prior also produces a marginal prior distribution on $\tilde{\Lambda}$ that goes to zero as $\tilde{\Lambda}$
(i.e., $p(\tilde{\Lambda})\propto \tilde{\Lambda}$ at small $\tilde{\Lambda}$).   To facilitate visualization and
downstream calculation using posterior samples, other studies have reweighted the GW-only posterior distribution for
example to be uniform in $\tilde{\Lambda}$ or $R_{1.4}$; see, e.g., Figure 11 of \cite{LIGO-GW170817-SourceProperties}. These
two priors differ substantially: using Eq. (\ref{eq:ZL}), a uniform prior in $\tilde{\Lambda}$ (i.e., $P(<\Lambda)=\Lambda/\Lambda_{max}$) corresponds to a
cumulative prior distribution that strongly favors the largest radii (i.e.,  $P(<R_{1.4})\propto R_{1.4}^6$).  As our calculations
employ a directly-interpolated marginal likelihood $L(X)$, not a smoothed posterior, we do not produce such an intermediate data
product.    
Finally, even though the various model-agnostic GW priors used to construct the posterior distributions shown in Figs. \ref{fig:GW_parameter_posterior},
\ref{fig:GW_parameter_posterior:Aggressive} and \ref{fig:170817:R1p4_nominal} strongly disfavor extremely small NS radii, the GW
likelihood itself does not: very small  $\Lambda_i$ are nearly as consistent with GW170817 observations as
$\Lambda_i\simeq O(100)$.

\subsubsection{Joint EOS inference using GW170817}
\label{sec:sub:sub:GWanalysis_joint}
\begin{figure*}
\includegraphics[trim={0.5cm 0cm 2cm 1cm},clip,width=\columnwidth]{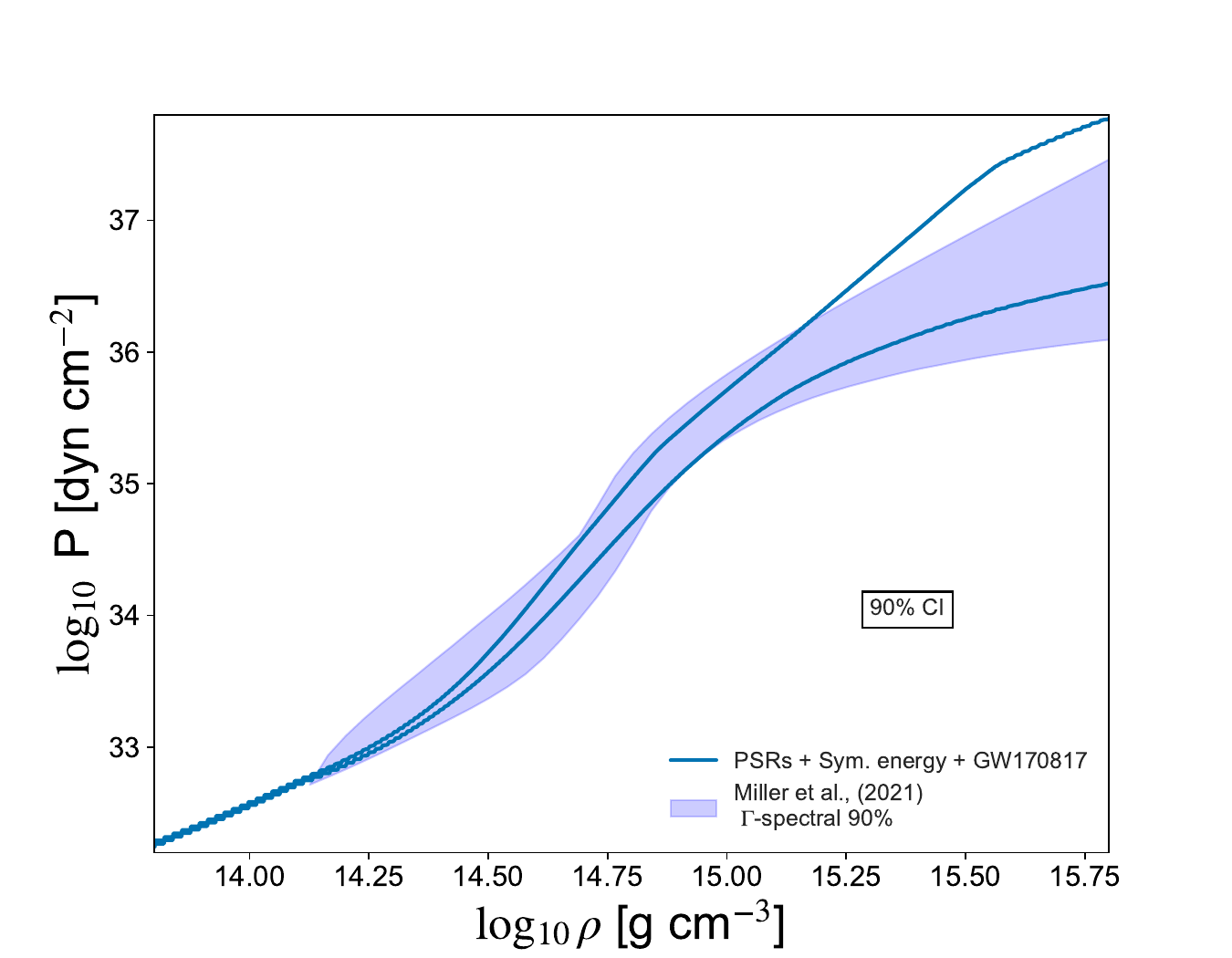}
\includegraphics[trim={0.5cm 0cm 2cm 1cm},clip,width=\columnwidth]{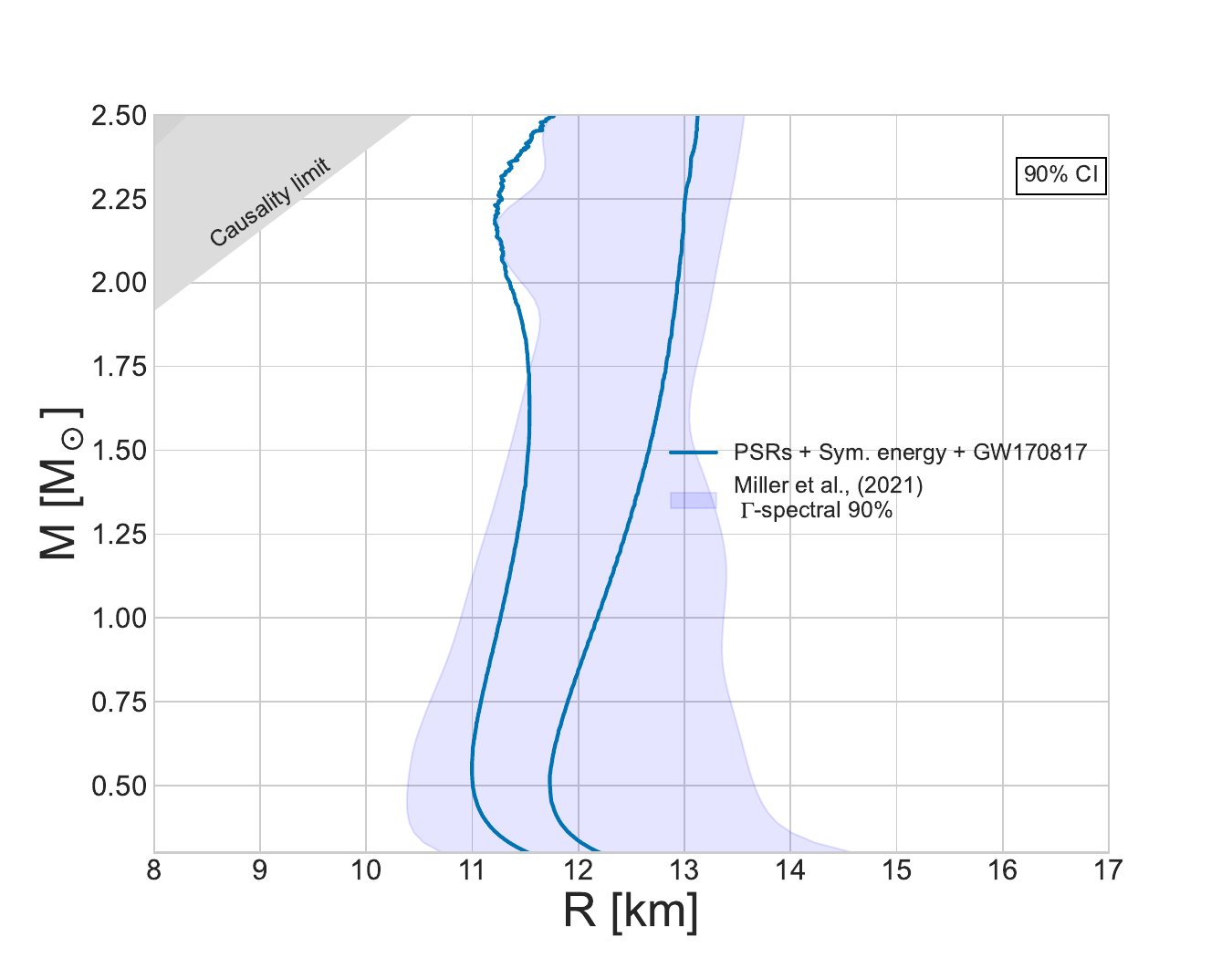}
\caption{\label{fig:joint_gw_psr}\textbf{Joint EoS constraint from galactic  pulsars, symmetry energy, and GW} using the
  $\Gamma$-spectral EoS parameterization. We also provide a comparison analysis which adopts the
  same PSRs, symmetry energy, 
  includes GW170817, and uses the same $\Gamma$-spectral parameterization for these constraints
  \citep{Miller:2021qha}. The calculations differ in their implicit prior along each EOS (our analysis is
    uniform in mass; the comparison analysis is quadratic in central density, strongly disfavoring low masses and therefore broadening their inherently two-dimensional posterior there); the  manner in which GW
    information is factored in to the analysis (RIFT in our analysis versus a KDE in the comparison, and we omit GW190425); and the inference technique used to sample the space of spectral EOS. 
    }
\end{figure*}

Gravitational wave measurements, currently favoring smaller radii, are by our prior's construction more informative than
NICER radius constraints.   As noted in Figure \ref{fig:p_rho_m_r_prior} and the associated discussion, our EoS family and prior enforce a reasonably high radius for low-mass NS
leading, through regularity of the mass-radius curve and a need for large maximum mass, to EoS models that favor
comparable or larger radii for typical NS. For context, 
comparing Figures \ref{fig:demo_gaussian} and \ref{fig:p_rho_m_r_prior}, our
prior favors NS radii largely consistent with the radius of J0030, including a wide range between $9~\unit{km}$ and
$16~\unit{km}$ for a fiducial $1.4~\unit{M_\odot}$ NS.  By contrast, as surveyed in Figure
\ref{fig:170817:R1p4_nominal}, GW observations of GW170817 support typically smaller radii for a similar fiducial NS, with little support above $13~\unit{km}$
in a direct interpretation of our fiducial GW analyses.
As a result and as seen in previous studies, a joint analysis incorporating both galactic pulsars and GW observations
finds fiducial pulsar radii comparable to the intersection of these two limits.  Figure \ref{fig:joint_gw_psr}
shows our fiducial calculation incorporating galactic  pulsar measurements, GW170817 (with our fiducial prior), and the
symmetry energy, following \citep{Miller:2021qha}.   Using similar parameterizations and observational constraints
as a previous study, we find comparable inferences for the nuclear equation of state \citep{Miller:2021qha}.  Some 
differences between our inferences and prior work are expected for technical reasons.   For example, our
  inference technique adaptively estimates the posterior distribution, carefully assessing convergence and coverage with
  an effective sample size. Also, we present 90\%
quantiles in radius at each fixed NS mass, while previous and other works adopt a NS central density prior distribution;
see Section \ref{sec:discussion} for further discussion.

\begin{figure}
\includegraphics[width=0.95\columnwidth]{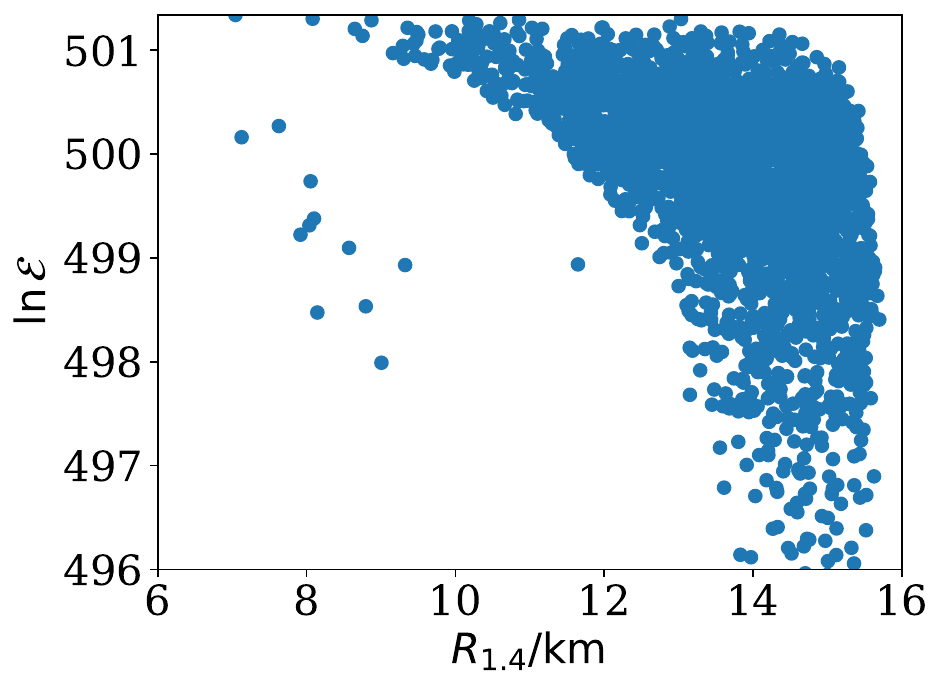}
\includegraphics[width=0.95\columnwidth]{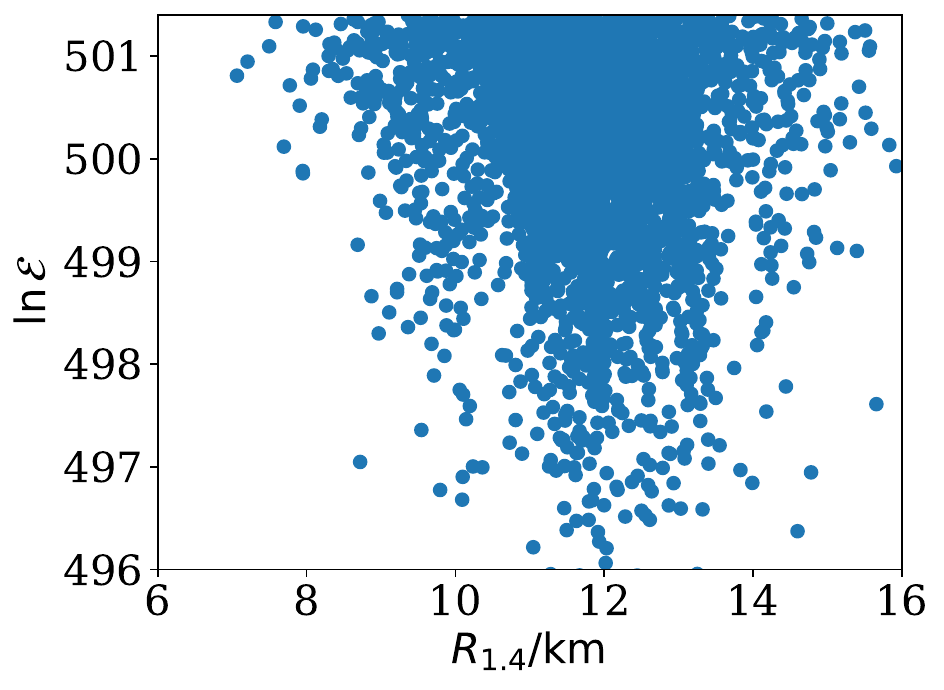}\\
\includegraphics[width=0.95\columnwidth]{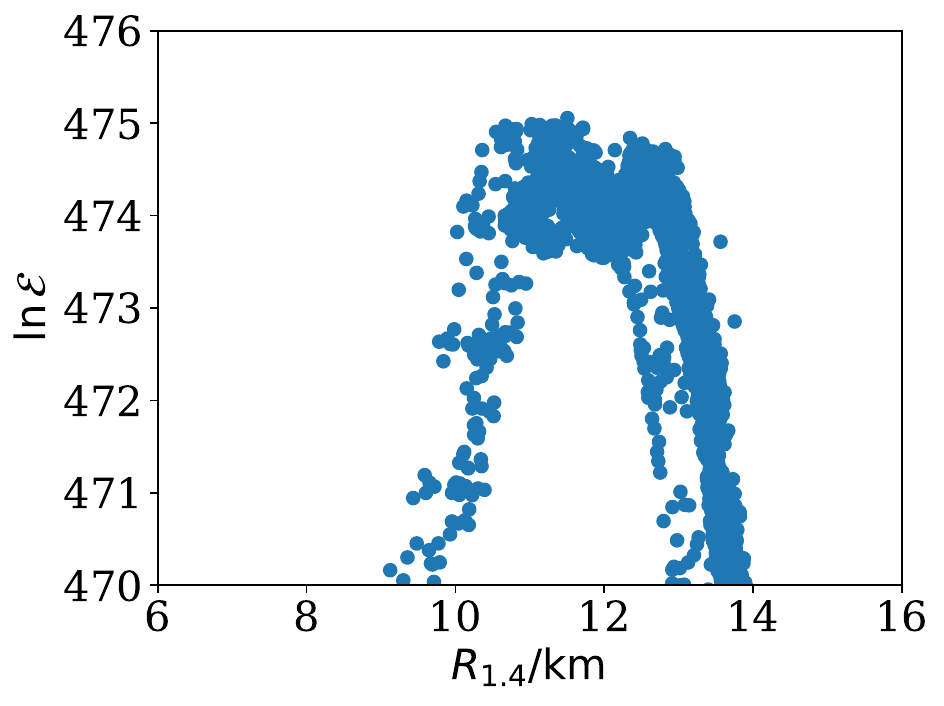}
\caption{\textbf{GW constraints for different astrophysical priors} Scatterplot of $\ln {\cal E}$ versus $R_{1.4}$
  evaluated for the same three  astrophysical priors for NS binaries illustrated in Figure
  \ref{fig:170817:R1p4_nominal}.  As in that figure, this representation uses evaluations drawn from the $\Gamma$-spectral EoS.  
\emph{Top panel}: Likelihood for the fiducial spin and mass ratio priors (\texttt{base}).
\emph{Center panel}: Likelihood for the high-mass ratio  prior (\texttt{highSpin\_lowQ}).
\emph{Bottom panel}: Likelihood for the comparable-mass and positive-$\chi_{i,z}$ prior (\texttt{pos\_highQ}).
\label{fig:GW:lnL_R1p4}
}
\end{figure}

That said, our GW analyses also demonstrate that astrophysical priors have a substantial impact on the
inferred NS radius.  To characterize this effect, Figure \ref{fig:GW:lnL_R1p4}  shows a one-dimensional scatterplot of
$\ln{\cal E}$ versus $R_{1.4}$ using the three different GW analyses presented in Figure \ref{fig:170817:R1p4_nominal}.
These three analyses draw on identical \texttt{RIFT} marginal likelihood data, differing only in the assumed
binary NS population model for mass ratio and spin.
We see that the GW signal places an upper bound on the NS radius, and no strong lower bound (\texttt{base}). 
High mass ratio case (\texttt{highSpin\_lowQ}) more strongly and more frequently penalizes higher NS radii, though some configurations with large radii are still highly likely; the overall likelihood shifts to the lowest radii.
The comparable masses and positive $\chi_{i,z}$ (\texttt{pos\_highQ}) strongly disfavours small radii and is bound on the higher radii showing most of the likelihood in a narrow range of radii on the higher end.
This affirms (directly within the context of EoS inference)
that astrophysical priors embedded in the GW-only analysis propagate directly into different conclusions
about the nuclear EoS.

\begin{figure*}
\centering
\includegraphics[trim={0.5cm 0cm 2cm 2cm},clip,width=\columnwidth]{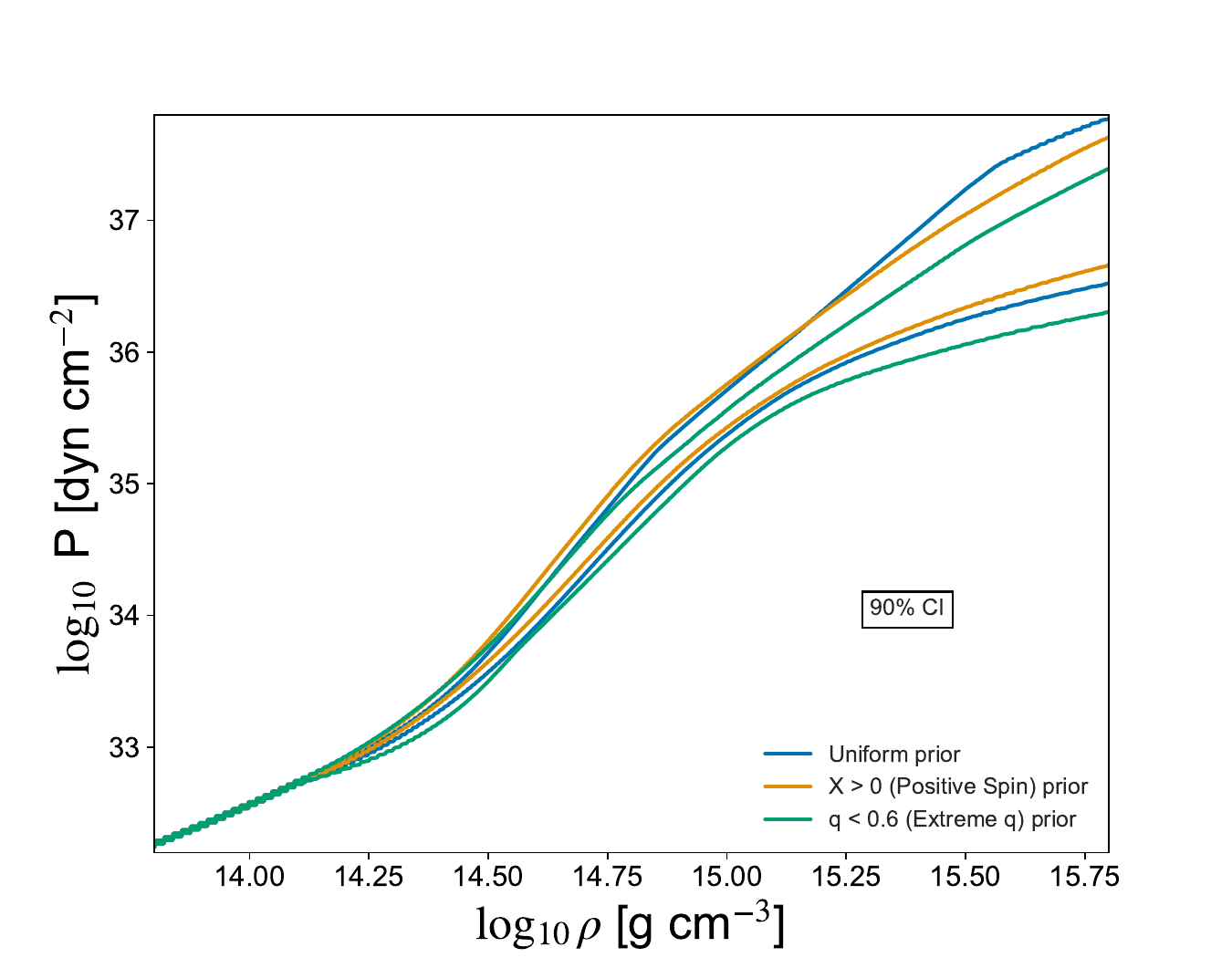}
\includegraphics[trim={0.5cm 0cm 2cm 1cm},clip,width=\columnwidth]{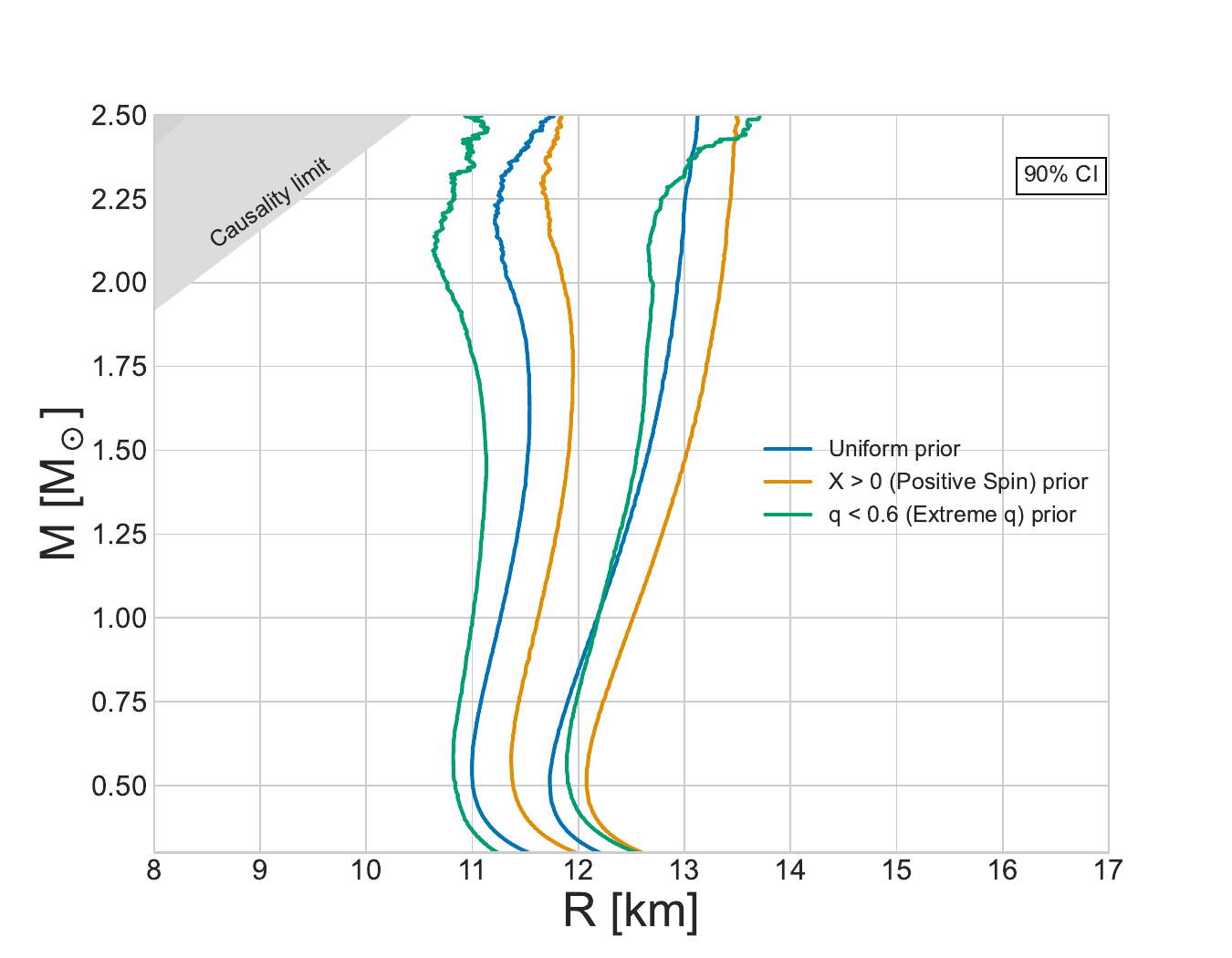}
\includegraphics[width=\columnwidth]{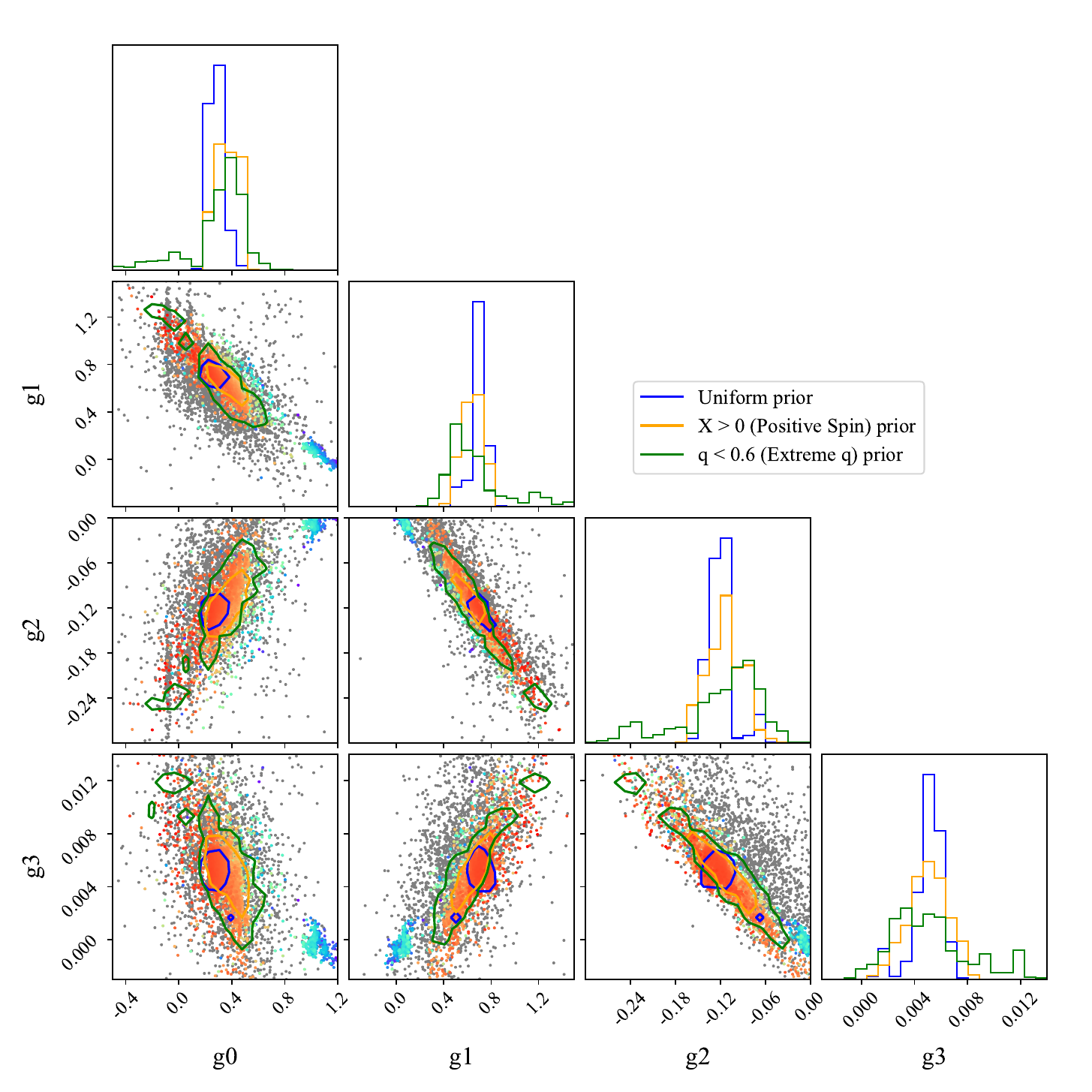}
    \caption{\label{fig:joint_gw_psr:ExtremeQ}
    Joint EoS constraint from GW, galactic  pulsars, and symmetry energy,
derived using $\Gamma$-spectral EoS parameterization, using three different prior assumptions about GW170817.  The blue
contours show results derived using the conventional 170817 prior (uniform masses); the orange requires $\chi_{i,z}>0$;
and the green requires $q<0.6$. The analyses shown in blue have also been presented in  Fig. \ref{fig:joint_gw_psr}.
Top-left panel:  90\% percentile pointwise confidence intervals for $p(\rho)$.
    Top-right panel: 90\% percentile pointwise confidence interval for radius at each mass.
    Bottom panel: $\Gamma$-spectral parameter posterior distribution; we show marginal likelihood
    evaluations accumulated for the fiducial analysis (corresponding to the blue contours).
}
\end{figure*}

The pulsar-only likelihoods represented in  Figures \ref{fig:demo_gaussian} and \ref{fig:lnL_1d} and the (multiple)
GW-only likelihoods characterized by Figure \ref{fig:GW:lnL_R1p4} show that the pulsar and GW measurements may or may not agree, with the amount of
tension depending strongly on astrophysical priors.  In one scenario,  the pulsar/GW results may be in modest tension, consistent with previously
presented work.  In another scenario, the two may be in good agreement, favoring radii of $13~\unit{km}$, if we assume GW170817 is similar to galactic
pulsars.  Finally, the two approaches could have strong tension, with HESS/GW favoring small radii and NICER favoring
large radii, if we assume the kilonova associated with GW170817 favors asymmetric NS binaries.
As an example, Figure \ref{fig:joint_gw_psr:ExtremeQ} shows an analysis comparable to Figure
\ref{fig:joint_gw_psr}, excepting only the mass ratio prior ($q<0.6$) applied for GW170817.  Despite adopting the same
raw GW observation and inference strategy, the same galactic pulsar measurements, and the same symmetry energy, in this
scenario we would infer a notably smaller  radius for  most pulsars, compared to previous work.

\subsection{On the maximum neutron star mass and pertinent information that limits it}
\label{sec:results:mmax_limit}

\begin{figure}
\includegraphics[width=\columnwidth]{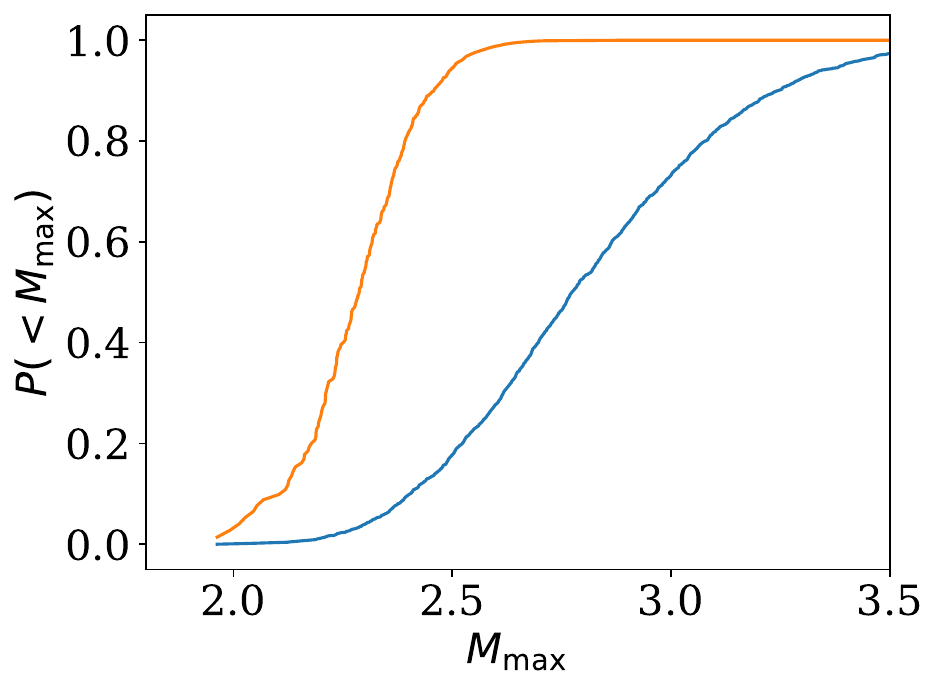}
\includegraphics[trim={0.5cm 0cm 1cm 1cm},clip,width=\columnwidth]{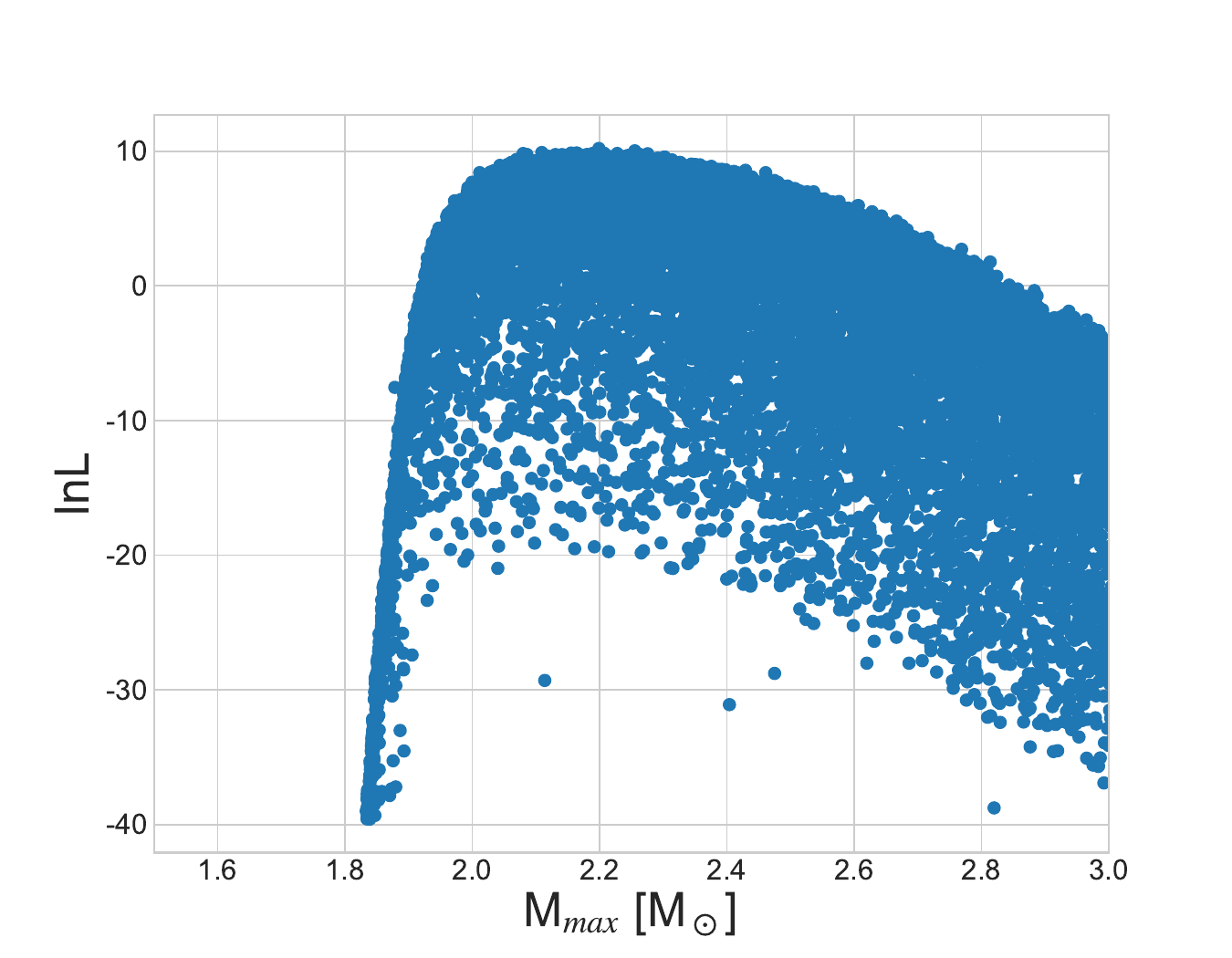}
\caption{\label{fig:Mmax_posterior}\textbf{Impact of maximum nonrotating NS mass}: \emph{Top panel}: Posterior cumulative distribution for maximum nonrotating NS mass: The fraction of EoS in our
  posterior with $M_{\rm max}$ less than a given threshold, versus that threshold.  The blue curve shows the result for
  our fiducial analysis with the $\Gamma$-spectral EoS with galactic pulsars and the symmetry energy, as shown in Figure
  \ref{fig:PSR_only}.  The orange curve shows the corresponding result after applying a nominal likelihood designed to
  disfavor large maximum masses, as shown in the bottom panel of Figure \ref{fig:lnL_1d}. 
\emph{Bottom panel}: Marginal likelihood for different EoS, versus Mmax.  These likelihood evaluations correspond
to the orange curve in the top panel.
}
\end{figure}

Despite being quite restricted in form and range, our two headline equation of state families allow for EoS with very
large maximum neutron star mass $M_{max}$; see, e.g., Figure  \ref{fig:Mmax_posterior}.  
The observations of galactic
pulsars described above do not identify an upper bound on the neutron star maximum mass, explicitly or implicitly; see,
e.g.,  \cite{2018ApJ...852L..25R} and references therein.
Within our framework, this degeneracy manifests as a broad plateau of the likelihood that covers EoS hyperparameters
including such extreme scenarios, as required by this observational constraint; see Figure \ref{fig:lnL_1d}.
In our framework these EoS usually exhibit extreme behavior at high density such as their $c_s\rightarrow 1$.

Several investigations have proposed theoretically or observationally motivated limits on the EoS which would disfavor
these high-$M_{max}$ EoS.  In the most extreme approach, \cite{2018ApJ...852L..25R} suggests that AT2017gfo's
energetics require a prompt collapse to form a black hole, limiting the neutron star maximum mass to be below
$2.16^{+0.17}_{-0.15}~\unit{M_\odot}$.  The bottom panel of Figure \ref{fig:lnL_1d} shows the specific likelihood we adopt to
impose this constraint, following previous work \citep{Dietrich:2020efo}.

\begin{figure}
\includegraphics[trim={0.5cm 0cm 1cm 1cm},clip,width=\columnwidth]{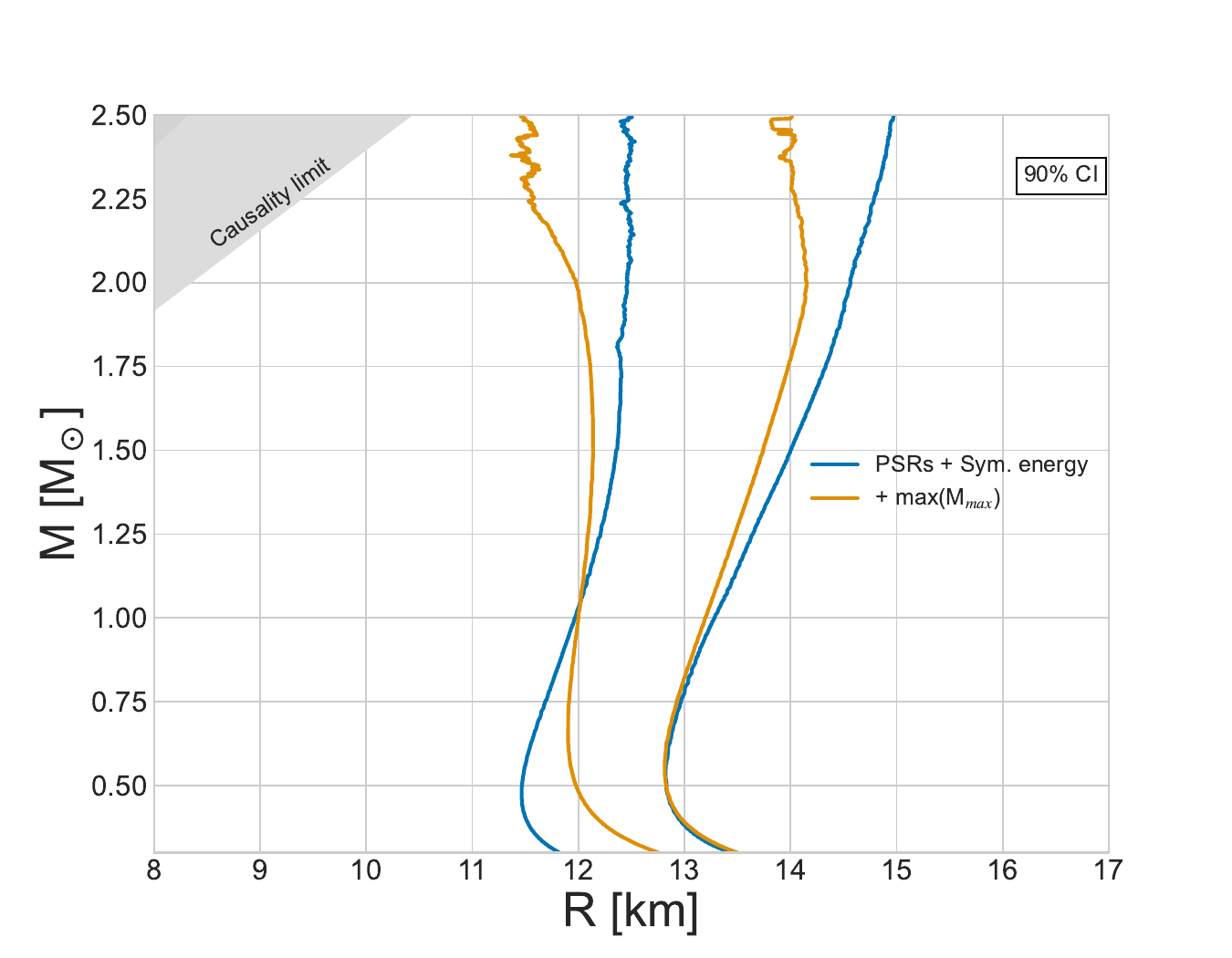}
\includegraphics[trim={0.5cm 0cm 1cm 1cm},clip,width=\columnwidth]{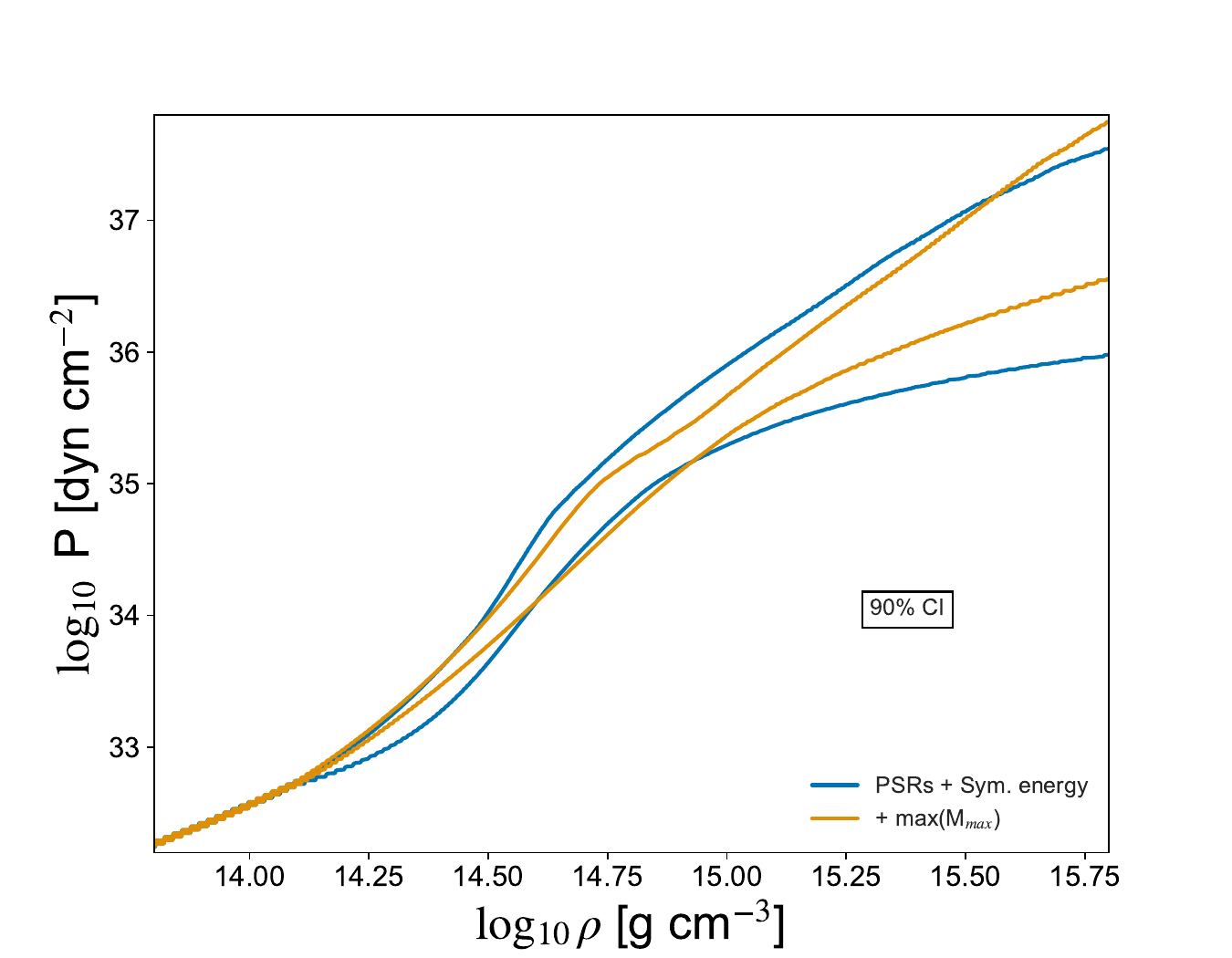}
\includegraphics[trim={0.5cm 0cm 1cm 1cm},clip,width=\columnwidth]{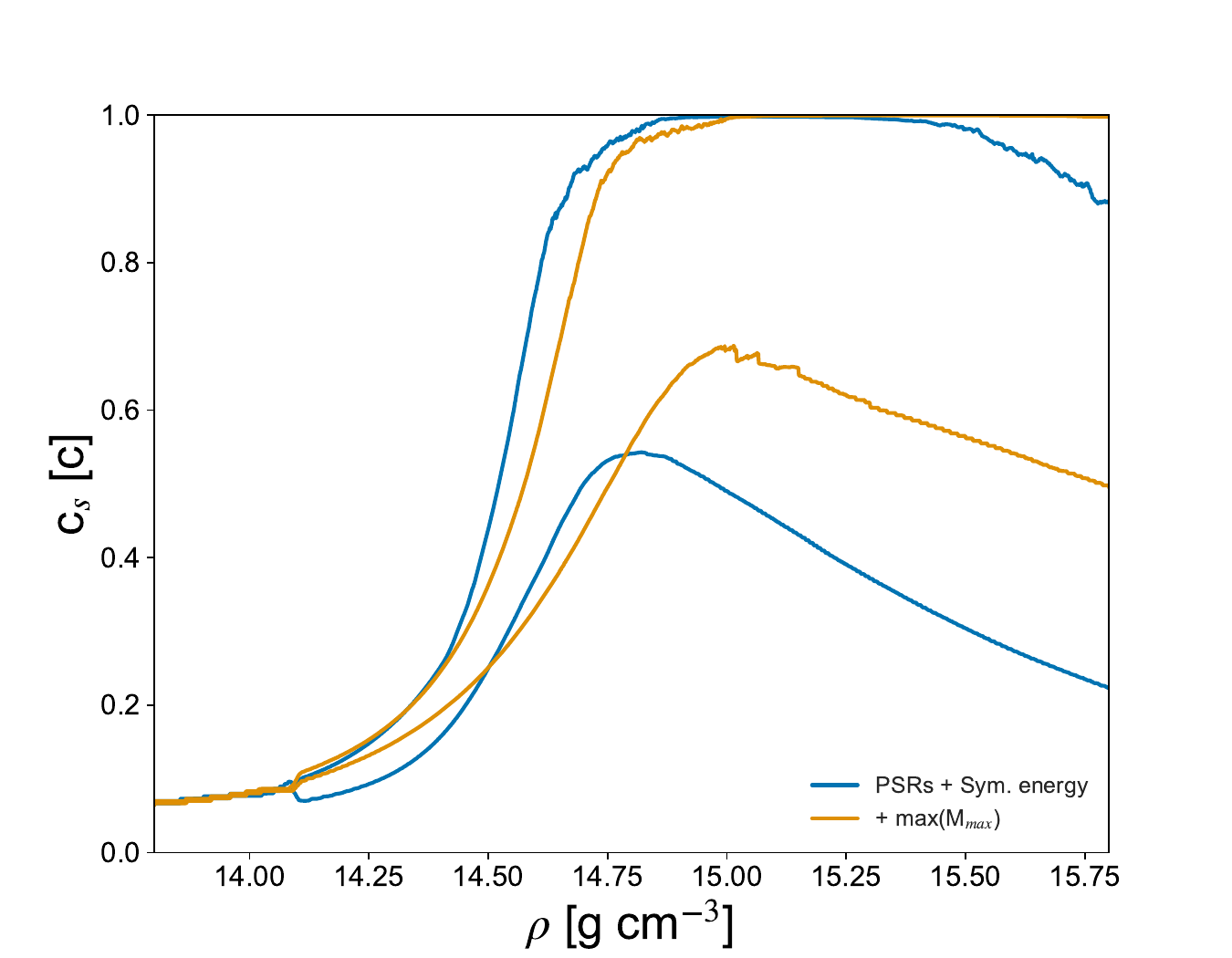}
    \centering
    \caption{\textbf{Impact of maximum mass upper limit}: Like Figure \ref{fig:PSR_only} -- an analysis of galactic
      pulsar observations and the symmetry energy using the
      $\Gamma$-spectral EoS --  but adding the upper limit on
      maximum mass proposed in  \cite{2018ApJ...852L..25R}. } 
    \label{fig:Mmax_posteriors}
\end{figure}

To illustrate the impact of this suggestion, Figure \ref{fig:Mmax_posteriors} shows updated results from our fiducial
NICER analysis, now incorporating this proposed upper limit.  As desired, these constraints disfavor large $M_{max}$.
That said, taken with our other constraints, our analysis still allows for the possibility that $M_{\rm max}$ is larger than the median estimate
proposed by \cite{2018ApJ...852L..25R} using this prior information alone, and conversely still disfavors the
smallest maximum masses allowed a priori due to the strongly implied minimums on $M_{max}$ through observed NSs.

We notice that there is a shift of the M-R curve at large masses towards smaller radii which indicates that favored EoSs have a M-R that bends towards the causality limit boundary at the top left, and hence implies restrictions on EoSs. Note that these boundaries are obtained by evaluating the 90\% percentile credible intervals at each mass, and hence will always have a posterior at high masses even though the EoSs with highest likelihoods do not reach such higher mass values.

As a critical side effect, these constraints on $M_{max}$ also inform the EoS over its full dynamic range, shifting
and narrowing our posterior distributions in $p(\rho)$ and $c_s(\rho)$, particularly at densities well above nuclear
saturation density.
Further, the soundspeed distribution narrows to higher speeds at high densities in order to compensate for the reduced stiffness at intermediate densities with lower sound speeds and lower pressures to provide with larger pressures that produce stable cores for the substantially compactified star. This results in an M-R that reach Masses of up to $\sim2~\unit{M_\odot}$ but does not surpass it substantially.

\subsection{Adding the HESS source}
\label{sec:results:hess_j1731}
\begin{figure}
\includegraphics[trim={0.5cm 0cm 1cm 1cm},clip,width=\columnwidth]{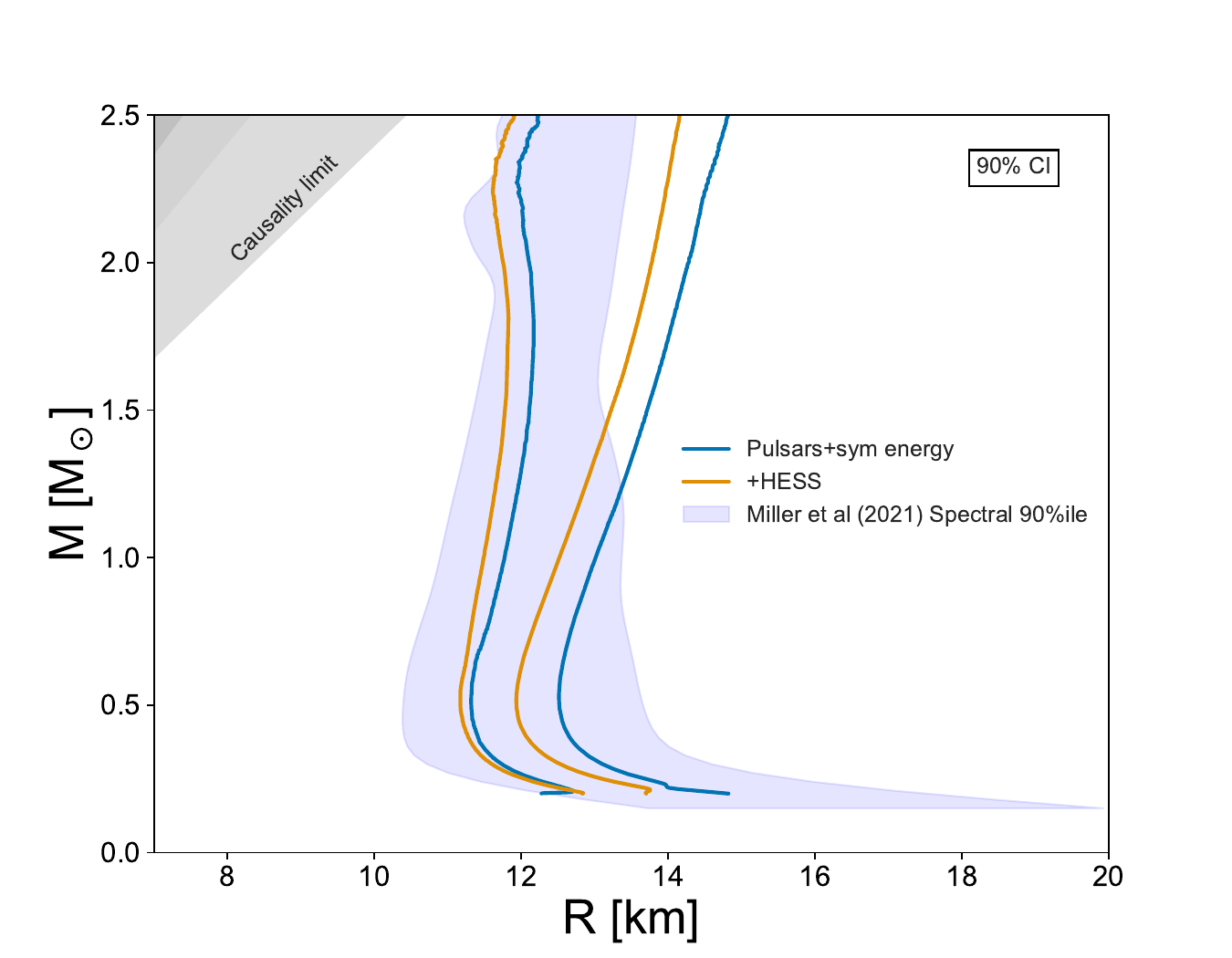}
\caption{\textbf{Incremental impact of HESS source}: Like Figure \ref{fig:PSR_only}, but adding the HESS source. Except
  for a modest change at small radius, the overall EoS  is only incrementally changed when the HESS source is included,
  largely at low mass. A comparison analysis with the same PSRs and symmetry energy and uses the same $\Gamma$-spectral parameterization (also includes GW constraint) is also provided \citep{Miller:2021qha}.
\label{fig:add_HESS}
}
\end{figure}

In the main text, we have emphasized EoS inferences using conventional galactic pulsars, explicitly omitting the  HESS source J1731
\citep{2022NatAs...6.1444D} shown in Figure \ref{fig:demo_gaussian}.  The inferred radius of this source is
  subject to strong systematic uncertainties \citep{2023ApJ...944...36A}. Motivated by previous investigations to assess the
impact of this event on EoS inference
\citep{2023ApJ...958...49S,2023PhLB..84438062L,2023PhRvC.108b5806B,2023A&A...672L..11H,2023ApJ...949...11J,PhysRevD.109.063017}, with Figure \ref{fig:add_HESS} we show the results of EoS inference which explicitly include this
event, compared with our previous results for galactic pulsars only.   Despite the nominal low maximum-likelihood mass and radius deduced for this
event, when accounting for the full marginal likelihood we find GW measurements in particular are very consistent with
the HESS likelihood. As a result, adding the HESS event produces only incremental changes in the EoS, both as shown here
and when adding in GW and other information.  In particular,
the HESS observation does not seem to require exceptional EoS choices.

\section{Discussion}
\label{sec:discussion}

\noindent\emph{Novel results and techniques}: Our most novel results systematically demonstrate that the
  assumed population model for GW170817 will substantially impact the inferred EOS, relative to the effect of an
  additional pulsar observation -- even an extreme observation like  HESS J1731.  However, in this work we make
  several additional technical contributions not previously presented elsewhere. First and foremost, we introduced the
  hyperpipeline technique and provided a distributed and ready-to-use implementation for iterative simulation-based
  inference with arbitrary user-supplied executables. Next and more specific to EOS inference, building on previous work
  \citep{2020arXiv200101747W}, we systematically
  demonstrate that astrophysical assumptions about the NS population model are implicitly present whenever using NS
  observations to constrain the nuclear EOS;  how to insert population assumptions
  self-consistently into calculations of EOS inference; and that the relevant astrophysical population priors generally
  depend on the formation channel (and observational mechanism) for the observation of interest and not held in common.  Third, we perform population inference accounting for both the
  mass \emph{and spin} distribution.  (Doing otherwise in effect assumes that the mass and spin priors adopted during initial GW
  parameter inference perfectly reflects the underlying astrophysical population.)  Addtionally, in our Appendix, we
  demonstrate the impact of alternative GW parameter inferences and model systematics.  Finally, in a Appendix \ref{sec:ap:fisher}, we
  investigate how alternative mass and distance distributions impact inferences derived from multiple observations, using a novel
  Fisher-matrix based approach to characterize the population implications of many measurements.

\noindent \emph{Likelihood model}: In this work, we adopt non-constant single-observation likelihoods and marginal
likelihoods, adopting consistent probabilistic Bayesian models for the measurement uncertainty associated with each
observation.  By contrast, as discussed in \citet{2023ApJ...949...11J}, many other investigations instead
adopt a likelihood which is constant within some allowed domain and zero elsewhere. Their analysis demonstrates that
these two approaches can produce surprisingly similar results when applied to the low-dimensional parametric models they
employ.

\noindent \emph{How does our isolated NS model differ from others?}
For context, other groups have adopted a wide range of different implicit or explicit priors for the neutron star mass
distribution.
Other priors will inherit different priori mass distributions from their prior on the assumed
parameterization $s$.
For example, some adopt priors on the neutron star central density, not the neutron star mass  (e.g., a
quadratic prior \citep{PhysRevD.110.123009,Miller:2021qha} or a log-uniform prior \citep{2019MNRAS.485.5363G} between the
maximum central density and some reference density).   
 If instead one follows \cite{PhysRevD.110.123009} and chooses $s$ to be the central
density and $p(\rho_c)$ to be some prior on that central density, then the implied mass distribution is proportional to
$p(\rho_c) /dM/d\rho_c$.  Equivalently, the cumulative distribution function for any individual equation of state satisfies
$P(<\rho_c) = P(<M(\rho_c))$, for simplicity assuming a single-branched relationship between mass and central density.

Alternatively, even within the context of nominally uniform mass distributions, others \citep{2020PhRvD.101l3007L} have specifically assumed
$p(m|{\cal Y})$ is uniform between a fixed minimum mass and the EoS-dependent maximum TOV mass $M_{max}({\cal Y})$, where
${\cal Y}$ here and henceforth denotes the EoS or its parameterization. In other words, this model asserts all NSs that can exist under an EoS do exist.   In terms of the unit step function $\Theta_I(x)=1$ for
$x\in I$ ($I$ being the interval) and 0 otherwise, their one- and two-dimensional priors are 
$p(m)=\Theta_I(m)/|I|$  and $p(m_1,m_2) = 2 \Theta_I(m_1)\Theta_I(m_2)\Theta(m_1-m_2)/|I|^2$, where
$I=[m_{min},M_{max}({\cal Y})]$ is the interval between $m_{min}$ and the EoS-dependent maximum mass.  
Compared to our fiducial approach, both of these aggressive  population models introduce  EoS-dependent normalization factors, which weakly penalizes models with large $M_{\rm
  max}({\cal Y})$ for single observations and strongly penalize them for many observations.  Additionally, these aggressive population models should be strongly constrained by the absence of
massive neutron stars; see, e.g., \citet{Alsing:2017bbc,2020PhRvD.102f4063C} and references therein.
Though a joint analysis was not performed in this work, 
\citet{2020arXiv200101747W} demonstrated that, if not known a priori, the neutron star mass
distribution and EoS must be jointly inferred, to avoid biasing the deduced EoS.  

{\changed \noindent\emph{Implicit mass priors and the EOS posterior}: Many other studies present mass-radius
 distributions nominally for the nuclear EOS alone.  
}
These discussions of
the NS mass-radius distribution adopt priors \emph{along} the family of stable NS by, for example, adopting a prior on the NS
central density; see, e.g., \cite{2019MNRAS.485.5363G,PhysRevD.110.123009,Miller:2021qha}.  These
analyses are in effect adopting a prior on the NS mass distribution.  As a result, in order to replicate their results,
we must present our results in a comparable fashion.  Specifically, rather than construct a 90\% credible interval for
the NS radius at each fixed mass, we must instead construct a two-dimensional marginal distribution in both mass and
radius.  This two-dimensional posterior distribution will explicitly depend on the choice of NS mass distribution (or
equivalently NS central density distribution).

Rather than invoke an implicit mass distribution, we perform Bayesian inference on the EOS posterior alone and
  represent it accordingly. Bayesian inference frameworks provide a likelihood and hence (with a prior) probability for each equation of state
function.  We can therefore interpret observations as providing information about a credible interval on the
space of functions: for example, deducing a volume of EoS which contain  90\%  of all (joint) posterior probability. Figure
\ref{fig:p_rho_m_r_prior} shows our 100 percentile prior EoS distribution. This
interpretation, however, conflicts with the customary approach to rendering and reporting conclusions about the EoS,
which are presented as \emph{marginal distributions for} (uncorrelated) \emph{point measurements}: for example, the distribution of pressure at a given density, or the
radius of NS of a fiducial mass.  As highlighted in previous work \citep{Landry:2020vaw}, the superposition of point
measurements neglects important information including correlations between different densities (for $p(\rho)$) or masses
(for $R(M)$).   As a concrete example for any two densities $\rho_i$, the posterior pressures $\ln p(\rho_1),\ln p(\rho_2)$ will
 be correlated by measurements, above and beyond any correlations introduced by our prior.
Nonetheless, in this work to simplify comparison with previously-reported results, except for Figure
\ref{fig:p_rho_m_r_prior} we render our conclusions
using this point-measurement approach.  Thus for example a confidence interval on $p(\rho)$ follows by taking all weighted EoS
realizations $(p_k(\rho),w_k)$ and, for each reference density $\rho_i$, constructing a confidence interval for
$p(\rho_i)$ using the weighted sample at that density.  Similarly but not identical are confidence intervals on
$\rho(p)$, constructed similarly but using the inverse relationship.
%

%

\noindent \emph{Assumptions disguise sources of tension}:
The individual underlying inferences involve many often subtle steps which disguise their implicit and explicit
priors, particularly when combining very
heterogeneous sources of external information.    Unsurprisingly and closely related,  looking more closely at individual measurements and their systematics suggests
tension between these different observational approaches to the EoS. For example, several studies have pointed out that
the amount and nature of material ejected from a merger like GW170817 is inconsistent with the matter needed to
reproduce the details of the kilonova emission
\citep{2021ApJ...906...98N,2020ARNPS..7013120R,2021ApJ...910..116K,PhysRevResearch.5.013168,Ristic22,2025arXiv250312320R}.
%
Conversely, several investigations have wholeheartedly embraced combining all available measurements to produce overall
estimates, despite the incremental impact most of these measurements have on overall inferences, owing to large
statistical measurement errors (let alone 
poorly understood systematics); see, e.g., the discussion in \citep{Dietrich:2020efo}.
Finally, as highlighted by previous work \citep{2017ApJ...844..156R,2021PhRvL.126f1101A,2021ApJ...908..103R,2022PhRvD.105d3016L}, the relative impact of all measurements derives strongly
from their explicit prior assumptions, both on the small and large scale.  For example, compared to theoretically well-motivated
priors  \citep{2021ARNPS..71..403D,2021ARNPS..71..433L}, adopting an extremely conservative phenomenological approach with much wider priors enables
 individual measurements to strongly rule out individual realizations (e.g., \cite{PhysRevD.110.123009}).   Conversely, a too-tightly-parameterized approach could conceivably miss subtle
degeneracies associated with putative phase transitions; see, e.g., \cite{2023PhRvL.130t1403R} and references therein.

\noindent \emph{Unexpected features in source parameter inference with mass ratio constraints}:
Practitioners of GW source parameter inference often worry that posteriors which exhibit upper or lower limits other
than that parameters' natural limit (e.g., $m_2/m_1 \in [0,1]$)  may reflect user error.  By construction, however, our
source parameter inferences for GW170817 conditioned on our two alternative prior distributions (b) and (c) described in
Section \ref{sec:sub:prior_binary_gw_pop} have either upper or lower limits on $m_2/m_1$ imposed by our adopted prior.
In other words, the sharp edges in the  posterior distributions seen in Figure \ref{fig:GW_parameter_posterior} as desired reflect the appropriate prior limits adopted
in each analysis respectively.


\section{Conclusions}
\label{sec:conclusion}


In this study we have investigated how implicit assumptions propagate
into statements about the nuclear equation of state.  
In particular, we have demonstrated that astrophysical
population assumptions about the source responsible for GW170817/AT2017gfo directly propagate into substantially
different conclusions, both about the radius of a fiducial neutron star and consequently the nuclear equation of
state.
In doing so, we have shown how EoS inference requires and explicitly depends on strong assumptions
about each pertinent neutron star population.
The impact of astrophysical population assumptions is of shifting the $R_{1.4~\unit{M_\odot}}$ by $\pm0.5~\unit{km}$, whereas individual observations only shift $R_{1.4~\unit{M_\odot}}$ by $\pm0.1~\unit{km}$.
Using two EoS parameterizations, we show how severely conclusions
about the nuclear EoS can depend on parameterization prior choices.
To illuminate the impact of the prior, the hyperparameter plots indicate marginal EoS likelihoods; in every
analysis, many EoSs with the highest marginal likelihood are not within the hyperparameter posterior, being disfavored by
the hyperparameter prior (i.e. the priors on the two spectral parameterizations).
Using the marginal likelihood to separate the impact of EoS priors, we show many EoS in our two model families are both consistent with current pulsar and
GW observations but support very large  NS masses.  
Accounting for these systematics, we argue that for the first few observational constraints,  which previously
  seemed to establish a consensus about the EoS,  the nuclear equation of state is much less well constrained  than previously reported, particularly by investigations which stack multiple heterogeneous measurements whose
relative systematics and particular population assumptions are not well understood.

The most surprising new result in our study is the strong dependence of GW-inferred tidal deformability on astrophysical
prior assumptions.    In addition to a fiducial analysis of GW170817, we also analyze this event using two well-motivated  binary NS
population models.  In one model, we require GW170817 be more similar to the expected binary neutron star population,
based on theoretical models for their formation: a comparable mass ratio, and with spins preferentially positively
aligned with the orbital angular momentum.   In the other model,   designed to produce (much) larger amounts of
radioactive ejecta, we require a highly asymmetric binary (potentially rapidly spinning).  These two prior assumptions
notably change the conclusions derived from GW170817 and hence the conclusions derived when combining this with
other events.
Though the latter of our two scenarios seems in tension with the long-lived galactic BNS population, we note that a
rapidly-merging subpopulation could have different population properties, and offers an attractive resolution to the
long-term challenge of forming r-process elements.
Conversely,  if we adopt a conventional galactic BNS origin for 170817 and incorporate all early-era measurements reviewed in this paper, we
conclude $R_{1.4}\simeq 11.8~\unit{km} - 12.9~\unit{km}$.

The $\Upsilon$-spectral model for EoS attains a higher marginal likelihood than $\Gamma$-spectral model does, implying that specific EOS samples produced by the $\Upsilon$-spectral parameterization fits the observational constraint better than $\Gamma$-spectral. This is because $\Upsilon$-spectral is able to produce EoSs with rapid variation in pressure with density increase as can be seen in Fig. \ref{fig:causal}. Other parameterizations such as $\Gamma$-spectral and even piecewise-polytropic do not form EoSs that can show rapid shift in slope due to the specific functional forms, and hence produce EoSs with a weaker fit. The implication of $\Upsilon$-spectral parameterization is that it infers a strong phase-transition-like feature at $\sim 10^{14.5}$ g cm$^{-3}$.

Our conclusions are compatible with other recent reassessments of EoS-inference-adjacent observational and theoretical
systematics \citep{2023ApJ...944....7S}. Notably, recent re-examination of NICER observations of J0030+4051
\citep{2024ApJ...961...62V} unveils a wider range of possible interpretations than
the widely-used unimodal likelihood  employed as input in this work. Our conclusions are also compatible with
a recent galactic pulsar observations of J0437, as summarized in an approximate analysis in Appendix \ref{sec:sub:revised_0437} \citep{2024ApJ...971L..20C}.

We have methodically investigated the impact on EoS inference by several external sources of information:
galactic pulsar measurements of mass, mass and radius, including the HESS measurement of PSR J1731, GW measurements from
GW170817, the nuclear symmetry energy, an upper bound on the NS maximum mass motivated by energetics of AT2017gfo, and
discussed proposed lower bounds on $\tilde{\Lambda}$ also motivated by this event.
Using the (range of the) marginal log-likelihoods involved as a diagnostic of information provided by each measurement
we argue that the primary measurements (galactic pulsars, including the HESS source; GW measurements; and the symmetry
energy) are remarkably consistent for our model family, particularly when allowing for systematic uncertainty in the interpretation of
GW170817 due to astrophysical population assumptions.
This consistency suggests observations presently only coarsely constrain the EoS space, leaving limited opportunity for
more flexible EoS to arrive at qualitatively different conclusions.
Thus, rather than use modest tension between different observations (e.g., fidicual GW versus NICER) as a mechanism to
draw narrower conclusions about the NS radius, which requires well-controlled systematics, we instead argue these
tensions point to wider posterior conclusions than previously reported.  Specifically, taking the union of our credible
intervals between the most optimistic and pessimistic scenarios allowing for both choices for EoS family, we argue
$R_{1.4}$ is between $11~\unit{km}$ and $14~\unit{km}$; see Figures \ref{fig:joint_gw_psr} and \ref{fig:causal}.  This range is
qualitatively consistent with a simple, pessimistic, heuristic interpretation of how systematics limit our observations,
when analyzed jointly and accounting for systematics: individual measurements' maximum likelihood
estimates bracket the available range. Preliminary interpretations of J0614 are qualitatively consistent with our pessimistic conclusion.
We emphasize that these optimistic and pessimistic conclusions are driven in no small part by assumptions:  which events
and interpretations are used, with what EoS family and priors.

In Appendix \ref{sec:ap:fisher}, we outline a simple estimate of how well future GW surveys can constrain the nuclear
equation of state.  This estimate demonstrates that most individual NS observations will directly inform the nuclear
EoS, not just the loudest few.  This estimate also demonstrates that, due to the strong trend of tidal deformability
with radius, the least massive NS should dominate conclusions about the EoS, excepting only that nature produces them
too infrequently.
Additionally, the appendix illustrates how EOS constraints will depend on two key assumptions held fixed in
the main text: the assumption of a fixed, uniform NS mass distribution and the number and distance distribution.

For simplicity, in this analysis we have omitted several pertinent areas of investigations.  We have
notably adopted a single GW waveform model, eliding the effect of waveform modeling systematics; see
Appendix \ref{ap:170817} for an analysis using TEOBResumS showing that changing the waveform used for EOS inference has a small effect on baseline conclusions, in contrast to changes in the prior assumptions used.  Of course, GW waveform models are only calibrated over a certain range of mass ratio, spins, and tides; for the purposes of illustration, at least one of our calculations (the use of high-mass-ratio binaries) stretches the regime of validity of the NRTidalV2 approximation. That said, our headline result -- that NS population priors can strongly impact EOS inference, by changing the interpretation of GW170817 -- also follows from another example, using comparable masses and positive spins, well within the calibration regime of the NRTidal method.  We therefore anticipate the effect size of NS population priors to remain large even as GW waveform models improve. Also, given substantial
modeling systematics, we have also only adopted toy models for the information provided from kilonova observations which
limit the possible BNS mass, spin, and EoS.  We have taken all existing fiducial galactic pulsar measurements at face
value; reanalyses of these observations may point to larger systematics and broader posteriors than applied here \citep{2024ApJ...961...62V}.
Finally, we have adopted an intentionally simple EoS model family and prior, omitting often-discussed 
constraints available at low density (e.g., from chiral effective field theory and from state-of-the-art crust models) and at high density (e.g, that
high-density EoS should asymptotically attain the perturbative-QCD limit of $c_s\rightarrow 1/\sqrt{3}$ approaching from
below \citep{2010PhRvD..81j5021K, 2020NatPh..16..907A}).  Our choice of these simplifications enables us to focus
exclusively on the often strong and impact that astrophysical priors produce.

The multifactor iterative inference algorithm employed in this work is available as the \texttt{HyperPipe} component of
the \texttt{RIFT} python software
package, available at \url{https://pypi.org/project/RIFT/}.  Instructions for this approach are available at
\citep{rift-docs}.
Appendix \ref{ap:hyperpipe} illustrates how the hyperpipeline accumulates information iteratively in a manner
similar to \texttt{RIFT}.

\begin{acknowledgements}
We thank Jocelyn Read for constructive feedback during the development of this work.
A.K. thanks Wolfgang Kastaun for help with using \texttt{RePrimAnd}. We thank the NICER collaboration, in particular
M. Coleman Miller, for providing us with data for certain figures. A.K. acknowledges support from National Science Foundation (NSF)
Grant No. AST-1909534. ROS acknowledges support from NSF Grant No. AST-1909534, NSF Grant No. PHY-2012057, and the
Simons Foundation. A.B.Y. acknowledges support from NSF Grant No. PHY-2012057. 
The EoS parameterizations were developed under my NSF Grant No. PHYS-2011874.
We are grateful for
the computational resources provided by the LIGO Laboratory, supported via National Science Foundation Grants PHY-0757058 and PHY-0823459.
This material is based upon work supported by NSF’s LIGO Laboratory which is a major facility fully funded by the
National Science Foundation.
The code settings needed to reproduce and validate our results are available on \url{https://pypi.org/project/RIFT/} and \url{https://github.com/oshaughn/research-projects-RIT/}.
\end{acknowledgements}

\bibliographystyle{aa}
\bibliography{bibliography,LIGO-publications,gw-astronomy-mergers-ns-gw170817,gw-astronomy-mergers-approximations,references}

@ARTICLE{LIGO-GW170817-bns,
   author = {{The LIGO Scientific Collaboration} and {the Virgo Collaboration} and 
	{Abbott}, B.~P. and {Abbott}, R. and {Abbott}, T.~D. and {Acernese}, F. and 
	{Ackley}, K. and {Adams}, C. and {Adams}, T. and {Addesso}, P. and et al.},
    title = "{GW170817: Observation of gravitational waves from a binary neutron star inspiral}",
   journal = {\prl},
   year=2017,
  volume = {119},
  pages = {161101},
   month=oct,
  doi = {10.1103/PhysRevLett.119.161101}
}

@ARTICLE{LIGO-GW170817-SourceProperties,
  bibcode={2019PhRvX...9a1001A},
   author = {{The LIGO Scientific Collaboration} and {the Virgo Collaboration} and 
	{Abbott}, B.~P. and {Abbott}, R. and {Abbott}, T.~D. and {Acernese}, F. and 
	{Ackley}, K. and {Adams}, C. and {Adams}, T. and {Addesso}, P. and et al.},
  title = "{Properties of the Binary Neutron Star Merger GW170817}",
  journal = {\prx},
archivePrefix = "arXiv",
   eprint = {1805.11579},
 primaryClass = "gr-qc",
     year = 2019,
    month = jan,
   volume = 9,
   number = 1,
      eid = {011001},
    pages = {011001},
      doi = {10.1103/PhysRevX.9.011001},
   adsurl = {https://ui.adsabs.harvard.edu/abs/2019PhRvX...9a1001A},
  eprint-url={https://arxiv.org/abs/1805.11579},
  dcc-url={https://dcc.ligo.org/LIGO-P1800061}
}

@ARTICLE{LIGO-GW170817-EOS,
   author = {{The LIGO Scientific Collaboration} and {the Virgo Collaboration} and 
	{Abbott}, B.~P. and {Abbott}, R. and {Abbott}, T.~D. and {Acernese}, F. and 
	{Ackley}, K. and {Adams}, C. and {Adams}, T. and {Addesso}, P. and et al.},
    title = "{GW170817: Measurements of neutron star radii and equation of state}",
   year=2018,
   month=oct,
   volume=121,
  pages=161101,
  journal ={\prl},
  eprint-url={https://arxiv.org/abs/1805.11581},
  dcc-url={https://dcc.ligo.org/LIGO-P1800115},
  doi={10.1103/PhysRevLett.121.161101}
}

@ARTICLE{LIGO-O2-Catalog,
   author = {{The LIGO Scientific Collaboration} and {The Virgo Collaboration} and 
	{Abbott}, B.~P. and {Abbott}, R. and {Abbott}, T.~D. and {Acernese}, F. and 
	{Ackley}, K. and {Adams}, C. and {Adams}, T. and {Addesso}, P. and et al.},
  title ="{GWTC-1: A Gravitational-Wave Transient Catalog of Compact Binary Mergers Observed by LIGO and Virgo during the First and Second Observing Runs}",
  journal = {\prx},
  year = 2019,
    month = jul,
   volume = 9,
   number = 3,
      eid = {031040},
    pages = {031040},
      doi = {10.1103/PhysRevX.9.031040},
   adsurl = {https://ui.adsabs.harvard.edu/abs/2019PhRvX...9c1040A},
  eprint-url = {https://arxiv.org/abs/1811.12907},
  dcc-url = {https://dcc.ligo.org/LIGO-P1800307},
  journal-url={https://journals.aps.org/prx/accepted/5c07bK1cM7211b02c3bb33c4baf76d8481781e2e0}
}

@misc{lalsuite,
       author         = "{LIGO Scientific Collaboration}",
       title          = "{LIGO} {A}lgorithm {L}ibrary - {LALS}uite",
       howpublished   = "free software (GPL)",
       doi            = "10.7935/GT1W-FZ16",
       year           = "2018"
}

@ARTICLE{LIGO-GW170817-EOSrank,
  bibcode={2019arXiv190801012T},
  bibcode2={2020CQGra..37d5006A},
     author = {{The LIGO Scientific Collaboration} and {the Virgo Collaboration} and 
	{Abbott}, B.~P. and {Abbott}, R. and {Abbott}, T.~D. and {et al}},
  title = "{Model comparison from LIGO-Virgo data on GW170817's binary components and consequences for the merger remnant}",
      journal = {\cqg},
         year = 2020,
        month = feb,
       volume = {37},
       number = {4},
          eid = {045006},
        pages = {045006},
          doi = {10.1088/1361-6382/ab5f7c},
archivePrefix = {arXiv},
       eprint = {1908.01012},
 primaryClass = {gr-qc},
       adsurl = {https://ui.adsabs.harvard.edu/abs/2020CQGra..37d5006A}
}

@ARTICLE{ligo-O1O2-opendata,
  bibcode={2019arXiv191211716T},
 bibcode2={2021SoftX..1300658A},
       author = {{Abbott}, R. and {Abbott}, T.~D. and {Abraham}, S. and {Acernese}, F. and {Ackley}, K. and {Adams}, C. and {Adhikari}, R.~X. and {Adya}, V.~B. and {Affeldt}, C. and {Agathos}, M. and et al.},
        title = "{Open data from the first and second observing runs of Advanced LIGO and Advanced Virgo}",
      journal = {SoftwareX},
         year = 2021,
        month = jan,
       volume = {13},
          eid = {100658},
        pages = {100658},
          doi = {10.1016/j.softx.2021.100658},
archivePrefix = {arXiv},
       eprint = {1912.11716},
 primaryClass = {gr-qc},
       adsurl = {https://ui.adsabs.harvard.edu/abs/2021SoftX..1300658A}
}

@ARTICLE{2020ApJ...892L...3A,
       author = {{Abbott}, B.~P. and {Abbott}, R. and {Abbott}, T.~D. and {Abraham}, S. and
         {Acernese}, F. and {Ackley}, K. and {Adams}, C. and {Adhikari}, R.~X. and
         {Adya}, V.~B. and {Affeldt}, C. and et al.},
        title = "{GW190425: Observation of a Compact Binary Coalescence with Total Mass {\ensuremath{\sim}} 3.4 M$_{\odot}$}",
      journal = {\apjl},
         year = 2020,
        month = mar,
       volume = {892},
       number = {1},
          eid = {L3},
        pages = {L3},
          doi = {10.3847/2041-8213/ab75f5},
archivePrefix = {arXiv},
       eprint = {2001.01761},
 primaryClass = {astro-ph.HE},
       adsurl = {https://ui.adsabs.harvard.edu/abs/2020ApJ...892L...3A}
}

@ARTICLE{LIGO-O3-NSBH,
   bibcode={2021ApJ...915L...5A},
       author = {{Abbott}, R. and {Abbott}, T.~D. and {Abraham}, S. and {Acernese}, F. and {Ackley}, K. and {Adams}, A. and {Adams}, C. and {Adhikari}, R.~X. and {Adya}, V.~B. and {Affeldt}, C. and et al.},
         title = "{Observation of Gravitational Waves from Two Neutron Star-Black Hole Coalescences}",
      journal = {\apjl},
         year = 2021,
        month = jul,
       volume = {915},
       number = {1},
          eid = {L5},
        pages = {L5},
          doi = {10.3847/2041-8213/ac082e},
archivePrefix = {arXiv},
       eprint = {2106.15163},
 primaryClass = {astro-ph.HE},
       adsurl = {https://ui.adsabs.harvard.edu/abs/2021ApJ...915L...5A}
}

@ARTICLE{LIGO-O3-O3bpop,
   bibcode={2021arXiv211103634T},
 bibcode2={PhysRevX.13.011048},
  bibcode3={2023PhRvX..13a1048A},
       author = {{The LIGO Scientific Collaboration} and {The Virgo Collaboration} and {The KAGRA Scientific Collaboration} and {Abbott}, R. and {Abbott}, T.~D. and {Acernese}, F. and {Ackley}, K. and {Adams}, C. and {Adhikari}, N. and {Adhikari}, R.~X. and et al.},
        title = "{The population of merging compact binaries inferred using gravitational waves through GWTC-3}",
  collaboration = {LIGO Scientific Collaboration, Virgo Collaboration, and KAGRA Collaboration},
  journal = {Phys. Rev. X},
  volume = {13},
  issue = {1},
  pages = {011048},
  numpages = {75},
  year = {2023},
  month = {Mar},
  publisher = {American Physical Society},
  doi = {10.1103/PhysRevX.13.011048},
  url = {https://link.aps.org/doi/10.1103/PhysRevX.13.011048}
}

@misc{LIGO-O2-Catalog-PSDRelease,
    title={Parameter estimation sample release for GWTC-1},
    url={https://dcc.ligo.org/LIGO-P1800370/public},
    DOI={10.7935/KSX7-QQ51},
    publisher={LIGO Scientific Collaboration},
    author={LIGO Scientific Collaboration},
    year={2018}
}

@ARTICLE{2018Natur.554..207M,
       author = {{Mooley}, K.~P. and {Nakar}, E. and {Hotokezaka}, K. and {Hallinan}, G. and
         {Corsi}, A. and {Frail}, D.~A. and {Horesh}, A. and {Murphy}, T. and
         {Lenc}, E. and {Kaplan}, D.~L. and {de}, K. and {Dobie}, D. and {Chand
        ra}, P. and {Deller}, A. and {Gottlieb}, O. and {Kasliwal}, M.~M. and
         {Kulkarni}, S.~R. and {Myers}, S.~T. and {Nissanke}, S. and
         {Piran}, T. and {Lynch}, C. and {Bhalerao}, V. and {Bourke}, S. and
         {Bannister}, K.~W. and {Singer}, L.~P.},
        title = "{A mildly relativistic wide-angle outflow in the neutron-star merger event GW170817}",
      journal = {\nat},
         year = 2018,
        month = feb,
       volume = {554},
       number = {7691},
        pages = {207-210},
          doi = {10.1038/nature25452},
archivePrefix = {arXiv},
       eprint = {1711.11573},
 primaryClass = {astro-ph.HE},
       adsurl = {https://ui.adsabs.harvard.edu/abs/2018Natur.554..207M}
}

@ARTICLE{2019NatAs...3..940H,
       author = {{Hotokezaka}, K. and {Nakar}, E. and {Gottlieb}, O. and {Nissanke}, S. and
         {Masuda}, K. and {Hallinan}, G. and {Mooley}, K.~P. and {Deller}, A.~T.},
        title = "{A Hubble constant measurement from superluminal motion of the jet in GW170817}",
      journal = {\nata},
         year = 2019,
        month = jul,
       volume = {3},
        pages = {940-944},
          doi = {10.1038/s41550-019-0820-1},
       adsurl = {https://ui.adsabs.harvard.edu/abs/2019NatAs...3..940H}
}

@ARTICLE{gwastro-PENR-RIFT,
       author = {{Lange}, Jacob and {O'Shaughnessy}, Richard and {Rizzo}, Monica},
        title = "{Rapid and accurate parameter inference for coalescing, precessing compact binaries}",
      journal = {arXiv:1805.10457},
         year = 2018,
        month = may,
          url = {https://arxiv.org/abs/1805.10457},
        dcc-url={https://dcc.ligo.org/LIGO-P1800084}
}

@ARTICLE{gwastro-PENR-RIFT-GPU,
   bibcode={2019PhRvD..99h4026W},
   author = {{Wysocki}, D. and {O'Shaughnessy}, R. and {Lange}, J. and {Fang}, Y.-L.~L.
	},
    title = "{Accelerating parameter inference with graphics processing units}",
  journal = {\prd},
archivePrefix = "arXiv",
   eprint = {1902.04934},
 primaryClass = "astro-ph.IM",
     year = 2019,
    month = apr,
   volume = 99,
   number = 8,
      eid = {084026},
    pages = {084026},
      doi = {10.1103/PhysRevD.99.084026},
   adsurl = {https://ui.adsabs.harvard.edu/abs/2019PhRvD..99h4026W},
  dcc-url={https://dcc.ligo.org/LIGO-P1900036}
}

@ARTICLE{2020ARNPS..7013120R,
       author = {{Radice}, David and {Bernuzzi}, Sebastiano and {Perego}, Albino},
        title = "{The Dynamics of Binary Neutron Star Mergers and GW170817}",
      journal = {\arnps},
         year = 2020,
        month = oct,
       volume = {70},
       number = {1},
          eid = {annurev},
        pages = {annurev},
          doi = {10.1146/annurev-nucl-013120-114541},
       adsurl = {https://ui.adsabs.harvard.edu/abs/2020ARNPS..7013120R}
}

@ARTICLE{2020GReGr..52..109C,
       author = {{Chatziioannou}, Katerina},
        title = "{Neutron-star tidal deformability and equation-of-state constraints}",
      journal = {\grg},
         year = 2020,
        month = nov,
       volume = {52},
       number = {11},
          eid = {109},
        pages = {109},
          doi = {10.1007/s10714-020-02754-3},
}

@ARTICLE{2021ApJ...910..116K,
       author = {{Korobkin}, Oleg and {Wollaeger}, Ryan T. and {Fryer}, Christopher L. and {Hungerford}, Aimee L. and {Rosswog}, Stephan and {Fontes}, Christopher J. and {Mumpower}, Matthew R. and {Chase}, Eve A. and {Even}, Wesley P. and {Miller}, Jonah and {Misch}, G. Wendell and {Lippuner}, Jonas},
        title = "{Axisymmetric Radiative Transfer Models of Kilonovae}",
      journal = {\apj},
         year = 2021,
        month = apr,
       volume = {910},
       number = {2},
          eid = {116},
        pages = {116},
          doi = {10.3847/1538-4357/abe1b5},
       adsurl = {https://ui.adsabs.harvard.edu/abs/2021ApJ...910..116K}
}

@article{PhysRevD.107.024040,
  title = {Improving performance for gravitational-wave parameter inference with an efficient and highly-parallelized algorithm},
  author = {Wofford, J. and Yelikar, A. B. and Gallagher, Hannah and Champion, E. and Wysocki, D. and Delfavero, V. and Lange, J. and Rose, C. and Valsan, V. and Morisaki, S. and Read, J. and Henshaw, C. and O'Shaughnessy, R.},
  journal = {Phys. Rev. D},
  volume = {107},
  issue = {2},
  pages = {024040},
  numpages = {33},
  year = {2023},
  month = {Jan},
  publisher = {American Physical Society},
  doi = {10.1103/PhysRevD.107.024040},
  url = {https://link.aps.org/doi/10.1103/PhysRevD.107.024040}
}

@article{Henkel:2022naw,
    author = "Henkel, Amelia and Foucart, Francois and Raaijmakers, Geert and Nissanke, Samaya",
    title = "{Study of the agreement between binary neutron star ejecta models derived from numerical relativity simulations}",
    eprint = "2207.07658",
    archivePrefix = "arXiv",
    primaryClass = "astro-ph.HE",
    doi = "10.1103/PhysRevD.107.063028",
    journal = "Phys. Rev. D",
    volume = "107",
    number = "6",
    pages = "063028",
    year = "2023"
}

@article{PhysRevResearch.5.013168,
  title = {Surrogate light curve models for kilonovae with comprehensive wind ejecta outflows and parameter estimation for AT2017gfo},
  author = {Kedia, Atul and Ristic, Marko and O'Shaughnessy, Richard and Yelikar, Anjali B. and Wollaeger, Ryan T. and Korobkin, Oleg and Chase, Eve A. and Fryer, Christopher L. and Fontes, Christopher J.},
  journal = {Phys. Rev. Res.},
  volume = {5},
  issue = {1},
  pages = {013168},
  numpages = {15},
  year = {2023},
  month = {Mar},
  publisher = {American Physical Society},
  doi = {10.1103/PhysRevResearch.5.013168},
  url = {https://link.aps.org/doi/10.1103/PhysRevResearch.5.013168}
}

@ARTICLE{2018MNRAS.481.4009V,
       author = {{Vigna-G{\'o}mez}, Alejandro and {Neijssel}, Coenraad J. and {Stevenson}, Simon and {Barrett}, Jim W. and {Belczynski}, Krzysztof and {Justham}, Stephen and {de Mink}, Selma E. and {M{\"u}ller}, Bernhard and {Podsiadlowski}, Philipp and {Renzo}, Mathieu and {Sz{\'e}csi}, Dorottya and {Mandel}, Ilya},
        title = "{On the formation history of Galactic double neutron stars}",
      journal = {\mnras},
         year = 2018,
        month = dec,
       volume = {481},
       number = {3},
        pages = {4009-4029},
          doi = {10.1093/mnras/sty2463},
       adsurl = {https://ui.adsabs.harvard.edu/abs/2018MNRAS.481.4009V}
}

@ARTICLE{2023MNRAS.521.4669Y,
       author = {{Yang}, Y. -Y. and {Zhang}, C. -M. and {Li}, D. and {Chen}, L. and {Zhang}, J. -W. and {Wang}, D. -H. and {Jiang}, L. -Y. and {Cui}, X. -H.},
        title = "{Investigating the distribution of double neutron stars and unconventional component mass}",
      journal = {\mnras},
         year = 2023,
        month = may,
       volume = {521},
       number = {3},
        pages = {4669-4678},
          doi = {10.1093/mnras/stad754},
       adsurl = {https://ui.adsabs.harvard.edu/abs/2023MNRAS.521.4669Y}
}

@ARTICLE{2020PASA...37...38V,
       author = {{Vigna-G{\'o}mez}, Alejandro and {MacLeod}, Morgan and {Neijssel}, Coenraad J. and {Broekgaarden}, Floor S. and {Justham}, Stephen and {Howitt}, George and {de Mink}, Selma E. and {Vinciguerra}, Serena and {Mandel}, Ilya},
        title = "{Common envelope episodes that lead to double neutron star formation}",
      journal = {\pasa},
         year = 2020,
        month = sep,
       volume = {37},
          eid = {e038},
        pages = {e038},
          doi = {10.1017/pasa.2020.31},
archivePrefix = {arXiv},
       eprint = {2001.09829},
 primaryClass = {astro-ph.SR},
       adsurl = {https://ui.adsabs.harvard.edu/abs/2020PASA...37...38V}
}

@ARTICLE{2020ApJ...900L..41A,
       author = {{Andrews}, Jeff J.},
        title = "{Mass Ratios of Merging Double Neutron Stars as Implied by the Milky Way Population}",
      journal = {\apjl},
         year = 2020,
        month = sep,
       volume = {900},
       number = {2},
          eid = {L41},
        pages = {L41},
          doi = {10.3847/2041-8213/abb1bf},
archivePrefix = {arXiv},
       eprint = {2007.06560},
 primaryClass = {astro-ph.HE},
       adsurl = {https://ui.adsabs.harvard.edu/abs/2020ApJ...900L..41A}
}

@ARTICLE{2000ApJ...541..319K,
       author = {{Kalogera}, Vassiliki},
        title = "{Spin-Orbit Misalignment in Close Binaries with Two Compact Objects}",
      journal = {\apj},
         year = 2000,
        month = sep,
       volume = {541},
       number = {1},
        pages = {319-328},
          doi = {10.1086/309400},
archivePrefix = {arXiv},
       eprint = {astro-ph/9911417},
 primaryClass = {astro-ph},
       adsurl = {https://ui.adsabs.harvard.edu/abs/2000ApJ...541..319K}
}

@ARTICLE{2019MNRAS.486.1086M,
       author = {{Mandel}, Ilya and {Farr}, Will M. and {Gair}, Jonathan R.},
        title = "{Extracting distribution parameters from multiple uncertain observations with selection biases}",
      journal = {\mnras},
         year = 2019,
        month = jun,
       volume = {486},
       number = {1},
        pages = {1086-1093},
          doi = {10.1093/mnras/stz896},
archivePrefix = {arXiv},
       eprint = {1809.02063},
 primaryClass = {physics.data-an},
       adsurl = {https://ui.adsabs.harvard.edu/abs/2019MNRAS.486.1086M}
}

@ARTICLE{2019PhRvD.100d3012W,
       author = {{Wysocki}, Daniel and {Lange}, Jacob and {O'Shaughnessy}, Richard},
        title = "{Reconstructing phenomenological distributions of compact binaries via gravitational wave observations}",
      journal = {\prd},
         year = 2019,
        month = aug,
       volume = {100},
       number = {4},
          eid = {043012},
        pages = {043012},
          doi = {10.1103/PhysRevD.100.043012},
archivePrefix = {arXiv},
       eprint = {1805.06442},
 primaryClass = {gr-qc},
       adsurl = {https://ui.adsabs.harvard.edu/abs/2019PhRvD.100d3012W}
}

@ARTICLE{2019RNAAS...3...66F,
       author = {{Farr}, Will M.},
        title = "{Accuracy Requirements for Empirically Measured Selection Functions}",
      journal = {Research Notes of the American Astronomical Society},
         year = 2019,
        month = may,
       volume = {3},
       number = {5},
          eid = {66},
        pages = {66},
          doi = {10.3847/2515-5172/ab1d5f},
archivePrefix = {arXiv},
       eprint = {1904.10879},
 primaryClass = {astro-ph.IM},
       adsurl = {https://ui.adsabs.harvard.edu/abs/2019RNAAS...3...66F}
}

@ARTICLE{2023ApJ...944...36A,
       author = {{Alford}, J.~A.~J. and {Halpern}, J.~P.},
        title = "{Do Central Compact Objects have Carbon Atmospheres?}",
      journal = {\apj},
         year = 2023,
        month = feb,
       volume = {944},
       number = {1},
          eid = {36},
        pages = {36},
          doi = {10.3847/1538-4357/acaf55},
archivePrefix = {arXiv},
       eprint = {2302.05893},
 primaryClass = {astro-ph.HE},
       adsurl = {https://ui.adsabs.harvard.edu/abs/2023ApJ...944...36A}
}

@ARTICLE{2018PhRvD..98j4052N,
       author = {{Nagar}, Alessandro and {Bernuzzi}, Sebastiano and {Del Pozzo}, Walter and {Riemenschneider}, Gunnar and {Akcay}, Sarp and {Carullo}, Gregorio and {Fleig}, Philipp and {Babak}, Stanislav and {Tsang}, Ka Wa and {Colleoni}, Marta and {Messina}, Francesco and {Pratten}, Geraint and {Radice}, David and {Rettegno}, Piero and {Agathos}, Michalis and {Fauchon-Jones}, Edward and {Hannam}, Mark and {Husa}, Sascha and {Dietrich}, Tim and {Cerd{\'a}-Duran}, Pablo and {Font}, Jos{\'e} A. and {Pannarale}, Francesco and {Schmidt}, Patricia and {Damour}, Thibault},
        title = "{Time-domain effective-one-body gravitational waveforms for coalescing compact binaries with nonprecessing spins, tides, and self-spin effects}",
      journal = {\prd},
         year = 2018,
        month = nov,
       volume = {98},
       number = {10},
          eid = {104052},
        pages = {104052},
          doi = {10.1103/PhysRevD.98.104052},
archivePrefix = {arXiv},
       eprint = {1806.01772},
 primaryClass = {gr-qc},
       adsurl = {https://ui.adsabs.harvard.edu/abs/2018PhRvD..98j4052N}
}

@ARTICLE{2019PhRvD.100d4003D,
       author = {{Dietrich}, Tim and {Samajdar}, Anuradha and {Khan}, Sebastian and {Johnson-McDaniel}, Nathan K. and {Dudi}, Reetika and {Tichy}, Wolfgang},
        title = "{Improving the NRTidal model for binary neutron star systems}",
      journal = {\prd},
         year = 2019,
        month = aug,
       volume = {100},
       number = {4},
          eid = {044003},
        pages = {044003},
          doi = {10.1103/PhysRevD.100.044003},
       adsurl = {https://ui.adsabs.harvard.edu/abs/2019PhRvD.100d4003D}
}

@ARTICLE{2019PhRvD..99b4029D,
       author = {{Dietrich}, Tim and {Khan}, Sebastian and {Dudi}, Reetika and {Kapadia}, Shasvath J. and {Kumar}, Prayush and {Nagar}, Alessandro and {Ohme}, Frank and {Pannarale}, Francesco and {Samajdar}, Anuradha and {Bernuzzi}, Sebastiano and {Carullo}, Gregorio and {Del Pozzo}, Walter and {Haney}, Maria and {Markakis}, Charalampos and {P{\"u}rrer}, Michael and {Riemenschneider}, Gunnar and {Setyawati}, Yoshinta Eka and {Tsang}, Ka Wa and {Van Den Broeck}, Chris},
        title = "{Matter imprints in waveform models for neutron star binaries: Tidal and self-spin effects}",
      journal = {\prd},
         year = 2019,
        month = jan,
       volume = {99},
       number = {2},
          eid = {024029},
        pages = {024029},
          doi = {10.1103/PhysRevD.99.024029},
       adsurl = {https://ui.adsabs.harvard.edu/abs/2019PhRvD..99b4029D}
}

@ARTICLE{2020PhRvD.102b4031B,
       author = {{Barkett}, Kevin and {Chen}, Yanbei and {Scheel}, Mark A. and {Varma}, Vijay},
        title = "{Gravitational waveforms of binary neutron star inspirals using post-Newtonian tidal splicing}",
      journal = {\prd},
         year = 2020,
        month = jul,
       volume = {102},
       number = {2},
          eid = {024031},
        pages = {024031},
          doi = {10.1103/PhysRevD.102.024031},
       adsurl = {https://ui.adsabs.harvard.edu/abs/2020PhRvD.102b4031B}
}

@ARTICLE{2018PhRvD..97b1501R,
   author = {{Ruiz}, M. and {Shapiro}, S.~L. and {Tsokaros}, A.},
    title = "{GW170817, general relativistic magnetohydrodynamic simulations, and the neutron star maximum mass}",
  journal = {\prd},
archivePrefix = "arXiv",
   eprint = {1711.00473},
 primaryClass = "astro-ph.HE",
     year = 2018,
    month = jan,
   volume = 97,
   number = 2,
      eid = {021501},
    pages = {021501},
      doi = {10.1103/PhysRevD.97.021501},
   adsurl = {http://adsabs.harvard.edu/abs/2018PhRvD..97b1501R},
}

@ARTICLE{2018ApJ...852L..25R,
   author = {{Rezzolla}, L. and {Most}, E.~R. and {Weih}, L.~R.},
    title = "{Using Gravitational-wave Observations and Quasi-universal Relations to Constrain the Maximum Mass of Neutron Stars}",
  journal = {\apjl},
archivePrefix = "arXiv",
   eprint = {1711.00314},
 primaryClass = "astro-ph.HE",
     year = 2018,
    month = jan,
   volume = 852,
      eid = {L25},
    pages = {L25},
      doi = {10.3847/2041-8213/aaa401},
   adsurl = {http://adsabs.harvard.edu/abs/2018ApJ...852L..25R}
}

@ARTICLE{2018ApJ...852L..29R,
   author = {{Radice}, D. and {Perego}, A. and {Zappa}, F. and {Bernuzzi}, S.
	},
    title = "{GW170817: Joint Constraint on the Neutron Star Equation of State from Multimessenger Observations}",
  journal = {\apjl},
archivePrefix = "arXiv",
   eprint = {1711.03647},
 primaryClass = "astro-ph.HE",
     year = 2018,
    month = jan,
   volume = 852,
      eid = {L29},
    pages = {L29},
      doi = {10.3847/2041-8213/aaa402},
   adsurl = {http://adsabs.harvard.edu/abs/2018ApJ...852L..29R},
}

@ARTICLE{2018ApJ...866...60P,
   author = {{Pankow}, C.},
    title = "{On GW170817 and the Galactic Binary Neutron Star Population}",
  journal = {\apj},
archivePrefix = "arXiv",
   eprint = {1806.05097},
 primaryClass = "astro-ph.HE",
     year = 2018,
    month = oct,
   volume = 866,
      eid = {60},
    pages = {60},
      doi = {10.3847/1538-4357/aadc66},
   adsurl = {http://adsabs.harvard.edu/abs/2018ApJ...866...60P}
}

@ARTICLE{Ristic22,
       author = {{Ristic}, M. and {Champion}, E. and {O'Shaughnessy}, R. and {Wollaeger}, R. and {Korobkin}, O. and {Chase}, E.~A. and {Fryer}, C.~L. and {Hungerford}, A.~L. and {Fontes}, C.~J.},
        title = "Interpolating detailed simulations of kilonovae: Adaptive learning and parameter inference applications",
      journal = {\prr},
         year = 2022,
        month = jan,
       volume = {4},
       number = {1},
          eid = {013046},
        pages = {013046},
          doi = {10.1103/PhysRevResearch.4.013046},
}

@ARTICLE{2021ApJ...906...98N,
       author = {{Nedora}, Vsevolod and {Bernuzzi}, Sebastiano and {Radice}, David and {Daszuta}, Boris and {Endrizzi}, Andrea and {Perego}, Albino and {Prakash}, Aviral and {Safarzadeh}, Mohammadtaher and {Schianchi}, Federico and {Logoteta}, Domenico},
        title = "{Numerical Relativity Simulations of the Neutron Star Merger GW170817: Long-term Remnant Evolutions, Winds, Remnant Disks, and Nucleosynthesis}",
      journal = {\apj},
         year = 2021,
        month = jan,
       volume = {906},
       number = {2},
          eid = {98},
        pages = {98},
          doi = {10.3847/1538-4357/abc9be},
}

@ARTICLE{2025arXiv250312320R,
       author = {{Risti{\'c}}, Marko and {O'Shaughnessy}, Richard and {Wagner}, Kate and {Fontes}, Christopher J. and {Fryer}, Chris L. and {Korobkin}, Oleg and {Mumpower}, Matthew R. and {Wollaeger}, Ryan T.},
        title = "{Joint Electromagnetic and Gravitational Wave Inference of Binary Neutron Star Merger GW170817 Using Forward-Modeling Ejecta Predictions}",
      journal = {arXiv e-prints},
         year = 2025,
        month = mar,
          eid = {arXiv:2503.12320},
        pages = {arXiv:2503.12320},
          doi = {10.48550/arXiv.2503.12320},
archivePrefix = {arXiv},
       eprint = {2503.12320},
 primaryClass = {astro-ph.HE},
       adsurl = {https://ui.adsabs.harvard.edu/abs/2025arXiv250312320R}
}

@article{Landry:2020vaw,
    author = "Landry, Philippe and Essick, Reed and Chatziioannou, Katerina",
    title = "{Nonparametric constraints on neutron star matter with existing and upcoming gravitational wave and pulsar observations}",
    eprint = "2003.04880",
    archivePrefix = "arXiv",
    primaryClass = "astro-ph.HE",
    doi = "10.1103/PhysRevD.101.123007",
    journal = "Phys. Rev. D",
    volume = "101",
    number = "12",
    pages = "123007",
    year = "2020"
}

@article{Landry:2021hvl,
    author = "Landry, Philippe and Read, Jocelyn S.",
    title = "{The Mass Distribution of Neutron Stars in Gravitational-wave Binaries}",
    eprint = "2107.04559",
    archivePrefix = "arXiv",
    primaryClass = "astro-ph.HE",
    doi = "10.3847/2041-8213/ac2f3e",
    journal = "Astrophys. J. Lett.",
    volume = "921",
    number = "2",
    pages = "L25",
    year = "2021"
}

@article{Oppenheimer:1939ne,
    author = "Oppenheimer, J.R. and Volkoff, G.M.",
    title = "{On Massive neutron cores}",
    doi = "10.1103/PhysRev.55.374",
    journal = "Phys. Rev.",
    volume = "55",
    pages = "374--381",
    year = "1939"
}

@article{Tolman:1939jz,
    author = "Tolman, Richard C.",
    title = "{Static solutions of Einstein's field equations for spheres of fluid}",
    doi = "10.1103/PhysRev.55.364",
    journal = "Phys. Rev.",
    volume = "55",
    pages = "364--373",
    year = "1939"
}

@ARTICLE{2020ApJ...893...61Z,
       author = {{Zhang}, Nai-Bo and {Li}, Bao-An},
        title = "{Constraints on the Muon Fraction and Density Profile in Neutron Stars}",
      journal = {\apj},
         year = 2020,
        month = apr,
       volume = {893},
       number = {1},
          eid = {61},
        pages = {61},
          doi = {10.3847/1538-4357/ab7dbc},
archivePrefix = {arXiv},
       eprint = {2002.06446},
 primaryClass = {astro-ph.HE},
       adsurl = {https://ui.adsabs.harvard.edu/abs/2020ApJ...893...61Z},
}

@ARTICLE{2020NatAs...4..625C,
       author = {{Capano}, Collin D. and {Tews}, Ingo and {Brown}, Stephanie M. and {Margalit}, Ben and {De}, Soumi and {Kumar}, Sumit and {Brown}, Duncan A. and {Krishnan}, Badri and {Reddy}, Sanjay},
        title = "{Stringent constraints on neutron-star radii from multimessenger observations and nuclear theory}",
      journal = {Nature Astronomy},
         year = 2020,
        month = mar,
       volume = {4},
        pages = {625-632},
          doi = {10.1038/s41550-020-1014-6},
archivePrefix = {arXiv},
       eprint = {1908.10352},
 primaryClass = {astro-ph.HE},
       adsurl = {https://ui.adsabs.harvard.edu/abs/2020NatAs...4..625C},
}

@article{TheVirgo:2014hva,
      author         = "Acernese, F. and others",
      title          = "{Advanced Virgo: a second-generation interferometric
                        gravitational wave detector}",
      collaboration = {Virgo Collaboration},
      journal        = "Class. Quant. Grav.",
      volume         = "32",
      year           = "2015",
      number         = "2",
      pages          = "024001",
      doi            = "10.1088/0264-9381/32/2/024001",
      eprint         = "1408.3978",
      archivePrefix  = "arXiv",
      primaryClass   = "gr-qc",
      SLACcitation   = "%%CITATION = ARXIV:1408.3978;%%"
}

@article{TheLIGOScientific:2014jea,
      author         = "Aasi, J. and others",
      title          = "{Advanced LIGO}",
      collaboration = {LIGO Scientific Collaboration},
      journal        = "Class. Quant. Grav.",
      volume         = "32",
      year           = "2015",
      pages          = "074001",
      doi            = "10.1088/0264-9381/32/7/074001",
      eprint         = "1411.4547",
      archivePrefix  = "arXiv",
      primaryClass   = "gr-qc",
      SLACcitation   = "%%CITATION = ARXIV:1411.4547;%%"
}

@article{Pang:2021jta,
    author = "Pang, Peter T. H. and Tews, Ingo and Coughlin, Michael W. and Bulla, Mattia and Van Den Broeck, Chris and Dietrich, Tim",
    title = "{Nuclear Physics Multimessenger Astrophysics Constraints on the Neutron Star Equation of State: Adding NICER\textquoteright{}s PSR J0740+6620 Measurement}",
    eprint = "2105.08688",
    archivePrefix = "arXiv",
    primaryClass = "astro-ph.HE",
    reportNumber = "LA-UR-21-20534",
    doi = "10.3847/1538-4357/ac19ab",
    journal = "Astrophys. J.",
    volume = "922",
    number = "1",
    pages = "14",
    year = "2021"
}

@article{Miller:2019cac,
    author = "Miller, M. C. and others",
    title = "{PSR J0030+0451 Mass and Radius from $NICER$ Data and Implications for the Properties of Neutron Star Matter}",
    eprint = "1912.05705",
    archivePrefix = "arXiv",
    primaryClass = "astro-ph.HE",
    doi = "10.3847/2041-8213/ab50c5",
    journal = "Astrophys. J. Lett.",
    volume = "887",
    number = "1",
    pages = "L24",
    year = "2019"
}

@article{Riley:2019yda,
    author = "Riley, Thomas E. and others",
    title = "{A $NICER$ View of PSR J0030+0451: Millisecond Pulsar Parameter Estimation}",
    eprint = "1912.05702",
    archivePrefix = "arXiv",
    primaryClass = "astro-ph.HE",
    doi = "10.3847/2041-8213/ab481c",
    journal = "Astrophys. J. Lett.",
    volume = "887",
    number = "1",
    pages = "L21",
    year = "2019"
}

@article{Raaijmakers:2019qny,
      author         = "Raaijmakers, G. and others",
      title          = "{A NICER view of PSR J0030+0451: Implications for the
                        dense matter equation of state}",
      journal        = "Astrophys. J. Lett.",
      volume         = "887",
      year           = "2019",
      pages          = "L22",
      doi            = "10.3847/2041-8213/ab451a",
      eprint         = "1912.05703",
      archivePrefix  = "arXiv",
      primaryClass   = "astro-ph.HE",
      SLACcitation   = "%%CITATION = ARXIV:1912.05703;%%"
}

@article{Bilous:2019knh,
      author         = "Bilous, Anna V. and others",
      title          = "{A NICER view of PSR J0030+0451: evidence for a
                        global-scale multipolar magnetic field}",
      journal        = "Astrophys. J. Lett.",
      volume         = "887",
      year           = "2019",
      pages          = "L23",
      doi            = "10.3847/2041-8213/ab53e7",
      eprint         = "1912.05704",
      archivePrefix  = "arXiv",
      primaryClass   = "astro-ph.HE",
      SLACcitation   = "%%CITATION = ARXIV:1912.05704;%%"
}

@ARTICLE{2019ApJ...887L..27G,
       author = {{Guillot}, Sebastien and {Kerr}, Matthew and {Ray}, Paul S. and {Bogdanov}, Slavko and {Ransom}, Scott and {Deneva}, Julia S. and {Arzoumanian}, Zaven and {Bult}, Peter and {Chakrabarty}, Deepto and {Gendreau}, Keith C. and {Ho}, Wynn C.~G. and {Jaisawal}, Gaurava K. and {Malacaria}, Christian and {Miller}, M. Coleman and {Strohmayer}, Tod E. and {Wolff}, Michael T. and {Wood}, Kent S. and {Webb}, Natalie A. and {Guillemot}, Lucas and {Cognard}, Ismael and {Theureau}, Gilles},
        title = "{NICER X-Ray Observations of Seven Nearby Rotation-powered Millisecond Pulsars}",
      journal = {\apjl},
         year = 2019,
        month = dec,
       volume = {887},
       number = {1},
          eid = {L27},
        pages = {L27},
          doi = {10.3847/2041-8213/ab511b},
archivePrefix = {arXiv},
       eprint = {1912.05708},
 primaryClass = {astro-ph.HE},
}

@article{Bogdanov:2019ixe,
    author = "Bogdanov, Slavko and others",
    title = "{Constraining the Neutron Star Mass\textendash{}Radius Relation and Dense Matter Equation of State with $NICER$. I. The Millisecond Pulsar X-Ray Data Set}",
    eprint = "1912.05706",
    archivePrefix = "arXiv",
    primaryClass = "astro-ph.HE",
    doi = "10.3847/2041-8213/ab53eb",
    journal = "Astrophys. J. Lett.",
    volume = "887",
    number = "1",
    pages = "L25",
    year = "2019"
}

@ARTICLE{2021ApJ...914L..15B,
       author = {{Bogdanov}, Slavko and {Dittmann}, Alexander J. and {Ho}, Wynn C.~G. and {Lamb}, Frederick K. and {Mahmoodifar}, Simin and {Miller}, M. Coleman and {Morsink}, Sharon M. and {Riley}, Thomas E. and {Strohmayer}, Tod E. and {Watts}, Anna L. and {Choudhury}, Devarshi and {Guillot}, Sebastien and {Harding}, Alice K. and {Ray}, Paul S. and {Wadiasingh}, Zorawar and {Wolff}, Michael T. and {Markwardt}, Craig B. and {Arzoumanian}, Zaven and {Gendreau}, Keith C.},
        title = "{Constraining the Neutron Star Mass-Radius Relation and Dense Matter Equation of State with NICER. III. Model Description and Verification of Parameter Estimation Codes}",
      journal = {\apjl},
         year = 2021,
        month = jun,
       volume = {914},
       number = {1},
          eid = {L15},
        pages = {L15},
          doi = {10.3847/2041-8213/abfb79},
archivePrefix = {arXiv},
       eprint = {2104.06928},
 primaryClass = {astro-ph.HE},
}

@article{Bogdanov:2019qjb,
    author = "Bogdanov, Slavko and others",
    title = "{Constraining the Neutron Star Mass\textendash{}Radius Relation and Dense Matter Equation of State with $NICER$. II. Emission from Hot Spots on a Rapidly Rotating Neutron Star}",
    eprint = "1912.05707",
    archivePrefix = "arXiv",
    primaryClass = "astro-ph.HE",
    doi = "10.3847/2041-8213/ab5968",
    journal = "Astrophys. J. Lett.",
    volume = "887",
    number = "1",
    pages = "L26",
    year = "2019"
}

@article{Miller:2021qha,
    author = "Miller, M. C. and others",
    title = "{The Radius of PSR J0740+6620 from NICER and XMM-Newton Data}",
    eprint = "2105.06979",
    archivePrefix = "arXiv",
    primaryClass = "astro-ph.HE",
    doi = "10.3847/2041-8213/ac089b",
    journal = "Astrophys. J. Lett.",
    volume = "918",
    number = "2",
    pages = "L28",
    year = "2021"
}

@article{Riley:2021pdl,
    author = "Riley, Thomas E. and others",
    title = "{A NICER View of the Massive Pulsar PSR J0740+6620 Informed by Radio Timing and XMM-Newton Spectroscopy}",
    eprint = "2105.06980",
    archivePrefix = "arXiv",
    primaryClass = "astro-ph.HE",
    doi = "10.3847/2041-8213/ac0a81",
    journal = "Astrophys. J. Lett.",
    volume = "918",
    number = "2",
    pages = "L27",
    year = "2021"
}

@ARTICLE{2021ApJ...918L..29R,
       author = {{Raaijmakers}, G. and {Greif}, S.~K. and {Hebeler}, K. and {Hinderer}, T. and {Nissanke}, S. and {Schwenk}, A. and {Riley}, T.~E. and {Watts}, A.~L. and {Lattimer}, J.~M. and {Ho}, W.~C.~G.},
        title = "{Constraints on the Dense Matter Equation of State and Neutron Star Properties from NICER's Mass-Radius Estimate of PSR J0740+6620 and Multimessenger Observations}",
      journal = {\apjl},
         year = 2021,
        month = sep,
       volume = {918},
       number = {2},
          eid = {L29},
        pages = {L29},
          doi = {10.3847/2041-8213/ac089a},
archivePrefix = {arXiv},
       eprint = {2105.06981},
 primaryClass = {astro-ph.HE},
}

@dataset{riley_thomas_e_2021_4697625,
  author       = {Riley, Thomas E. and
                  Watts, Anna L. and
                  Ray, Paul S. and
                  Bogdanov, Slavko and
                  Guillot, Sebastien and
                  Morsink, Sharon M. and
                  Bilous, Anna V. and
                  Arzoumanian, Zaven and
                  Choudhury, Devarshi and
                  Deneva, Julia S. and
                  Gendreau, Keith C. and
                  Harding, Alice K. and
                  Ho, Wynn C.G. and
                  Lattimer, James M. and
                  Loewenstein, Michael and
                  Ludlam, Renee M. and
                  Markwardt, Craig B. and
                  Okajima, Takashi and
                  Prescod-Weinstein, Chanda and
                  Remillard, Ronald A. and
                  Wolff, Michael T. and
                  Fonseca, Emanuel and
                  Cromartie, H. Thankful and
                  Kerr, Matthew and
                  Pennucci, Timothy T. and
                  Parthasarathy, Aditya and
                  Ransom, Scott and
                  Stairs, Ingrid and
                  Guillemot, Lucas and
                  Cognard, Ismael},
  title        = {{A NICER View of the Massive Pulsar PSR J0740+6620 
                   Informed by Radio Timing and XMM-Newton
                   Spectroscopy: Nested Samples for Millisecond
                   Pulsar Parameter Estimation}},
  month        = apr,
  year         = 2021,
  publisher    = {Zenodo},
  version      = {v1.0.0},
  doi          = {10.5281/zenodo.4697625},
  url          = {https://doi.org/10.5281/zenodo.4697625}
}

@dataset{miller_m_c_2021_4670689,
  author       = {Miller, M.C. and
                  Lamb, F. K. and
                  Dittmann, A. J. and
                  Bogdanov, S. and
                  Arzoumanian, Z. and
                  Gendreau, K. C. and
                  Guillot, S. and
                  Ho, W. C. G. and
                  Lattimer, J. M. and
                  Morsink, S. M. and
                  Ray, P. S. and
                  Wolff, M. T. and
                  Baker, C. L. and
                  Cazeau, T. and
                  Manthripragada, S. and
                  Markwardt, C. B. and
                  Okajima, T. and
                  Pollard, S. and
                  Cognard, I. and
                  Cromartie, H. T. and
                  Fonseca, E. and
                  Guillemot, L. and
                  Kerr, M. and
                  Parthasarathy, A. and
                  Pennucci, T. T. and
                  Ransom, S. and
                  Stairs, I. and
                  Loewenstein, M.},
  title        = {{NICER PSR J0740+6620 Illinois-Maryland MCMC 
                   Samples}},
  month        = apr,
  year         = 2021,
  publisher    = {Zenodo},
  doi          = {10.5281/zenodo.4670689},
  url          = {https://doi.org/10.5281/zenodo.4670689}
}

@ARTICLE{2020ApJ...894L...8C,
       author = {{Christian}, Jan-Erik and {Schaffner-Bielich}, J{\"u}rgen},
        title = "{Twin Stars and the Stiffness of the Nuclear Equation of State: Ruling Out Strong Phase Transitions below 1.7 n$_{0}$ with the New NICER Radius Measurements}",
      journal = {\apjl},
         year = 2020,
        month = may,
       volume = {894},
       number = {1},
          eid = {L8},
        pages = {L8},
          doi = {10.3847/2041-8213/ab8af4},
archivePrefix = {arXiv},
       eprint = {1912.09809},
 primaryClass = {astro-ph.HE},
}

@article{Agathos:2015uaa,
      author         = "Agathos, Michalis and Meidam, Jeroen and Del Pozzo,
                        Walter and Li, Tjonnie G. F. and Tompitak, Marco and
                        Veitch, John and Vitale, Salvatore and Van Den Broeck,
                        Chris",
      title          = "{Constraining the neutron star equation of state with
                        gravitational wave signals from coalescing binary neutron
                        stars}",
      journal        = "Phys. Rev. D",
      volume         = "92",
      year           = "2015",
      number         = "2",
      pages          = "023012",
      doi            = "10.1103/PhysRevD.92.023012",
      eprint         = "1503.05405",
      archivePrefix  = "arXiv",
      primaryClass   = "gr-qc",
      SLACcitation   = "%%CITATION = ARXIV:1503.05405;%%"
}

@article{Lackey:2014fwa,
     author         = "Lackey, Benjamin D. and Wade, Leslie",
      title          = "{Reconstructing the neutron-star equation of state with
                        gravitational-wave detectors from a realistic population
                        of inspiralling binary neutron stars}",
      journal        = "Phys. Rev. D",
      number         = "4",
      volume         = "91",
      pages          = "043002",
      doi            = "10.1103/PhysRevD.91.043002",
      year           = "2015",
      eprint         = "1410.8866",
      archivePrefix  = "arXiv",
      primaryClass   = "gr-qc",
      SLACcitation   = "%%CITATION = ARXIV:1410.8866;%%",
}

@article{Raaijmakers:2019dks,
    author = "Raaijmakers, G. and others",
    title = "{Constraining the dense matter equation of state with joint analysis of NICER and LIGO/Virgo measurements}",
    eprint = "1912.11031",
    archivePrefix = "arXiv",
    primaryClass = "astro-ph.HE",
    doi = "10.3847/2041-8213/ab822f",
    journal = "Astrophys. J. Lett.",
    volume = "893",
    number = "1",
    pages = "L21",
    year = "2020"
}

@article{Flanagan:2007ix,
    author = "Flanagan, Eanna E. and Hinderer, Tanja",
    title = "{Constraining neutron star tidal Love numbers with gravitational wave detectors}",
    eprint = "0709.1915",
    archivePrefix = "arXiv",
    primaryClass = "astro-ph",
    doi = "10.1103/PhysRevD.77.021502",
    journal = "Phys. Rev. D",
    volume = "77",
    pages = "021502",
    year = "2008"
}

@article{Radice:2017lry,
      author         = "Radice, David and Perego, Albino and Zappa, Francesco and
                        Bernuzzi, Sebastiano",
      title          = "{GW170817: Joint Constraint on the Neutron Star Equation
                        of State from Multimessenger Observations}",
      journal        = "Astrophys. J.",
      volume         = "852",
      year           = "2018",
      number         = "2",
      pages          = "L29",
      doi            = "10.3847/2041-8213/aaa402",
      eprint         = "1711.03647",
      archivePrefix  = "arXiv",
      primaryClass   = "astro-ph.HE",
      reportNumber   = "LIGO-P1700421-AND-VIR-0894A-17",
      SLACcitation   = "%%CITATION = ARXIV:1711.03647;%%"
}

@article{Margalit:2017dij,
      author         = "Margalit, Ben and Metzger, Brian D.",
      title          = "{Constraining the Maximum Mass of Neutron Stars From
                        Multi-Messenger Observations of GW170817}",
      journal        = "Astrophys. J.",
      volume         = "850",
      year           = "2017",
      number         = "2",
      pages          = "L19",
      doi            = "10.3847/2041-8213/aa991c",
      eprint         = "1710.05938",
      archivePrefix  = "arXiv",
      primaryClass   = "astro-ph.HE",
      SLACcitation   = "%%CITATION = ARXIV:1710.05938;%%"
}

@article{Read:2009iy,
      author         = "Read, Jocelyn S. and Lackey, Benjamin D. and Owen,
                        Benjamin J. and Friedman, John L.",
      title          = "{Constraints on a phenomenologically parameterized
                        neutron-star equation of state}",
      journal        = "Phys. Rev. D",
      volume         = "79",
      year           = "2009",
      pages          = "124032",
      doi            = "10.1103/PhysRevD.79.124032",
}

@article{Dietrich:2020efo,
    author = "Dietrich, Tim and Coughlin, Michael W. and Pang, Peter T. H. and Bulla, Mattia and Heinzel, Jack and Issa, Lina and Tews, Ingo and Antier, Sarah",
    title = "{Multimessenger constraints on the neutron-star equation of state and the Hubble constant}",
    eprint = "2002.11355",
    archivePrefix = "arXiv",
    primaryClass = "astro-ph.HE",
    reportNumber = "LA-UR-20-21470",
    doi = "10.1126/science.abb4317",
    journal = "Science",
    volume = "370",
    number = "6523",
    pages = "1450--1453",
    year = "2020"
}

@article{Rezzolla:2017aly,
    author = "Rezzolla, Luciano and Most, Elias R. and Weih, Lukas R.",
    title = "{Using gravitational-wave observations and quasi-universal relations to constrain the maximum mass of neutron stars}",
    eprint = "1711.00314",
    archivePrefix = "arXiv",
    primaryClass = "astro-ph.HE",
    doi = "10.3847/2041-8213/aaa401",
    journal = "Astrophys. J. Lett.",
    volume = "852",
    number = "2",
    pages = "L25",
    year = "2018"
}

@article{Antoniadis:2013pzd,
      author         = "Antoniadis, John and Freire, Paulo C.C. and Wex, Norbert
                        and Tauris, Thomas M. and Lynch, Ryan S. and others",
      title          = "{A Massive Pulsar in a Compact Relativistic Binary}",
      journal        = "Science",
      volume         = "340",
      number         = "6131",
      pages          = "1233232",
      doi            = "10.1126/science.1233232",
      year           = "2013",
      eprint         = "1304.6875",
      archivePrefix  = "arXiv",
      primaryClass   = "astro-ph.HE",
      SLACcitation   = "%%CITATION = ARXIV:1304.6875;%%",
}

@article{Carney:2018sdv,
      author         = "Carney, Matthew F. and Wade, Leslie E. and Irwin, Burke
                        S.",
      title          = "{Comparing two models for measuring the neutron star
                        equation of state from gravitational-wave signals}",
      journal        = "Phys. Rev.",
      volume         = "D98",
      year           = "2018",
      number         = "6",
      pages          = "063004",
      doi            = "10.1103/PhysRevD.98.063004",
      eprint         = "1805.11217",
      archivePrefix  = "arXiv",
      primaryClass   = "gr-qc",
      SLACcitation   = "%%CITATION = ARXIV:1805.11217;%%"
}

@article{Landry:2018prd,
    author = "Landry, Philippe and Essick, Reed",
    title = "{Nonparametric inference of the neutron star equation of state from gravitational wave observations}",
    eprint = "1811.12529",
    archivePrefix = "arXiv",
    primaryClass = "gr-qc",
    doi = "10.1103/PhysRevD.99.084049",
    journal = "Phys. Rev. D",
    volume = "99",
    number = "8",
    pages = "084049",
    year = "2019"
}

@article{Cromartie:2019kug,
    author = "Cromartie, H. Thankful and others",
    title = "{Relativistic Shapiro delay measurements of an extremely massive millisecond pulsar}",
    eprint = "1904.06759",
    archivePrefix = "arXiv",
    primaryClass- = "astro-ph.HE",
    doi = "10.1038/s41550-019-0880-2",
    journal = "Nature Astron.",
    volume = "4",
    number = "1",
    pages = "72--76",
    year = "2019"
}

@dataset{miller_m_c_2019_3473466,
  author       = {Miller, M. C. and
                  Lamb, F. K. and
                  Dittmann, A. J. and
                  Bogdanov, S. and
                  Arzoumanian, Z. and
                  Gendreau, K. C. and
                  Guillot, S. and
                  Harding, A. K. and
                  Ho, W. C. G. and
                  Lattimer, J. M. and
                  Ludlam, R. M. and
                  Mahmoodifar, S. and
                  Morsink, S. M. and
                  Ray, P. S. and
                  Strohmayer, T. E. and
                  Wood, K. S. and
                  Enoto, T. and
                  Foster, R. and
                  Okajima, T. and
                  Prigozhin, G. and
                  Soong, Y.},
  title        = {{NICER PSR J0030+0451 Illinois-Maryland MCMC 
                   Samples}},
  month        = dec,
  year         = 2019,
  publisher    = {Zenodo},
  version      = {1.0.0},
  doi          = {10.5281/zenodo.3473466},
  url          = {https://doi.org/10.5281/zenodo.3473466}
}

@dataset{riley_thomas_e_2019_3386449,
  author       = {Riley, Thomas E. and
                  Watts, Anna L. and
                  Bogdanov, Slavko and
                  Ray, Paul S. and
                  Ludlam, Renee M. and
                  Guillot, Sebastien and
                  Arzoumanian, Zaven and
                  Baker, Charles L. and
                  Bilous, Anna V. and
                  Chakrabarty, Deepto and
                  Gendreau, Keith C. and
                  Harding, Alice K. and
                  Ho, Wynn C. G. and
                  Lattimer, James M. and
                  Morsink, Sharon M. and
                  Strohmayer, Tod E.},
  title        = {{A NICER View of PSR J0030+0451: Nested Samples for 
                   Millisecond Pulsar Parameter Estimation}},
  month        = dec,
  year         = 2019,
  publisher    = {Zenodo},
  version      = {v1.0.0},
  doi          = {10.5281/zenodo.3386449},
  url          = {https://doi.org/10.5281/zenodo.3386449}
}

@article{De:2018uhw,
      author         = "De, Soumi and Finstad, Daniel and Lattimer, James M. and
                        Brown, Duncan A. and Berger, Edo and Biwer, Christopher
                        M.",
      title          = "{Tidal Deformabilities and Radii of Neutron Stars from
                        the Observation of GW170817}",
      journal        = "Phys. Rev. Lett.",
      volume         = "121",
      year           = "2018",
      number         = "9",
      pages          = "091102",
      doi            = "10.1103/PhysRevLett.121.259902,
                        10.1103/PhysRevLett.121.091102",
      note           = "[Erratum: Phys. Rev. Lett.121,no.25,259902(2018)]",
      eprint         = "1804.08583",
      archivePrefix  = "arXiv",
      primaryClass   = "astro-ph.HE",
      SLACcitation   = "%%CITATION = ARXIV:1804.08583;%%"
}

@article{Zhao:2018nyf,
    author = "Zhao, Tianqi and Lattimer, James M.",
    title = "{Tidal Deformabilities and Neutron Star Mergers}",
    eprint = "1808.02858",
    archivePrefix = "arXiv",
    primaryClass = "astro-ph.HE",
    doi = "10.1103/PhysRevD.98.063020",
    journal = "Phys. Rev. D",
    volume = "98",
    number = "6",
    pages = "063020",
    year = "2018"
}

@article{Essick:2020flb,
    author = "Essick, Reed and Tews, Ingo and Landry, Philippe and Reddy, Sanjay and Holz, Daniel E.",
    title = "{Direct Astrophysical Tests of Chiral Effective Field Theory at Supranuclear Densities}",
    eprint = "2004.07744",
    archivePrefix = "arXiv",
    primaryClass = "astro-ph.HE",
    reportNumber = "LA-UR-20-22615",
    doi = "10.1103/PhysRevC.102.055803",
    journal = "Phys. Rev. C",
    volume = "102",
    number = "5",
    pages = "055803",
    year = "2020"
}

@article{Essick:2021kjb,
    author = "Essick, Reed and Tews, Ingo and Landry, Philippe and Schwenk, Achim",
    title = "{Astrophysical Constraints on the Symmetry Energy and the Neutron Skin of Pb208 with Minimal Modeling Assumptions}",
    eprint = "2102.10074",
    archivePrefix = "arXiv",
    primaryClass = "nucl-th",
    reportNumber = "LA-UR-21-20527",
    doi = "10.1103/PhysRevLett.127.192701",
    journal = "Phys. Rev. Lett.",
    volume = "127",
    number = "19",
    pages = "192701",
    year = "2021"
}

@article{Reed:2021nqk,
    author = "Reed, Brendan T. and Fattoyev, F. J. and Horowitz, C. J. and Piekarewicz, J.",
    title = "{Implications of PREX-2 on the Equation of State of Neutron-Rich Matter}",
    eprint = "2101.03193",
    archivePrefix = "arXiv",
    primaryClass = "nucl-th",
    doi = "10.1103/PhysRevLett.126.172503",
    journal = "Phys. Rev. Lett.",
    volume = "126",
    number = "17",
    pages = "172503",
    year = "2021"
}

@article{Annala:2017llu,
      author         = "Annala, Eemeli and Gorda, Tyler and Kurkela, Aleksi and
                        Vuorinen, Aleksi",
      title          = "{Gravitational-wave constraints on the
                        neutron-star-matter Equation of State}",
      journal        = "Phys. Rev. Lett.",
      volume         = "120",
      year           = "2018",
      number         = "17",
      pages          = "172703",
      doi            = "10.1103/PhysRevLett.120.172703",
      eprint         = "1711.02644",
      archivePrefix  = "arXiv",
      primaryClass   = "astro-ph.HE",
      reportNumber   = "CERN-TH-2017-236",
      SLACcitation   = "%%CITATION = ARXIV:1711.02644;%%"
}

@article{Raithel:2018ncd,
      author         = "Raithel, Carolyn and {\"O}zel, Feryal and Psaltis,
                        Dimitrios",
      title          = "{Tidal deformability from GW170817 as a direct probe of
                        the neutron star radius}",
      journal        = "Astrophys. J.",
      volume         = "857",
      year           = "2018",
      number         = "2",
      pages          = "L23",
      doi            = "10.3847/2041-8213/aabcbf",
      eprint         = "1803.07687",
      archivePrefix  = "arXiv",
      primaryClass   = "astro-ph.HE",
      SLACcitation   = "%%CITATION = ARXIV:1803.07687;%%"
}

@article{Coughlin:2018miv,
      author         = "Coughlin, Michael W. and others",
      title          = "{Constraints on the neutron star equation of state from
                        AT2017gfo using radiative transfer simulations}",
      journal        = "Mon. Not. Roy. Astron. Soc.",
      volume         = "480",
      year           = "2018",
      number         = "3",
      pages          = "3871-3878",
      doi            = "10.1093/mnras/sty2174",
      eprint         = "1805.09371",
      archivePrefix  = "arXiv",
      primaryClass   = "astro-ph.HE",
      SLACcitation   = "%%CITATION = ARXIV:1805.09371;%%"
}

@article{Coughlin:2018fis,
    author = "Coughlin, Michael W. and Dietrich, Tim and Margalit, Ben and Metzger, Brian D.",
    title = "{Multimessenger Bayesian parameter inference of a binary neutron star merger}",
    eprint = "1812.04803",
    archivePrefix = "arXiv",
    primaryClass = "astro-ph.HE",
    doi = "10.1093/mnrasl/slz133",
    journal = "Mon. Not. Roy. Astron. Soc.",
    volume = "489",
    number = "1",
    pages = "L91--L96",
    year = "2019"
}

@article{Most:2018hfd,
      author         = "Most, Elias R. and Weih, Lukas R. and Rezzolla, Luciano
                        and Schaffner-Bielich, JÃÅrgen",
      title          = "{New constraints on radii and tidal deformabilities of
                        neutron stars from GW170817}",
      journal        = "Phys. Rev. Lett.",
      volume         = "120",
      year           = "2018",
      number         = "26",
      pages          = "261103",
      doi            = "10.1103/PhysRevLett.120.261103",
      eprint         = "1803.00549",
      archivePrefix  = "arXiv",
      primaryClass   = "gr-qc",
      SLACcitation   = "%%CITATION = ARXIV:1803.00549;%%"
}

@article{Shibata:2019ctb,
    author = "Shibata, Masaru and Zhou, Enping and Kiuchi, Kenta and Fujibayashi, Sho",
    title = "{Constraint on the maximum mass of neutron stars using GW170817 event}",
    eprint = "1905.03656",
    archivePrefix = "arXiv",
    primaryClass = "astro-ph.HE",
    doi = "10.1103/PhysRevD.100.023015",
    journal = "Phys. Rev. D",
    volume = "100",
    number = "2",
    pages = "023015",
    year = "2019"
}

@article{Lindblom:2010bb,
    author = "Lindblom, Lee",
    title = "{Spectral Representations of Neutron-Star Equations of State}",
    eprint = "1009.0738",
    archivePrefix = "arXiv",
    primaryClass = "astro-ph.HE",
    doi = "10.1103/PhysRevD.82.103011",
    journal = "Phys. Rev. D",
    volume = "82",
    pages = "103011",
    year = "2010"
}

@article{Lindblom:2018rfr,
      author         = "Lindblom, Lee",
      title          = "{Causal Representations of Neutron-Star Equations of
                        State}",
      journal        = "Phys. Rev.",
      volume         = "D97",
      year           = "2018",
      number         = "12",
      pages          = "123019",
      doi            = "10.1103/PhysRevD.97.123019",
      eprint         = "1804.04072",
      archivePrefix  = "arXiv",
      primaryClass   = "astro-ph.HE",
      SLACcitation   = "%%CITATION = ARXIV:1804.04072;%%"
}

@article{Alsing:2017bbc,
    author = "Alsing, Justin and Silva, Hector O. and Berti, Emanuele",
    title = "{Evidence for a maximum mass cut-off in the neutron star mass distribution and constraints on the equation of state}",
    eprint = "1709.07889",
    archivePrefix = "arXiv",
    primaryClass = "astro-ph.HE",
    doi = "10.1093/mnras/sty1065",
    journal = "Mon. Not. Roy. Astron. Soc.",
    volume = "478",
    number = "1",
    pages = "1377--1391",
    year = "2018"
}

@ARTICLE{2010PhRvD..81j5021K,
       author = {{Kurkela}, Aleksi and {Romatschke}, Paul and {Vuorinen}, Aleksi},
        title = "{Cold quark matter}",
      journal = {\prd},
         year = 2010,
        month = may,
       volume = {81},
       number = {10},
          eid = {105021},
        pages = {105021},
          doi = {10.1103/PhysRevD.81.105021},
archivePrefix = {arXiv},
       eprint = {0912.1856},
 primaryClass = {hep-ph},
       adsurl = {https://ui.adsabs.harvard.edu/abs/2010PhRvD..81j5021K}
}

@article{Alvarez-Castillo:2016oln,
      author         = "Alvarez-Castillo, D. and Ayriyan, A. and Benic, S. and
                        Blaschke, D. and Grigorian, H. and Typel, S.",
      title          = "{New class of hybrid EoS and Bayesian M-R data analysis}",
      journal        = "Eur. Phys. J.",
      volume         = "A52",
      year           = "2016",
      number         = "3",
      pages          = "69",
      doi            = "10.1140/epja/i2016-16069-2",
      eprint         = "1603.03457",
      archivePrefix  = "arXiv",
      primaryClass   = "nucl-th",
      SLACcitation   = "%%CITATION = ARXIV:1603.03457;%%"
}

@ARTICLE{2018ApJS..235...37A,
       author = {{Arzoumanian}, Zaven and {Brazier}, Adam and {Burke-Spolaor}, Sarah and {Chamberlin}, Sydney and {Chatterjee}, Shami and {Christy}, Brian and {Cordes}, James M. and {Cornish}, Neil J. and {Crawford}, Fronefield and {Thankful Cromartie}, H. and {Crowter}, Kathryn and {DeCesar}, Megan E. and {Demorest}, Paul B. and {Dolch}, Timothy and {Ellis}, Justin A. and {Ferdman}, Robert D. and {Ferrara}, Elizabeth C. and {Fonseca}, Emmanuel and {Garver-Daniels}, Nathan and {Gentile}, Peter A. and {Halmrast}, Daniel and {Huerta}, E.~A. and {Jenet}, Fredrick A. and {Jessup}, Cody and {Jones}, Glenn and {Jones}, Megan L. and {Kaplan}, David L. and {Lam}, Michael T. and {Lazio}, T. Joseph W. and {Levin}, Lina and {Lommen}, Andrea and {Lorimer}, Duncan R. and {Luo}, Jing and {Lynch}, Ryan S. and {Madison}, Dustin and {Matthews}, Allison M. and {McLaughlin}, Maura A. and {McWilliams}, Sean T. and {Mingarelli}, Chiara and {Ng}, Cherry and {Nice}, David J. and {Pennucci}, Timothy T. and {Ransom}, Scott M. and {Ray}, Paul S. and {Siemens}, Xavier and {Simon}, Joseph and {Spiewak}, Ren{\'e}e and {Stairs}, Ingrid H. and {Stinebring}, Daniel R. and {Stovall}, Kevin and {Swiggum}, Joseph K. and {Taylor}, Stephen R. and {Vallisneri}, Michele and {van Haasteren}, Rutger and {Vigeland}, Sarah J. and {Zhu}, Weiwei and {NANOGrav Collaboration}},
        title = "{The NANOGrav 11-year Data Set: High-precision Timing of 45 Millisecond Pulsars}",
      journal = {\apjs},
         year = 2018,
        month = apr,
       volume = {235},
       number = {2},
          eid = {37},
        pages = {37},
          doi = {10.3847/1538-4365/aab5b0},
archivePrefix = {arXiv},
       eprint = {1801.01837},
 primaryClass = {astro-ph.HE},
       adsurl = {https://ui.adsabs.harvard.edu/abs/2018ApJS..235...37A}
}

@article{Kiuchi:2019lls,
    author = "Kiuchi, Kenta and Kyutoku, Koutarou and Shibata, Masaru and Taniguchi, Keisuke",
    title = "{Revisiting the lower bound on tidal deformability derived by AT 2017gfo}",
    eprint = "1903.01466",
    archivePrefix = "arXiv",
    primaryClass = "astro-ph.HE",
    doi = "10.3847/2041-8213/ab1e45",
    journal = "Astrophys. J. Lett.",
    volume = "876",
    number = "2",
    pages = "L31",
    year = "2019"
}

@article{Greif:2018njt,
      author         = "Greif, S. K. and Raaijmakers, G. and Hebeler, K. and
                        Schwenk, A. and Watts, A. L.",
      title          = "{Equation of state sensitivities when inferring neutron
                        star and dense matter properties}",
      journal        = "Mon. Not. Roy. Astron. Soc.",
      volume         = "485",
      year           = "2019",
      number         = "4",
      pages          = "5363-5376",
      doi            = "10.1093/mnras/stz654",
      eprint         = "1812.08188",
      archivePrefix  = "arXiv",
      primaryClass   = "astro-ph.HE",
      SLACcitation   = "%%CITATION = ARXIV:1812.08188;%%"
}

@ARTICLE{2020NatPh..16..907A,
       author = {{Annala}, Eemeli and {Gorda}, Tyler and {Kurkela}, Aleksi and {N{\"a}ttil{\"a}}, Joonas and {Vuorinen}, Aleksi},
        title = "{Evidence for quark-matter cores in massive neutron stars}",
      journal = {Nature Physics},
         year = 2020,
        month = jun,
       volume = {16},
       number = {9},
        pages = {907-910},
          doi = {10.1038/s41567-020-0914-9},
archivePrefix = {arXiv},
       eprint = {1903.09121},
 primaryClass = {astro-ph.HE},
       adsurl = {https://ui.adsabs.harvard.edu/abs/2020NatPh..16..907A}
}

@article{Farrow:2019xnc,
    author = "Farrow, Nicholas and Zhu, Xing-Jiang and Thrane, Eric",
    title = "{The mass distribution of Galactic double neutron stars}",
    eprint = "1902.03300",
    archivePrefix = "arXiv",
    primaryClass = "astro-ph.HE",
    doi = "10.3847/1538-4357/ab12e3",
    journal = "Astrophys. J.",
    volume = "876",
    number = "1",
    pages = "18",
    year = "2019"
}

@article{Steiner:2017vmg,
    author = "Steiner, A. W. and Heinke, C. O. and Bogdanov, S. and Li, C. and Ho, W. C. G. and Bahramian, A. and Han, S.",
    title = "{Constraining the Mass and Radius of Neutron Stars in Globular Clusters}",
    eprint = "1709.05013",
    archivePrefix = "arXiv",
    primaryClass = "astro-ph.HE",
    doi = "10.1093/mnras/sty215",
    journal = "Mon. Not. Roy. Astron. Soc.",
    volume = "476",
    number = "1",
    pages = "421--435",
    year = "2018"
}

@article{Most:2018eaw,
      author         = "Most, Elias R. and Papenfort, L. Jens and Dexheimer,
                        Veronica and Hanauske, Matthias and Schramm, Stefan and
                        St{\:o}cker, Horst and Rezzolla, Luciano",
      title          = "{Signatures of quark-hadron phase transitions in
                        general-relativistic neutron-star mergers}",
      journal        = "Phys. Rev. Lett.",
      volume         = "122",
      year           = "2019",
      number         = "6",
      pages          = "061101",
      doi            = "10.1103/PhysRevLett.122.061101",
      eprint         = "1807.03684",
      archivePrefix  = "arXiv",
      primaryClass   = "astro-ph.HE",
      SLACcitation   = "%%CITATION = ARXIV:1807.03684;%%"
}

@article{Tews:2018iwm,
    author = "Tews, I. and Margueron, J. and Reddy, S.",
    title = "{Critical examination of constraints on the equation of state of dense matter obtained from GW170817}",
    eprint = "1804.02783",
    archivePrefix = "arXiv",
    primaryClass = "nucl-th",
    reportNumber = "INT-PUB-18-014",
    doi = "10.1103/PhysRevC.98.045804",
    journal = "Phys. Rev. C",
    volume = "98",
    number = "4",
    pages = "045804",
    year = "2018"
}

@article{Riley:2018ekf,
    author = "Riley, Thomas E. and Raaijmakers, Geert and Watts, Anna L.",
    title = "{On parametrized cold dense matter equation-of-state inference}",
    eprint = "1804.09085",
    archivePrefix = "arXiv",
    primaryClass = "astro-ph.HE",
    doi = "10.1093/mnras/sty1051",
    journal = "Mon. Not. Roy. Astron. Soc.",
    volume = "478",
    number = "1",
    pages = "1093--1131",
    year = "2018"
}

@article{Legred:2021,
  title = {Impact of the PSR $\mathrm{J}0740+6620$ radius constraint on the properties of high-density matter},
  author = {Legred, Isaac and Chatziioannou, Katerina and Essick, Reed and Han, Sophia and Landry, Philippe},
  journal = {Phys. Rev. D},
  volume = {104},
  issue = {6},
  pages = {063003},
  numpages = {20},
  year = {2021},
  month = {Sep},
  publisher = {American Physical Society},
  doi = {10.1103/PhysRevD.104.063003},
  url = {https://link.aps.org/doi/10.1103/PhysRevD.104.063003}
}

@article{PhysRevD.106.103027,
  title = {Binary neutron star mergers as a probe of quark-hadron crossover equations of state},
  author = {Kedia, Atul and Kim, Hee Il and Suh, In-Saeng and Mathews, Grant J.},
  journal = {Phys. Rev. D},
  volume = {106},
  issue = {10},
  pages = {103027},
  numpages = {9},
  year = {2022},
  month = {Nov},
  publisher = {American Physical Society},
  doi = {10.1103/PhysRevD.106.103027},
  url = {https://link.aps.org/doi/10.1103/PhysRevD.106.103027}
}

@ARTICLE{2022NatAs...6.1444D,
       author = {{Doroshenko}, Victor and {Suleimanov}, Valery and {P{\"u}hlhofer}, Gerd and {Santangelo}, Andrea},
        title = "{A strangely light neutron star within a supernova remnant}",
      journal = {Nature Astronomy},
         year = 2022,
        month = dec,
       volume = {6},
        pages = {1444-1451},
          doi = {10.1038/s41550-022-01800-1},
}

@article{PhysRevD.105.063031,
  title = {Improved spectral representations of neutron-star equations of state},
  author = {Lindblom, Lee},
  journal = {Phys. Rev. D},
  volume = {105},
  issue = {6},
  pages = {063031},
  numpages = {6},
  year = {2022},
  month = {Mar},
  publisher = {American Physical Society},
  doi = {10.1103/PhysRevD.105.063031},
  url = {https://link.aps.org/doi/10.1103/PhysRevD.105.063031}
}

@ARTICLE{2023PhRvD.107l3017L,
       author = {{Legred}, Isaac and {Kim}, Yoonsoo and {Deppe}, Nils and {Chatziioannou}, Katerina and {Foucart}, Francois and {H{\'e}bert}, Fran{\c{c}}ois and {Kidder}, Lawrence E.},
        title = "{Simulating neutron stars with a flexible enthalpy-based equation of state parametrization in SpECTRE}",
      journal = {\prd},
         year = 2023,
        month = jun,
       volume = {107},
       number = {12},
          eid = {123017},
        pages = {123017},
          doi = {10.1103/PhysRevD.107.123017},
archivePrefix = {arXiv},
       eprint = {2301.13818},
 primaryClass = {astro-ph.HE},
       adsurl = {https://ui.adsabs.harvard.edu/abs/2023PhRvD.107l3017L}
}

@article{2024arXiv240411346K,
       author = {{Kastaun}, Wolfgang and {Ohme}, Frank},
        title = "{Modern tools for computing neutron star properties}",
      journal = {arXiv e-prints},
         year = 2024,
        month = apr,
          eid = {arXiv:2404.11346},
        pages = {arXiv:2404.11346},
          doi = {10.48550/arXiv.2404.11346},
archivePrefix = {arXiv},
       eprint = {2404.11346},
 primaryClass = {gr-qc},
       adsurl = {https://ui.adsabs.harvard.edu/abs/2024arXiv240411346K},
}

@misc{wolfgang_kastaun_2023_7700296,
  author       = {Wolfgang Kastaun and Roland Haas},
  title        = {wokast/RePrimAnd: Release 1.6},
  month        = apr,
  year         = 2024,
  publisher    = {Zenodo},
  version      = {v1.6},
  doi          = {10.5281/zenodo.3785074},
  url          = {https://doi.org/10.5281/zenodo.3785074}
}

@ARTICLE{2016ARA&A..54..401O,
       author = {{{\"O}zel}, Feryal and {Freire}, Paulo},
        title = "{Masses, Radii, and the Equation of State of Neutron Stars}",
      journal = {\araa},
         year = 2016,
        month = sep,
       volume = {54},
        pages = {401-440},
          doi = {10.1146/annurev-astro-081915-023322},
archivePrefix = {arXiv},
       eprint = {1603.02698},
 primaryClass = {astro-ph.HE},
       adsurl = {https://ui.adsabs.harvard.edu/abs/2016ARA&A..54..401O}
}

@ARTICLE{2021ApJ...915L..12F,
       author = {{Fonseca}, E. and {Cromartie}, H.~T. and {Pennucci}, T.~T. and {Ray}, P.~S. and {Kirichenko}, A. Yu. and {Ransom}, S.~M. and {Demorest}, P.~B. and {Stairs}, I.~H. and {Arzoumanian}, Z. and {Guillemot}, L. and {Parthasarathy}, A. and {Kerr}, M. and {Cognard}, I. and {Baker}, P.~T. and {Blumer}, H. and {Brook}, P.~R. and {DeCesar}, M. and {Dolch}, T. and {Dong}, F.~A. and {Ferrara}, E.~C. and {Fiore}, W. and {Garver-Daniels}, N. and {Good}, D.~C. and {Jennings}, R. and {Jones}, M.~L. and {Kaspi}, V.~M. and {Lam}, M.~T. and {Lorimer}, D.~R. and {Luo}, J. and {McEwen}, A. and {McKee}, J.~W. and {McLaughlin}, M.~A. and {McMann}, N. and {Meyers}, B.~W. and {Naidu}, A. and {Ng}, C. and {Nice}, D.~J. and {Pol}, N. and {Radovan}, H.~A. and {Shapiro-Albert}, B. and {Tan}, C.~M. and {Tendulkar}, S.~P. and {Swiggum}, J.~K. and {Wahl}, H.~M. and {Zhu}, W.~W.},
        title = "{Refined Mass and Geometric Measurements of the High-mass PSR J0740+6620}",
      journal = {\apjl},
         year = 2021,
        month = jul,
       volume = {915},
       number = {1},
          eid = {L12},
        pages = {L12},
          doi = {10.3847/2041-8213/ac03b8},
archivePrefix = {arXiv},
       eprint = {2104.00880},
 primaryClass = {astro-ph.HE},
       adsurl = {https://ui.adsabs.harvard.edu/abs/2021ApJ...915L..12F}
}

@ARTICLE{2013Sci...340..448A,
       author = {{Antoniadis}, John and {Freire}, Paulo C.~C. and {Wex}, Norbert and {Tauris}, Thomas M. and {Lynch}, Ryan S. and {van Kerkwijk}, Marten H. and {Kramer}, Michael and {Bassa}, Cees and {Dhillon}, Vik S. and {Driebe}, Thomas and {Hessels}, Jason W.~T. and {Kaspi}, Victoria M. and {Kondratiev}, Vladislav I. and {Langer}, Norbert and {Marsh}, Thomas R. and {McLaughlin}, Maura A. and {Pennucci}, Timothy T. and {Ransom}, Scott M. and {Stairs}, Ingrid H. and {van Leeuwen}, Joeri and {Verbiest}, Joris P.~W. and {Whelan}, David G.},
        title = "{A Massive Pulsar in a Compact Relativistic Binary}",
      journal = {Science},
         year = 2013,
        month = apr,
       volume = {340},
       number = {6131},
        pages = {448},
          doi = {10.1126/science.1233232},
archivePrefix = {arXiv},
       eprint = {1304.6875},
 primaryClass = {astro-ph.HE},
       adsurl = {https://ui.adsabs.harvard.edu/abs/2013Sci...340..448A}
}

@ARTICLE{2017ApJ...850L..34B,
       author = {{Bauswein}, Andreas and {Just}, Oliver and {Janka}, Hans-Thomas and {Stergioulas}, Nikolaos},
        title = "{Neutron-star Radius Constraints from GW170817 and Future Detections}",
      journal = {\apjl},
         year = 2017,
        month = dec,
       volume = {850},
       number = {2},
          eid = {L34},
        pages = {L34},
          doi = {10.3847/2041-8213/aa9994},
archivePrefix = {arXiv},
       eprint = {1710.06843},
 primaryClass = {astro-ph.HE},
       adsurl = {https://ui.adsabs.harvard.edu/abs/2017ApJ...850L..34B},
}

@ARTICLE{2010Natur.467.1081D,
       author = {{Demorest}, P.~B. and {Pennucci}, T. and {Ransom}, S.~M. and {Roberts}, M.~S.~E. and {Hessels}, J.~W.~T.},
        title = "{A two-solar-mass neutron star measured using Shapiro delay}",
      journal = {\nat},
         year = 2010,
        month = oct,
       volume = {467},
       number = {7319},
        pages = {1081-1083},
          doi = {10.1038/nature09466},
archivePrefix = {arXiv},
       eprint = {1010.5788},
 primaryClass = {astro-ph.HE},
       adsurl = {https://ui.adsabs.harvard.edu/abs/2010Natur.467.1081D}
}

@ARTICLE{2020ApJ...893L..21R,
       author = {{Raaijmakers}, G. and {Greif}, S.~K. and {Riley}, T.~E. and {Hinderer}, T. and {Hebeler}, K. and {Schwenk}, A. and {Watts}, A.~L. and {Nissanke}, S. and {Guillot}, S. and {Lattimer}, J.~M. and {Ludlam}, R.~M.},
        title = "{Constraining the Dense Matter Equation of State with Joint Analysis of NICER and LIGO/Virgo Measurements}",
      journal = {\apjl},
         year = 2020,
        month = apr,
       volume = {893},
       number = {1},
          eid = {L21},
        pages = {L21},
          doi = {10.3847/2041-8213/ab822f},
archivePrefix = {arXiv},
       eprint = {1912.11031},
 primaryClass = {astro-ph.HE},
       adsurl = {https://ui.adsabs.harvard.edu/abs/2020ApJ...893L..21R}
}

@ARTICLE{2023PhRvD.107d3034M,
       author = {{Most}, Elias R. and {Motornenko}, Anton and {Steinheimer}, Jan and {Dexheimer}, Veronica and {Hanauske}, Matthias and {Rezzolla}, Luciano and {Stoecker}, Horst},
        title = "{Probing neutron-star matter in the lab: Similarities and differences between binary mergers and heavy-ion collisions}",
      journal = {\prd},
         year = 2023,
        month = feb,
       volume = {107},
       number = {4},
          eid = {043034},
        pages = {043034},
          doi = {10.1103/PhysRevD.107.043034},
archivePrefix = {arXiv},
       eprint = {2201.13150},
 primaryClass = {nucl-th},
       adsurl = {https://ui.adsabs.harvard.edu/abs/2023PhRvD.107d3034M}
}

@ARTICLE{2021PhRvL.126f1101A,
       author = {{Al-Mamun}, Mohammad and {Steiner}, Andrew W. and {N{\"a}ttil{\"a}}, Joonas and {Lange}, Jacob and {O'Shaughnessy}, Richard and {Tews}, Ingo and {Gandolfi}, Stefano and {Heinke}, Craig and {Han}, Sophia},
        title = "{Combining Electromagnetic and Gravitational-Wave Constraints on Neutron-Star Masses and Radii}",
      journal = {\prl},
         year = 2021,
        month = feb,
       volume = {126},
       number = {6},
          eid = {061101},
        pages = {061101},
          doi = {10.1103/PhysRevLett.126.061101},
archivePrefix = {arXiv},
       eprint = {2008.12817},
 primaryClass = {astro-ph.HE},
       adsurl = {https://ui.adsabs.harvard.edu/abs/2021PhRvL.126f1101A}
}

@article{PhysRevResearch.2.043039,
  title = {Reanalysis of the binary neutron star mergers GW170817 and GW190425 using numerical-relativity calibrated waveform models},
  author = {Narikawa, Tatsuya and Uchikata, Nami and Kawaguchi, Kyohei and Kiuchi, Kenta and Kyutoku, Koutarou and Shibata, Masaru and Tagoshi, Hideyuki},
  journal = {Phys. Rev. Res.},
  volume = {2},
  issue = {4},
  pages = {043039},
  numpages = {15},
  year = {2020},
  month = {Oct},
  publisher = {American Physical Society},
  doi = {10.1103/PhysRevResearch.2.043039},
  url = {https://link.aps.org/doi/10.1103/PhysRevResearch.2.043039}
}

@ARTICLE{2020EPJST.229.3663C,
       author = {{Cierniak}, Mateusz and {Blaschke}, David},
        title = "{The special point on the hybrid star mass-radius diagram and its multi-messenger implications}",
      journal = {European Physical Journal Special Topics},
         year = 2020,
        month = dec,
       volume = {229},
       number = {22-23},
        pages = {3663-3673},
          doi = {10.1140/epjst/e2020-000235-5},
archivePrefix = {arXiv},
       eprint = {2009.12353},
 primaryClass = {astro-ph.HE},
}

@ARTICLE{2020PhRvC.102d5807L,
       author = {{Li}, Bao-An and {Magno}, Macon},
        title = "{Curvature-slope correlation of nuclear symmetry energy and its imprints on the crust-core transition, radius, and tidal deformability of canonical neutron stars}",
      journal = {\prc},
         year = 2020,
        month = oct,
       volume = {102},
       number = {4},
          eid = {045807},
        pages = {045807},
          doi = {10.1103/PhysRevC.102.045807},
archivePrefix = {arXiv},
       eprint = {2008.11338},
 primaryClass = {nucl-th},
}

@ARTICLE{2021PhRvC.103c5802X,
       author = {{Xie}, Wen-Jie and {Li}, Bao-An},
        title = "{Bayesian inference of the dense-matter equation of state encapsulating a first-order hadron-quark phase transition from observables of canonical neutron stars}",
      journal = {\prc},
         year = 2021,
        month = mar,
       volume = {103},
       number = {3},
          eid = {035802},
        pages = {035802},
          doi = {10.1103/PhysRevC.103.035802},
archivePrefix = {arXiv},
       eprint = {2009.13653},
 primaryClass = {nucl-th},
}

@ARTICLE{2023JCAP...02..016F,
       author = {{Farrell}, Delaney and {Baldi}, Pierre and {Ott}, Jordan and {Ghosh}, Aishik and {Steiner}, Andrew W. and {Kavitkar}, Atharva and {Lindblom}, Lee and {Whiteson}, Daniel and {Weber}, Fridolin},
        title = "{Deducing neutron star equation of state parameters directly from telescope spectra with uncertainty-aware machine learning}",
      journal = {\jcap},
         year = 2023,
        month = feb,
       volume = {2023},
       number = {2},
          eid = {016},
        pages = {016},
          doi = {10.1088/1475-7516/2023/02/016},
archivePrefix = {arXiv},
       eprint = {2209.02817},
 primaryClass = {astro-ph.HE},
}

@ARTICLE{2016ApJ...831..184B,
       author = {{Bogdanov}, Slavko and {Heinke}, Craig O. and {{\"O}zel}, Feryal and {G{\"u}ver}, Tolga},
        title = "{Neutron Star Mass-Radius Constraints of the Quiescent Low-mass X-Ray Binaries X7 and X5 in the Globular Cluster 47 Tuc}",
      journal = {\apj},
         year = 2016,
        month = nov,
       volume = {831},
       number = {2},
          eid = {184},
        pages = {184},
          doi = {10.3847/0004-637X/831/2/184},
archivePrefix = {arXiv},
       eprint = {1603.01630},
 primaryClass = {astro-ph.HE},
       adsurl = {https://ui.adsabs.harvard.edu/abs/2016ApJ...831..184B},
}

@ARTICLE{2019PASA...36...10T,
       author = {{Thrane}, Eric and {Talbot}, Colm},
        title = "{An introduction to Bayesian inference in gravitational-wave astronomy: Parameter estimation, model selection, and hierarchical models}",
      journal = {\pasa},
         year = 2019,
        month = mar,
       volume = {36},
          eid = {e010},
        pages = {e010},
          doi = {10.1017/pasa.2019.2},
archivePrefix = {arXiv},
       eprint = {1809.02293},
 primaryClass = {astro-ph.IM},
       adsurl = {https://ui.adsabs.harvard.edu/abs/2019PASA...36...10T}
}

@ARTICLE{2021ApJ...908L..28N,
       author = {{Nathanail}, Antonios and {Most}, Elias R. and {Rezzolla}, Luciano},
        title = "{GW170817 and GW190814: Tension on the Maximum Mass}",
      journal = {\apjl},
         year = 2021,
        month = feb,
       volume = {908},
       number = {2},
          eid = {L28},
        pages = {L28},
          doi = {10.3847/2041-8213/abdfc6},
archivePrefix = {arXiv},
       eprint = {2101.01735},
 primaryClass = {astro-ph.HE},
       adsurl = {https://ui.adsabs.harvard.edu/abs/2021ApJ...908L..28N},
}

@article{PhysRev.116.1027,
  title = {General Relativistic Fluid Spheres},
  author = {Buchdahl, H. A.},
  journal = {Phys. Rev.},
  volume = {116},
  issue = {4},
  pages = {1027--1034},
  numpages = {0},
  year = {1959},
  month = {Nov},
  publisher = {American Physical Society},
  doi = {10.1103/PhysRev.116.1027},
  url = {https://link.aps.org/doi/10.1103/PhysRev.116.1027}
}

@ARTICLE{2012ARNPS..62..485L,
       author = {{Lattimer}, James M.},
        title = "{The Nuclear Equation of State and Neutron Star Masses}",
      journal = {\arnps},
         year = 2012,
        month = nov,
       volume = {62},
       number = {1},
        pages = {485-515},
          doi = {10.1146/annurev-nucl-102711-095018},
archivePrefix = {arXiv},
       eprint = {1305.3510},
 primaryClass = {nucl-th},
       adsurl = {https://ui.adsabs.harvard.edu/abs/2012ARNPS..62..485L},
}

@ARTICLE{2023A&A...676A..65S,
       author = {{Sotani}, H. and {Kokkotas}, K.~D. and {Stergioulas}, N.},
        title = "{Neutron star mass-radius constraints using the high-frequency quasi-periodic oscillations of GRB 200415A}",
      journal = {\aap},
         year = 2023,
        month = aug,
       volume = {676},
          eid = {A65},
        pages = {A65},
          doi = {10.1051/0004-6361/202346360},
archivePrefix = {arXiv},
       eprint = {2303.03150},
 primaryClass = {astro-ph.HE},
       adsurl = {https://ui.adsabs.harvard.edu/abs/2023A&A...676A..65S},
}

@ARTICLE{2023ApJ...951L...9A,
       author = {{Agazie}, Gabriella and {Alam}, Md Faisal and {Anumarlapudi}, Akash and {Archibald}, Anne M. and {Arzoumanian}, Zaven and {Baker}, Paul T. and {Blecha}, Laura and {Bonidie}, Victoria and {Brazier}, Adam and {Brook}, Paul R. and {Burke-Spolaor}, Sarah and {B{\'e}csy}, Bence and {Chapman}, Christopher and {Charisi}, Maria and {Chatterjee}, Shami and {Cohen}, Tyler and {Cordes}, James M. and {Cornish}, Neil J. and {Crawford}, Fronefield and {Cromartie}, H. Thankful and {Crowter}, Kathryn and {Decesar}, Megan E. and {Demorest}, Paul B. and {Dolch}, Timothy and {Drachler}, Brendan and {Ferrara}, Elizabeth C. and {Fiore}, William and {Fonseca}, Emmanuel and {Freedman}, Gabriel E. and {Garver-Daniels}, Nate and {Gentile}, Peter A. and {Glaser}, Joseph and {Good}, Deborah C. and {G{\"u}ltekin}, Kayhan and {Hazboun}, Jeffrey S. and {Jennings}, Ross J. and {Jessup}, Cody and {Johnson}, Aaron D. and {Jones}, Megan L. and {Kaiser}, Andrew R. and {Kaplan}, David L. and {Kelley}, Luke Zoltan and {Kerr}, Matthew and {Key}, Joey S. and {Kuske}, Anastasia and {Laal}, Nima and {Lam}, Michael T. and {Lamb}, William G. and {Lazio}, T. Joseph W. and {Lewandowska}, Natalia and {Lin}, Ye and {Liu}, Tingting and {Lorimer}, Duncan R. and {Luo}, Jing and {Lynch}, Ryan S. and {Ma}, Chung-Pei and {Madison}, Dustin R. and {Maraccini}, Kaleb and {McEwen}, Alexander and {McKee}, James W. and {McLaughlin}, Maura A. and {McMann}, Natasha and {Meyers}, Bradley W. and {Mingarelli}, Chiara M.~F. and {Mitridate}, Andrea and {Ng}, Cherry and {Nice}, David J. and {Ocker}, Stella Koch and {Olum}, Ken D. and {Panciu}, Elisa and {Pennucci}, Timothy T. and {Perera}, Benetge B.~P. and {Pol}, Nihan S. and {Radovan}, Henri A. and {Ransom}, Scott M. and {Ray}, Paul S. and {Romano}, Joseph D. and {Salo}, Laura and {Sardesai}, Shashwat C. and {Schmiedekamp}, Carl and {Schmiedekamp}, Ann and {Schmitz}, Kai and {Shapiro-Albert}, Brent J. and {Siemens}, Xavier and {Simon}, Joseph and {Siwek}, Magdalena S. and {Stairs}, Ingrid H. and {Stinebring}, Daniel R. and {Stovall}, Kevin and {Susobhanan}, Abhimanyu and {Swiggum}, Joseph K. and {Taylor}, Stephen R. and {Turner}, Jacob E. and {Unal}, Caner and {Vallisneri}, Michele and {Vigeland}, Sarah J. and {Wahl}, Haley M. and {Wang}, Qiaohong and {Witt}, Caitlin A. and {Young}, Olivia and {Nanograv Collaboration}},
        title = "{The NANOGrav 15 yr Data Set: Observations and Timing of 68 Millisecond Pulsars}",
      journal = {\apjl},
         year = 2023,
        month = jul,
       volume = {951},
       number = {1},
          eid = {L9},
        pages = {L9},
          doi = {10.3847/2041-8213/acda9a},
archivePrefix = {arXiv},
       eprint = {2306.16217},
 primaryClass = {astro-ph.HE},
       adsurl = {https://ui.adsabs.harvard.edu/abs/2023ApJ...951L...9A}
}

@article{PhysRevD.110.123009,
  title = {Nontrivial features in the speed of sound inside neutron stars},
  author = {Mroczek, D. and Miller, M. C. and Noronha-Hostler, J. and Yunes, N.},
  journal = {Phys. Rev. D},
  volume = {110},
  issue = {12},
  pages = {123009},
  numpages = {36},
  year = {2024},
  month = {Dec},
  publisher = {American Physical Society},
  doi = {10.1103/PhysRevD.110.123009},
  url = {https://link.aps.org/doi/10.1103/PhysRevD.110.123009}
}

@ARTICLE{2020arXiv200101747W,
       author = {{Wysocki}, Daniel and {O'Shaughnessy}, Richard and {Wade}, Leslie and {Lange}, Jacob},
        title = "{Inferring the neutron star equation of state simultaneously with the population of merging neutron stars}",
      journal = {arXiv},
         year = 2020,
        month = jan,
          eid = {arXiv:2001.01747},
        pages = {arXiv:2001.01747},
          doi = {10.48550/arXiv.2001.01747},
archivePrefix = {arXiv},
       eprint = {2001.01747},
 primaryClass = {gr-qc},
}

@ARTICLE{2020PhRvD.101l3007L,
       author = {{Landry}, Philippe and {Essick}, Reed and {Chatziioannou}, Katerina},
        title = "{Nonparametric constraints on neutron star matter with existing and upcoming gravitational wave and pulsar observations}",
      journal = {\prd},
         year = 2020,
        month = jun,
       volume = {101},
       number = {12},
          eid = {123007},
        pages = {123007},
          doi = {10.1103/PhysRevD.101.123007},
archivePrefix = {arXiv},
       eprint = {2003.04880},
 primaryClass = {astro-ph.HE},
       adsurl = {https://ui.adsabs.harvard.edu/abs/2020PhRvD.101l3007L}
}

@ARTICLE{2020PhRvD.102f4063C,
       author = {{Chatziioannou}, Katerina and {Farr}, Will M.},
        title = "{Inferring the maximum and minimum mass of merging neutron stars with gravitational waves}",
      journal = {\prd},
         year = 2020,
        month = sep,
       volume = {102},
       number = {6},
          eid = {064063},
        pages = {064063},
          doi = {10.1103/PhysRevD.102.064063},
archivePrefix = {arXiv},
       eprint = {2005.00482},
 primaryClass = {astro-ph.HE},
       adsurl = {https://ui.adsabs.harvard.edu/abs/2020PhRvD.102f4063C}
}

@ARTICLE{2020A&A...639A.123K,
       author = {{Kruckow}, Matthias U.},
        title = "{Masses of double neutron star mergers}",
      journal = {\aap},
         year = 2020,
        month = jul,
       volume = {639},
          eid = {A123},
        pages = {A123},
          doi = {10.1051/0004-6361/202037519},
archivePrefix = {arXiv},
       eprint = {2002.08011},
 primaryClass = {astro-ph.SR},
       adsurl = {https://ui.adsabs.harvard.edu/abs/2020A&A...639A.123K}
}

@ARTICLE{2020Natur.583..211F,
       author = {{Ferdman}, R.~D. and {Freire}, P.~C.~C. and {Perera}, B.~B.~P. and {Pol}, N. and {Camilo}, F. and {Chatterjee}, S. and {Cordes}, J.~M. and {Crawford}, F. and {Hessels}, J.~W.~T. and {Kaspi}, V.~M. and {McLaughlin}, M.~A. and {Parent}, E. and {Stairs}, I.~H. and {van Leeuwen}, J.},
        title = "{Asymmetric mass ratios for bright double neutron-star mergers}",
      journal = {\nat},
         year = 2020,
        month = jul,
       volume = {583},
       number = {7815},
        pages = {211-214},
          doi = {10.1038/s41586-020-2439-x},
archivePrefix = {arXiv},
       eprint = {2007.04175},
 primaryClass = {astro-ph.HE},
       adsurl = {https://ui.adsabs.harvard.edu/abs/2020Natur.583..211F}
}

@ARTICLE{2010ApJ...716..615O,
       author = {{O'Shaughnessy}, R. and {Kalogera}, V. and {Belczynski}, Krzysztof},
        title = "{Binary Compact Object Coalescence Rates: The Role of Elliptical Galaxies}",
      journal = {\apj},
         year = 2010,
        month = jun,
       volume = {716},
       number = {1},
        pages = {615-633},
          doi = {10.1088/0004-637X/716/1/615},
archivePrefix = {arXiv},
       eprint = {0908.3635},
 primaryClass = {astro-ph.CO},
       adsurl = {https://ui.adsabs.harvard.edu/abs/2010ApJ...716..615O}
}

@ARTICLE{2023arXiv230210350B,
       author = {{Bartos}, I. and {Rosswog}, S. and {Gayathri}, V. and {Miller}, M.~C. and {Veske}, D. and {Marka}, S.},
        title = "{Hierarchical Triples as Early Sources of $r$-process Elements}",
      journal = {arXiv e-prints},
         year = 2023,
        month = feb,
          eid = {arXiv:2302.10350},
        pages = {arXiv:2302.10350},
          doi = {10.48550/arXiv.2302.10350},
archivePrefix = {arXiv},
       eprint = {2302.10350},
 primaryClass = {astro-ph.HE},
       adsurl = {https://ui.adsabs.harvard.edu/abs/2023arXiv230210350B}
}

@ARTICLE{2023arXiv230705376H,
       author = {{Huxford}, Rachael and {Kashyap}, Rahul and {Borhanian}, Ssohrab and {Dhani}, Arnab and {Sathyaprakash}, B.~S.},
        title = "{The Accuracy of Neutron Star Radius Measurement with the Next Generation of Terrestrial Gravitational-Wave Observatories}",
      journal = {arXiv e-prints},
         year = 2023,
        month = jul,
          eid = {arXiv:2307.05376},
        pages = {arXiv:2307.05376},
          doi = {10.48550/arXiv.2307.05376},
archivePrefix = {arXiv},
       eprint = {2307.05376},
 primaryClass = {gr-qc},
       adsurl = {https://ui.adsabs.harvard.edu/abs/2023arXiv230705376H}
}

@ARTICLE{2023PhRvC.108b5811R,
       author = {{Rose}, Henrik and {Kunert}, Nina and {Dietrich}, Tim and {Pang}, Peter T.~H. and {Smith}, Rory and {Van Den Broeck}, Chris and {Gandolfi}, Stefano and {Tews}, Ingo},
        title = "{Revealing the strength of three-nucleon interactions with the proposed Einstein Telescope}",
      journal = {\prc},
         year = 2023,
        month = aug,
       volume = {108},
       number = {2},
          eid = {025811},
        pages = {025811},
          doi = {10.1103/PhysRevC.108.025811},
archivePrefix = {arXiv},
       eprint = {2303.11201},
 primaryClass = {astro-ph.HE},
       adsurl = {https://ui.adsabs.harvard.edu/abs/2023PhRvC.108b5811R}
}

@ARTICLE{2015PhRvD..91d3002L,
       author = {{Lackey}, Benjamin D. and {Wade}, Leslie},
        title = "{Reconstructing the neutron-star equation of state with gravitational-wave detectors from a realistic population of inspiralling binary neutron stars}",
      journal = {\prd},
         year = 2015,
        month = feb,
       volume = {91},
       number = {4},
          eid = {043002},
        pages = {043002},
          doi = {10.1103/PhysRevD.91.043002},
archivePrefix = {arXiv},
       eprint = {1410.8866},
 primaryClass = {gr-qc},
       adsurl = {https://ui.adsabs.harvard.edu/abs/2015PhRvD..91d3002L}
}

@article{PhysRevD.110.043013,
  title = {Precision constraints on the neutron star equation of state with third-generation gravitational-wave observatories},
  author = {Walker, Kris and Smith, Rory and Thrane, Eric and Reardon, Daniel J.},
  journal = {Phys. Rev. D},
  volume = {110},
  issue = {4},
  pages = {043013},
  numpages = {8},
  year = {2024},
  month = {Aug},
  publisher = {American Physical Society},
  doi = {10.1103/PhysRevD.110.043013},
  url = {https://link.aps.org/doi/10.1103/PhysRevD.110.043013}
}

@ARTICLE{2019MNRAS.485.5363G,
       author = {{Greif}, S.~K. and {Raaijmakers}, G. and {Hebeler}, K. and {Schwenk}, A. and {Watts}, A.~L.},
        title = "{Equation of state sensitivities when inferring neutron star and dense matter properties}",
      journal = {\mnras},
         year = 2019,
        month = jun,
       volume = {485},
       number = {4},
        pages = {5363-5376},
          doi = {10.1093/mnras/stz654},
archivePrefix = {arXiv},
       eprint = {1812.08188},
 primaryClass = {astro-ph.HE},
       adsurl = {https://ui.adsabs.harvard.edu/abs/2019MNRAS.485.5363G}
}

@ARTICLE{2020PhRvR...2c3514P,
       author = {{Pang}, Peter T.~H. and {Dietrich}, Tim and {Tews}, Ingo and {Van Den Broeck}, Chris},
        title = "{Parameter estimation for strong phase transitions in supranuclear matter using gravitational-wave astronomy}",
      journal = {\prr},
         year = 2020,
        month = sep,
       volume = {2},
       number = {3},
          eid = {033514},
        pages = {033514},
          doi = {10.1103/PhysRevResearch.2.033514},
archivePrefix = {arXiv},
       eprint = {2006.14936},
 primaryClass = {astro-ph.HE},
       adsurl = {https://ui.adsabs.harvard.edu/abs/2020PhRvR...2c3514P}
}

@ARTICLE{2024CQGra..41v5003B,
       author = {{Bandopadhyay}, Ananya and {Kacanja}, Keisi and {Somasundaram}, Rahul and {Nitz}, Alexander H. and {Brown}, Duncan A.},
        title = "{Measuring neutron star radius with second and third generation gravitational wave detector networks}",
      journal = {\cqg},
         year = 2024,
        month = nov,
       volume = {41},
       number = {22},
          eid = {225003},
        pages = {225003},
          doi = {10.1088/1361-6382/ad828a},
}

@ARTICLE{2023ApJ...955...45F,
       author = {{Finstad}, Daniel and {White}, Laurel V. and {Brown}, Duncan A.},
        title = "{Prospects for a Precise Equation of State Measurement from Advanced LIGO and Cosmic Explorer}",
      journal = {\apj},
         year = 2023,
        month = sep,
       volume = {955},
       number = {1},
          eid = {45},
        pages = {45},
          doi = {10.3847/1538-4357/acf12f},
archivePrefix = {arXiv},
       eprint = {2211.01396},
 primaryClass = {astro-ph.HE},
       adsurl = {https://ui.adsabs.harvard.edu/abs/2023ApJ...955...45F}
}

@ARTICLE{1993PhRvD..47.2198F,
       author = {{Finn}, Lee Samuel and {Chernoff}, David F.},
        title = "{Observing binary inspiral in gravitational radiation: One interferometer}",
      journal = {\prd},
         year = 1993,
        month = mar,
       volume = {47},
       number = {6},
        pages = {2198-2219},
          doi = {10.1103/PhysRevD.47.2198},
archivePrefix = {arXiv},
       eprint = {gr-qc/9301003},
 primaryClass = {gr-qc},
       adsurl = {https://ui.adsabs.harvard.edu/abs/1993PhRvD..47.2198F}
}

@ARTICLE{1994PhRvD..49.2658C,
       author = {{Cutler}, Curt and {Flanagan}, {\'E}anna E.},
        title = "{Gravitational waves from merging compact binaries: How accurately can one extract the binary's parameters from the inspiral waveform\textbackslash?}",
      journal = {\prd},
         year = 1994,
        month = mar,
       volume = {49},
       number = {6},
        pages = {2658-2697},
          doi = {10.1103/PhysRevD.49.2658},
archivePrefix = {arXiv},
       eprint = {gr-qc/9402014},
 primaryClass = {gr-qc},
       adsurl = {https://ui.adsabs.harvard.edu/abs/1994PhRvD..49.2658C}
}

@ARTICLE{2010PhRvD..82j4006O,
       author = {{O'Shaughnessy}, R. and {Vaishnav}, B. and {Healy}, J. and {Shoemaker}, D.},
        title = "{Intrinsic selection biases of ground-based gravitational wave searches for high-mass black hole-black hole mergers}",
      journal = {\prd},
         year = 2010,
        month = nov,
       volume = {82},
       number = {10},
          eid = {104006},
        pages = {104006},
          doi = {10.1103/PhysRevD.82.104006},
archivePrefix = {arXiv},
       eprint = {1007.4213},
 primaryClass = {gr-qc},
       adsurl = {https://ui.adsabs.harvard.edu/abs/2010PhRvD..82j4006O}
}

@ARTICLE{2017ApJ...844..156R,
       author = {{Raithel}, Carolyn A. and {{\"O}zel}, Feryal and {Psaltis}, Dimitrios},
        title = "{From Neutron Star Observables to the Equation of State. II. Bayesian Inference of Equation of State Pressures}",
      journal = {\apj},
         year = 2017,
        month = aug,
       volume = {844},
       number = {2},
          eid = {156},
        pages = {156},
          doi = {10.3847/1538-4357/aa7a5a},
archivePrefix = {arXiv},
       eprint = {1704.00737},
 primaryClass = {astro-ph.HE},
       adsurl = {https://ui.adsabs.harvard.edu/abs/2017ApJ...844..156R}
}

@ARTICLE{2021ApJ...908..103R,
       author = {{Raithel}, Carolyn A. and {{\"O}zel}, Feryal and {Psaltis}, Dimitrios},
        title = "{Optimized Statistical Approach for Comparing Multi-messenger Neutron Star Data}",
      journal = {\apj},
         year = 2021,
        month = feb,
       volume = {908},
       number = {1},
          eid = {103},
        pages = {103},
          doi = {10.3847/1538-4357/abd3a4},
archivePrefix = {arXiv},
       eprint = {2004.00656},
 primaryClass = {astro-ph.HE},
       adsurl = {https://ui.adsabs.harvard.edu/abs/2021ApJ...908..103R}
}

@ARTICLE{2022PhRvD.105d3016L,
       author = {{Legred}, Isaac and {Chatziioannou}, Katerina and {Essick}, Reed and {Landry}, Philippe},
        title = "{Implicit correlations within phenomenological parametric models of the neutron star equation of state}",
      journal = {\prd},
         year = 2022,
        month = feb,
       volume = {105},
       number = {4},
          eid = {043016},
        pages = {043016},
          doi = {10.1103/PhysRevD.105.043016},
archivePrefix = {arXiv},
       eprint = {2201.06791},
 primaryClass = {astro-ph.HE},
       adsurl = {https://ui.adsabs.harvard.edu/abs/2022PhRvD.105d3016L}
}

@ARTICLE{2021ARNPS..71..403D,
       author = {{Drischler}, C. and {Holt}, J.~W. and {Wellenhofer}, C.},
        title = "{Chiral Effective Field Theory and the High-Density Nuclear Equation of State}",
      journal = {\arnps},
         year = 2021,
        month = sep,
       volume = {71},
        pages = {403-432},
          doi = {10.1146/annurev-nucl-102419-041903},
archivePrefix = {arXiv},
       eprint = {2101.01709},
 primaryClass = {nucl-th},
       adsurl = {https://ui.adsabs.harvard.edu/abs/2021ARNPS..71..403D}
}

@ARTICLE{2021ARNPS..71..433L,
       author = {{Lattimer}, J.~M.},
        title = "{Neutron Stars and the Nuclear Matter Equation of State}",
      journal = {\arnps},
         year = 2021,
        month = sep,
       volume = {71},
        pages = {433-464},
          doi = {10.1146/annurev-nucl-102419-124827},
       adsurl = {https://ui.adsabs.harvard.edu/abs/2021ARNPS..71..433L}
}

@ARTICLE{2023PhRvL.130t1403R,
       author = {{Raithel}, Carolyn A. and {Most}, Elias R.},
        title = "{Degeneracy in the Inference of Phase Transitions in the Neutron Star Equation of State from Gravitational Wave Data}",
      journal = {\prl},
         year = 2023,
        month = may,
       volume = {130},
       number = {20},
          eid = {201403},
        pages = {201403},
          doi = {10.1103/PhysRevLett.130.201403},
archivePrefix = {arXiv},
       eprint = {2208.04294},
 primaryClass = {astro-ph.HE},
       adsurl = {https://ui.adsabs.harvard.edu/abs/2023PhRvL.130t1403R}
}

@ARTICLE{2003RvMP...75..121B,
       author = {{Bender}, Michael and {Heenen}, Paul-Henri and {Reinhard}, Paul-Gerhard},
        title = "{Self-consistent mean-field models for nuclear structure}",
      journal = {Reviews of Modern Physics},
         year = 2003,
        month = jan,
       volume = {75},
       number = {1},
        pages = {121-180},
          doi = {10.1103/RevModPhys.75.121},
}

@ARTICLE{2022PhRvC.106e5804A,
       author = {{Alford}, M.~G. and {Brodie}, L. and {Haber}, A. and {Tews}, I.},
        title = "{Relativistic mean-field theories for neutron-star physics based on chiral effective field theory}",
      journal = {\prc},
         year = 2022,
        month = nov,
       volume = {106},
       number = {5},
          eid = {055804},
        pages = {055804},
          doi = {10.1103/PhysRevC.106.055804},
archivePrefix = {arXiv},
       eprint = {2205.10283},
 primaryClass = {nucl-th},
       adsurl = {https://ui.adsabs.harvard.edu/abs/2022PhRvC.106e5804A}
}

@ARTICLE{2023PhRvC.108b5809Z,
       author = {{Zhu}, Zhenyu and {Li}, Ang and {Hu}, Jinniu and {Shen}, Hong},
        title = "{Equation of state of nuclear matter and neutron stars: Quark mean-field model versus relativistic mean-field model}",
      journal = {\prc},
         year = 2023,
        month = aug,
       volume = {108},
       number = {2},
          eid = {025809},
        pages = {025809},
          doi = {10.1103/PhysRevC.108.025809},
archivePrefix = {arXiv},
       eprint = {2305.16058},
 primaryClass = {nucl-th},
       adsurl = {https://ui.adsabs.harvard.edu/abs/2023PhRvC.108b5809Z}
}

@ARTICLE{2023ApJ...943..163Z,
       author = {{Zhu}, Zhenyu and {Li}, Ang and {Liu}, Tong},
        title = "{A Bayesian Inference of a Relativistic Mean-field Model of Neutron Star Matter from Observations of NICER and GW170817/AT2017gfo}",
      journal = {\apj},
         year = 2023,
        month = feb,
       volume = {943},
       number = {2},
          eid = {163},
        pages = {163},
          doi = {10.3847/1538-4357/acac1f},
archivePrefix = {arXiv},
       eprint = {2211.02007},
 primaryClass = {astro-ph.HE},
       adsurl = {https://ui.adsabs.harvard.edu/abs/2023ApJ...943..163Z}
}

@ARTICLE{2018PhRvC..97b5805M,
       author = {{Margueron}, J{\'e}r{\^o}me and {Hoffmann Casali}, Rudiney and {Gulminelli}, Francesca},
        title = "{Equation of state for dense nucleonic matter from metamodeling. I. Foundational aspects}",
      journal = {\prc},
         year = 2018,
        month = feb,
       volume = {97},
       number = {2},
          eid = {025805},
        pages = {025805},
          doi = {10.1103/PhysRevC.97.025805},
archivePrefix = {arXiv},
       eprint = {1708.06894},
 primaryClass = {nucl-th},
       adsurl = {https://ui.adsabs.harvard.edu/abs/2018PhRvC..97b5805M}
}

@ARTICLE{2024MNRAS.529.4650H,
       author = {{Huang}, Chun and {Raaijmakers}, Geert and {Watts}, Anna L. and {Tolos}, Laura and {Provid{\^e}ncia}, Constan{\c{c}}a},
        title = "{Constraining a relativistic mean field model using neutron star mass-radius measurements I: nucleonic models}",
      journal = {\mnras},
         year = 2024,
        month = apr,
       volume = {529},
       number = {4},
        pages = {4650-4665},
          doi = {10.1093/mnras/stae844},
}

@ARTICLE{2023PhRvD.107j3018M,
       author = {{Malik}, Tuhin and {Ferreira}, M{\'a}rcio and {Albino}, Milena Bastos and {Provid{\^e}ncia}, Constan{\c{c}}a},
        title = "{Spanning the full range of neutron star properties within a microscopic description}",
      journal = {\prd},
         year = 2023,
        month = may,
       volume = {107},
       number = {10},
          eid = {103018},
        pages = {103018},
          doi = {10.1103/PhysRevD.107.103018},
archivePrefix = {arXiv},
       eprint = {2301.08169},
 primaryClass = {nucl-th},
       adsurl = {https://ui.adsabs.harvard.edu/abs/2023PhRvD.107j3018M}
}

@ARTICLE{2024MNRAS.530.2336R,
       author = {{Rosswog}, S. and {Diener}, P. and {Torsello}, F. and {Tauris}, T.~M. and {Sarin}, N.},
        title = "{Mergers of double NSs with one high-spin component: brighter kilonovae and fallback accretion, weaker gravitational waves}",
      journal = {\mnras},
         year = 2024,
        month = may,
       volume = {530},
       number = {2},
        pages = {2336-2354},
          doi = {10.1093/mnras/stae454},
}

@ARTICLE{2022MNRAS.513.3646P,
       author = {{Papenfort}, L. Jens and {Most}, Elias R. and {Tootle}, Samuel and {Rezzolla}, Luciano},
        title = "{Impact of extreme spins and mass ratios on the post-merger observables of high-mass binary neutron stars}",
      journal = {\mnras},
         year = 2022,
        month = jul,
       volume = {513},
       number = {3},
        pages = {3646-3662},
          doi = {10.1093/mnras/stac964},
archivePrefix = {arXiv},
       eprint = {2201.03632},
 primaryClass = {astro-ph.HE},
       adsurl = {https://ui.adsabs.harvard.edu/abs/2022MNRAS.513.3646P}
}

@ARTICLE{1989RPPh...52..439R,
       author = {{Reinhard}, P. -G.},
        title = "{The relativistic mean-field description of nuclei and nuclear dynamics}",
      journal = {Reports on Progress in Physics},
         year = 1989,
        month = apr,
       volume = {52},
       number = {4},
        pages = {439-514},
          doi = {10.1088/0034-4885/52/4/002},
       adsurl = {https://ui.adsabs.harvard.edu/abs/1989RPPh...52..439R}
}

@ARTICLE{2024arXiv240203696C,
       author = {{Chen}, Hsin-Yu and {Landry}, Philippe and {Read}, Jocelyn S. and {Siegel}, Daniel M.},
        title = "{Inference of multi-channel r-process element enrichment in the Milky Way using binary neutron star merger observations}",
      journal = {arXiv e-prints},
         year = 2024,
        month = feb,
          eid = {arXiv:2402.03696},
        pages = {arXiv:2402.03696},
          doi = {10.48550/arXiv.2402.03696},
archivePrefix = {arXiv},
       eprint = {2402.03696},
 primaryClass = {astro-ph.HE},
       adsurl = {https://ui.adsabs.harvard.edu/abs/2024arXiv240203696C}
}

@ARTICLE{2017LRR....20....3M,
       author = {{Metzger}, Brian D.},
        title = "{Kilonovae}",
      journal = {Living Reviews in Relativity},
         year = 2017,
        month = may,
       volume = {20},
       number = {1},
          eid = {3},
        pages = {3},
          doi = {10.1007/s41114-017-0006-z},
archivePrefix = {arXiv},
       eprint = {1610.09381},
 primaryClass = {astro-ph.HE},
       adsurl = {https://ui.adsabs.harvard.edu/abs/2017LRR....20....3M}
}

@ARTICLE{2022PhRvD.105j3022K,
       author = {{Kashyap}, Rahul and {Das}, Abhishek and {Radice}, David and {Padamata}, Surendra and {Prakash}, Aviral and {Logoteta}, Domenico and {Perego}, Albino and {Godzieba}, Daniel A. and {Bernuzzi}, Sebastiano and {Bombaci}, Ignazio and {Fattoyev}, Farrukh J. and {Reed}, Brendan T. and {Schneider}, Andr{\'e} da Silva},
        title = "{Numerical relativity simulations of prompt collapse mergers: Threshold mass and phenomenological constraints on neutron star properties after GW170817}",
      journal = {\prd},
         year = 2022,
        month = may,
       volume = {105},
       number = {10},
          eid = {103022},
        pages = {103022},
          doi = {10.1103/PhysRevD.105.103022},
archivePrefix = {arXiv},
       eprint = {2111.05183},
 primaryClass = {astro-ph.HE},
       adsurl = {https://ui.adsabs.harvard.edu/abs/2022PhRvD.105j3022K}
}

@ARTICLE{2023MNRAS.519.2615E,
       author = {{Ecker}, Christian and {Rezzolla}, Luciano},
        title = "{Impact of large-mass constraints on the properties of neutron stars}",
      journal = {\mnras},
         year = 2023,
        month = feb,
       volume = {519},
       number = {2},
        pages = {2615-2622},
          doi = {10.1093/mnras/stac3755},
}

@ARTICLE{2024ApJ...961...62V,
       author = {{Vinciguerra}, Serena and {Salmi}, Tuomo and {Watts}, Anna L. and {Choudhury}, Devarshi and {Riley}, Thomas E. and {Ray}, Paul S. and {Bogdanov}, Slavko and {Kini}, Yves and {Guillot}, Sebastien and {Chakrabarty}, Deepto and {Ho}, Wynn C.~G. and {Huppenkothen}, Daniela and {Morsink}, Sharon M. and {Wadiasingh}, Zorawar and {Wolff}, Michael T.},
        title = "{An Updated Mass-Radius Analysis of the 2017-2018 NICER Data Set of PSR J0030+0451}",
      journal = {\apj},
         year = 2024,
        month = jan,
       volume = {961},
       number = {1},
          eid = {62},
        pages = {62},
          doi = {10.3847/1538-4357/acfb83},
}

@ARTICLE{2023ApJ...958...49S,
       author = {{Sagun}, Violetta and {Giangrandi}, Edoardo and {Dietrich}, Tim and {Ivanytskyi}, Oleksii and {Negreiros}, Rodrigo and {Provid{\^e}ncia}, Constan{\c{c}}a},
        title = "{What Is the Nature of the HESS J1731-347 Compact Object?}",
      journal = {\apj},
         year = 2023,
        month = nov,
       volume = {958},
       number = {1},
          eid = {49},
        pages = {49},
          doi = {10.3847/1538-4357/acfc9e},
}

@ARTICLE{2023PhLB..84438062L,
       author = {{Li}, Jia Jie and {Sedrakian}, Armen},
        title = "{Baryonic models of ultra-low-mass compact stars for the central compact object in HESS J1731-347}",
      journal = {Physics Letters B},
         year = 2023,
        month = sep,
       volume = {844},
          eid = {138062},
        pages = {138062},
          doi = {10.1016/j.physletb.2023.138062},
}

@ARTICLE{2023PhRvC.108b5806B,
       author = {{Brodie}, L. and {Haber}, A.},
        title = "{Nuclear and hybrid equations of state in light of the low-mass compact star in HESS J1731-347}",
      journal = {\prc},
         year = 2023,
        month = aug,
       volume = {108},
       number = {2},
          eid = {025806},
        pages = {025806},
          doi = {10.1103/PhysRevC.108.025806},
}

@ARTICLE{2023A&A...672L..11H,
       author = {{Horvath}, J.~E. and {Rocha}, L.~S. and {de S{\'a}}, L.~M. and {Moraes}, P.~H.~R.~S. and {Bar{\~a}o}, L.~G. and {de Avellar}, M.~G.~B. and {Bernardo}, A. and {Bachega}, R.~R.~A.},
        title = "{A light strange star in the remnant HESS J1731{\ensuremath{-}}347: Minimal consistency checks}",
      journal = {\aap},
         year = 2023,
        month = apr,
       volume = {672},
          eid = {L11},
        pages = {L11},
          doi = {10.1051/0004-6361/202345885},
}

@ARTICLE{2023ApJ...949...11J,
       author = {{Jiang}, Jin-Liang and {Ecker}, Christian and {Rezzolla}, Luciano},
        title = "{Bayesian Analysis of Neutron-star Properties with Parameterized Equations of State: The Role of the Likelihood Functions}",
      journal = {\apj},
         year = 2023,
        month = may,
       volume = {949},
       number = {1},
          eid = {11},
        pages = {11},
          doi = {10.3847/1538-4357/acc4be},
}

@article{PhysRevD.109.063017,
  title = {Hybrid stars in light of the HESS J1731-347 remnant and the PREX-II experiment},
  author = {Laskos-Patkos, P. and Koliogiannis, P. S. and Moustakidis, Ch. C.},
  journal = {Phys. Rev. D},
  volume = {109},
  issue = {6},
  pages = {063017},
  numpages = {16},
  year = {2024},
  month = {Mar},
  publisher = {American Physical Society},
  doi = {10.1103/PhysRevD.109.063017},
  url = {https://link.aps.org/doi/10.1103/PhysRevD.109.063017}
}

@ARTICLE{2023ApJ...944....7S,
       author = {{Shirke}, Swarnim and {Ghosh}, Suprovo and {Chatterjee}, Debarati},
        title = "{Constraining the Equation of State of Hybrid Stars Using Recent Information from Multidisciplinary Physics}",
      journal = {\apj},
         year = 2023,
        month = feb,
       volume = {944},
       number = {1},
          eid = {7},
        pages = {7},
          doi = {10.3847/1538-4357/acac31},
}

@ARTICLE{rift-docs,
   author={{Wagner}, K. and {O'Shaughnessy}, R. and {et al}},
   title = "{RIFT documentation}",
   journal={https://rift-documentation.readthedocs.io/en/latest/},
  year=2024,
  url={https://rift-documentation.readthedocs.io/en/latest/}
}

@ARTICLE{2024ApJ...971L..20C,
       author = {{Choudhury}, Devarshi and {Salmi}, Tuomo and {Vinciguerra}, Serena and {Riley}, Thomas E. and {Kini}, Yves and {Watts}, Anna L. and {Dorsman}, Bas and {Bogdanov}, Slavko and {Guillot}, Sebastien and {Ray}, Paul S. and {Reardon}, Daniel J. and {Remillard}, Ronald A. and {Bilous}, Anna V. and {Huppenkothen}, Daniela and {Lattimer}, James M. and {Rutherford}, Nathan and {Arzoumanian}, Zaven and {Gendreau}, Keith C. and {Morsink}, Sharon M. and {Ho}, Wynn C.~G.},
        title = "{A NICER View of the Nearest and Brightest Millisecond Pulsar: PSR J0437{\textendash}4715}",
      journal = {\apjl},
         year = 2024,
        month = aug,
       volume = {971},
       number = {1},
          eid = {L20},
        pages = {L20},
          doi = {10.3847/2041-8213/ad5a6f},
}

@ARTICLE{2023PASA...40...49Z,
       author = {{Zic}, Andrew and {Reardon}, Daniel J. and {Kapur}, Agastya and {Hobbs}, George and {Mandow}, Rami and {Cury{\l}o}, Ma{\l}gorzata and {Shannon}, Ryan M. and {Askew}, Jacob and {Bailes}, Matthew and {Bhat}, N.~D. Ramesh and {Cameron}, Andrew and {Chen}, Zu-Cheng and {Dai}, Shi and {Di Marco}, Valentina and {Feng}, Yi and {Kerr}, Matthew and {Kulkarni}, Atharva and {Lower}, Marcus E. and {Luo}, Rui and {Manchester}, Richard N. and {Miles}, Matthew T. and {Nathan}, Rowina S. and {Os{\l}owski}, Stefan and {Rogers}, Axl F. and {Russell}, Christopher J. and {Sarkissian}, John M. and {Shamohammadi}, Mohsen and {Spiewak}, Ren{\'e}e and {Thyagarajan}, Nithyanandan and {Toomey}, Lawrence and {Wang}, Shuangqiang and {Zhang}, Lei and {Zhang}, Songbo and {Zhu}, Xing-Jiang},
        title = "{The Parkes Pulsar Timing Array third data release}",
      journal = {\pasa},
         year = 2023,
        month = dec,
       volume = {40},
          eid = {e049},
        pages = {e049},
          doi = {10.1017/pasa.2023.36},
}

@article{PhysRevC.86.015803,
  title = {Constraints on the symmetry energy and neutron skins from experiments and theory},
  author = {Tsang, M. B. and Stone, J. R. and Camera, F. and Danielewicz, P. and Gandolfi, S. and Hebeler, K. and Horowitz, C. J. and Lee, Jenny and Lynch, W. G. and Kohley, Z. and Lemmon, R. and M\"oller, P. and Murakami, T. and Riordan, S. and Roca-Maza, X. and Sammarruca, F. and Steiner, A. W. and Vida\~na, I. and Yennello, S. J.},
  journal = {Phys. Rev. C},
  volume = {86},
  issue = {1},
  pages = {015803},
  numpages = {10},
  year = {2012},
  month = {Jul},
  publisher = {American Physical Society},
  doi = {10.1103/PhysRevC.86.015803},
  url = {https://link.aps.org/doi/10.1103/PhysRevC.86.015803}
}

@ARTICLE{2019EPJA...55..117L,
       author = {{Li}, Bao-An and {Krastev}, Plamen G. and {Wen}, De-Hua and {Zhang}, Nai-Bo},
        title = "{Towards understanding astrophysical effects of nuclear symmetry energy}",
      journal = {European Physical Journal A},
         year = 2019,
        month = jul,
       volume = {55},
       number = {7},
          eid = {117},
        pages = {117},
          doi = {10.1140/epja/i2019-12780-8},
}

@ARTICLE{2024ApJ...976...58S,
       author = {{Salmi}, Tuomo and {Deneva}, Julia S. and {Ray}, Paul S. and {Watts}, Anna L. and {Choudhury}, Devarshi and {Kini}, Yves and {Vinciguerra}, Serena and {Cromartie}, H. Thankful and {Wolff}, Michael T. and {Arzoumanian}, Zaven and {Bogdanov}, Slavko and {Gendreau}, Keith and {Guillot}, Sebastien and {Ho}, Wynn C.~G. and {Morsink}, Sharon M. and {Cognard}, Isma{\"e}l and {Guillemot}, Lucas and {Theureau}, Gilles and {Kerr}, Matthew},
        title = "{A NICER View of PSR J1231‑1411: A Complex Case}",
      journal = {\apj},
         year = 2024,
        month = nov,
       volume = {976},
       number = {1},
          eid = {58},
        pages = {58},
          doi = {10.3847/1538-4357/ad81d2},
archivePrefix = {arXiv},
       eprint = {2409.14923},
 primaryClass = {astro-ph.HE},
       adsurl = {https://ui.adsabs.harvard.edu/abs/2024ApJ...976...58S}
}

@ARTICLE{2025arXiv250614883M,
       author = {{Mauviard}, Lucien and {Guillot}, Sebastien and {Salmi}, Tuomo and {Choudhury}, Devarshi and {Dorsman}, Bas and {Gonz{\'a}lez-Caniulef}, Denis and {Hoogkamer}, Mariska and {Huppenkothen}, Daniela and {Kazantsev}, Christine and {Kini}, Yves and {Olive}, Jean-Francois and {Stammler}, Pierre and {Watts}, Anna L. and {Mendes}, Melissa and {Rutherford}, Nathan and {Schwenk}, Achim and {Svensson}, Isak and {Bogdanov}, Slavko and {Kerr}, Matthew and {Ray}, Paul S. and {Guillemot}, Lucas and {Cognard}, Isma{\"e}l and {Theureau}, Gilles},
        title = "{A NICER view of the 1.4 solar-mass edge-on pulsar PSR J0614--3329}",
      journal = {arXiv e-prints},
         year = 2025,
        month = jun,
          eid = {arXiv:2506.14883},
        pages = {arXiv:2506.14883},
          doi = {10.48550/arXiv.2506.14883},
archivePrefix = {arXiv},
       eprint = {2506.14883},
 primaryClass = {astro-ph.HE},
       adsurl = {https://ui.adsabs.harvard.edu/abs/2025arXiv250614883M}
}

@ARTICLE{2024ApJ...971L..19R,
       author = {{Rutherford}, Nathan and {Mendes}, Melissa and {Svensson}, Isak and {Schwenk}, Achim and {Watts}, Anna L. and {Hebeler}, Kai and {Keller}, Jonas and {Prescod-Weinstein}, Chanda and {Choudhury}, Devarshi and {Raaijmakers}, Geert and {Salmi}, Tuomo and {Timmerman}, Patrick and {Vinciguerra}, Serena and {Guillot}, Sebastien and {Lattimer}, James M.},
        title = "{Constraining the Dense Matter Equation of State with New NICER Mass{\textendash}Radius Measurements and New Chiral Effective Field Theory Inputs}",
      journal = {\apjl},
         year = 2024,
        month = aug,
       volume = {971},
       number = {1},
          eid = {L19},
        pages = {L19},
          doi = {10.3847/2041-8213/ad5f02},
archivePrefix = {arXiv},
       eprint = {2407.06790},
 primaryClass = {astro-ph.HE},
       adsurl = {https://ui.adsabs.harvard.edu/abs/2024ApJ...971L..19R}
}

@ARTICLE{2025PhLB..86539501L,
       author = {{Li}, Jia-Jie and {Tian}, Yu and {Sedrakian}, Armen},
        title = "{Bayesian constraints on covariant density functional equations of state of compact stars with new NICER mass-radius measurements}",
      journal = {Physics Letters B},
         year = 2025,
        month = jun,
       volume = {865},
          eid = {139501},
        pages = {139501},
          doi = {10.1016/j.physletb.2025.139501},
archivePrefix = {arXiv},
       eprint = {2412.16513},
 primaryClass = {hep-ph},
       adsurl = {https://ui.adsabs.harvard.edu/abs/2025PhLB..86539501L}
}

@ARTICLE{2025PhRvX..15b1014K,
       author = {{Koehn}, Hauke and {Rose}, Henrik and {Pang}, Peter T.~H. and {Somasundaram}, Rahul and {Reed}, Brendan T. and {Tews}, Ingo and {Abac}, Adrian and {Komoltsev}, Oleg and {Kunert}, Nina and {Kurkela}, Aleksi and {Coughlin}, Michael W. and {Healy}, Brian F. and {Dietrich}, Tim},
        title = "{From Existing and New Nuclear and Astrophysical Constraints to Stringent Limits on the Equation of State of Neutron-Rich Dense Matter}",
      journal = {Physical Review X},
         year = 2025,
        month = apr,
       volume = {15},
       number = {2},
          eid = {021014},
        pages = {021014},
          doi = {10.1103/PhysRevX.15.021014},
archivePrefix = {arXiv},
       eprint = {2402.04172},
 primaryClass = {astro-ph.HE},
       adsurl = {https://ui.adsabs.harvard.edu/abs/2025PhRvX..15b1014K}
}

@ARTICLE{2023PhRvD.107b4040W,
       author = {{Wofford}, J. and {Yelikar}, A.~B. and {Gallagher}, Hannah and {Champion}, E. and {Wysocki}, D. and {Delfavero}, V. and {Lange}, J. and {Rose}, C. and {Valsan}, V. and {Morisaki}, S. and {Read}, J. and {Henshaw}, C. and {O'Shaughnessy}, R.},
        title = "{Improving performance for gravitational-wave parameter inference with an efficient and highly-parallelized algorithm}",
      journal = {\prd},
         year = 2023,
        month = jan,
       volume = {107},
       number = {2},
          eid = {024040},
        pages = {024040},
          doi = {10.1103/PhysRevD.107.024040},
archivePrefix = {arXiv},
       eprint = {2210.07912},
 primaryClass = {gr-qc},
       adsurl = {https://ui.adsabs.harvard.edu/abs/2023PhRvD.107b4040W}
}

@ARTICLE{2024arXiv240715753V,
       author = {{Vilkha}, Askold and {Yelikar}, Anjali and {O'Shaughnessy}, Richard and {Read}, Jocelyn},
        title = "{Inference on neutron star parameters and the nuclear equation of state with RIFT, using prior EOS information}",
      journal = {arXiv e-prints},
         year = 2024,
        month = jul,
          eid = {arXiv:2407.15753},
        pages = {arXiv:2407.15753},
          doi = {10.48550/arXiv.2407.15753},
archivePrefix = {arXiv},
       eprint = {2407.15753},
 primaryClass = {gr-qc},
       adsurl = {https://ui.adsabs.harvard.edu/abs/2024arXiv240715753V}
}

@article{gwastro-RIFT_FinerNet,
  author={{Wagner}, K. and {O'Shaughnessy}, R. and {Yelikar}, A. and {Manning}, N. and {Fernando}, D. and {Lange}, J. and {Tiwari}, V. and {Fernando}, A.H.},
  title="{Narrowing RIFT: Focused simulation-based-inference for interpreting exceptional GW sources}",
  year=2025,
  month=may,
  journal={Submitted to PRD; available as arxiv:2505.11655 },
  eprint={arxiv:2505.11655 },
  url ={https://arxiv.org/abs/2505.11655},
  dcc-url={https://dcc.ligo.org/LIGO-P2500197}
  }

\appendix
\section{NICER data and Gaussian approximation}
\label{ap:nicer_gaussian}
The NICER data releases \citep{riley_thomas_e_2019_3386449,miller_m_c_2019_3473466,riley_thomas_e_2021_4697625,miller_m_c_2021_4670689} 
provide weighted posterior samples $(x_k,w_k)$ derived from comparing their pulse models and neutron star mass-radius combinations
with  observed X-ray photons. We fit a conventional two-dimensional normal distribution to these weighted samples.
For J0030, which lacks any hard edges introduced by causality constraints in the $(M,R)$ plane, we use the weighted
empirical mean
$\bar{x} = \sum_k w_k x_k/ \sum_k w_k$ and weighted empirical covariance matrix to infer the parameters of the
best-fitting gaussian.  More generally, however, we  optimize the mean $\mu$ and covariance matrix $\Sigma$ of a
standard normal distribution, according to
\begin{align}
\ell(\mu,\Sigma) = \sum_k w_k \ln p(x_k|\mu,\Sigma) ~~.
\end{align}
In the limit of many weighted samples constructed by importance sampling, with $w_k = L(x_k) $, the quantity $\ell$
will approach the KL divergence between the (truncated) Gaussian and the full posterior.
As a check on our results and their stability, we repeat our analysis with fair draws from the weighted samples, using
the  expression above with $w_k=1$.
The best fitting normal distribution in $(R/\unit{km},~M/\unit{M_\odot})$ has a mean $(13.34537396,~1.44968231)$ and a covariance matrix 
\[
\begin{bmatrix}
 1.52687831 &0.18343073 \\
 0.18343073 &0.02778826
\end{bmatrix} ~~~.
\]

\section{Understanding the impact of many observations}
\label{sec:ap:fisher}
Future instruments will  enable very high precision constraints on the nuclear equation of state, through high precision
constraints on $\Lambda(m)$.   Starting with the pioneering work using the Fisher matrix \citep{2015PhRvD..91d3002L}, several groups have
investigated how well EoS or radius constraints will improve as observations
accumulate, using various levels of approximation; see, e.g.,   \citet{2020PhRvD.101l3007L,2020PhRvR...2c3514P,2023arXiv230705376H,2024CQGra..41v5003B,2023ApJ...955...45F} and
references therein.  Unfortunately, many misconceptions about these constraints persist in the literature.  Notably,
most groups  assert that the loudest few observations dominate the overall constraint.  As discussed in \citet{2020arXiv200101747W} however, not only do the many weak observations make critical contributions to the overall EoS constraint,
but also we must use these  weak observations  to help simultaneously reconstruct the NS population.

Motivated by the need for practical projections for third-generation instruments and the many-detection era, in this
appendix we provide a simple ansatz allowing simple estimates showing how 
\emph{statistical measurement errors}  on the nuclear equation of state will improve as more observations accumulate.  
Our argument has five key ingredients.   First, we will assume all binaries are equal mass.   Second, following  \citet{Zhao:2018nyf} we will assume $\Lambda(m) =
\Lambda_*(m/m_*)^{-p}$ with $p=6$, where $\Lambda_*$ (the equation of state parameter to be measured with many
observations) is the value of the tidal deformability at a reference mass $m_*$ (assumed to be
at or close to $m_*=1.4~\unit{M_\odot}$).  Third, following the discussion in Section III.A of \citep{2020arXiv200101747W}, we will argue the marginal one-dimensional measurement accuracy for the tidal deformability
scales as $\sigma_\Lambda = \sigma_*
(m/m_*)^{-1}$ versus binary mass, adopting a fiducial uncertainty for the faintest source used in our analysis of $\sigma_*\simeq 200$; see also, e.g., Fig. 1 in \cite{2023arXiv230705376H}.  Fourth, we will assume that each independent observation's
marginal tidal deformability has a log-likelihood of the form
\begin{align}
\Gamma_{**,1} = \Big(\frac{\rho}{8}\Big)^2 \sigma_{\lambda}^{-2} \Big(\frac{d\Lambda}{d\Lambda_*}\Big)^2
 = \Big(\frac{\rho}{8}\Big)^2\sigma_*^{-2} \Big(\frac{m}{m_*}\Big)^{2-2p} ~~~.
\end{align}
Fifth, we will assume the signal amplitude $\rho$ scales with mass and luminosity distance as
$(m/m_*)^{5/6}d_{L,*}/d_L$ times geometrical angular factors of order unity, i.e.
\begin{align}
\rho=8w \Big(\frac{m}{m_*}\Big)^{5/6}d_{L,*}/d_L ~.
\end{align}
Combining these assumptions, the net marginal likelihood from many independent observations has the form
\begin{align}
\Gamma_{**,N} &= \sum_k \Gamma_{**,1,k} = N \E{\Gamma_{**,1}} \nonumber \\
\label{eq:tidal_simple_scaling}
&= N \E{w^2}\sigma_*^{-2}\E{\Big(\frac{m}{m_*}\Big)^{2/3+3-2p}}\E{\Big(\frac{d_{L,*}}{d_L}\Big)^2} 
\end{align}
where in the second line we have replaced the discrete sum by an ensemble average ($\sum_k \cdot = N \E{\cdot}$) and
for convenience in our ansatz we assume the angular, distance, and mass distributions of the $N$ detected sources are
independent.   The average $\E{w^2}$ of the
geometrical factors depends on the assumed network beampattern (and, for realistic heterogeneous networks, their duty
cycle), but is  well known to be $\E{w^2}=2/5$ for the
single-interferometer scenario;
see  \cite{1993PhRvD..47.2198F,1994PhRvD..49.2658C,2010PhRvD..82j4006O} for discussion. Unless otherwise noted, for the purposes of discussion we will assume the single-interferometer
beampattern factor, a uniform NS mass distribution between $1~\unit{M_\odot}$ and $2~\unit{M_\odot}$, and use $\E{\Big(\frac{d_{L,*}}{d_L}\Big)^2}\simeq 1$ as appropriate to
a third-generation network. 
A larger value of this expression corresponds to smaller marginal measurement accuracy $\sigma_{\Lambda_{*,N}}\equiv 1/\sqrt{\Gamma_{**,N}}$ on the
fiducial tidal deformability $\Lambda_*$.

The analytic estimate provided in Eq. (\ref{eq:tidal_simple_scaling}) lets us quickly assess the relative importance of different
contributions to the integral.  As a concrete example, the relative contribution of sources at different distances
scales as the relative contributions to $\E{1/d_L^2} $, which under the conventional assumption of an event rate
comparable to comoving volume has a critical contribution from the \emph{weakest}
sources available to the population average. For a Euclidean distribution of sources out to the
maximum detectable distance $d_{L,*}$, within our ansatz we find $\E{(d_{L,*}/d_L)^2}=3$.   Weak sources play an even more important role if sources occur more
frequently at higher redshift, as expected given the increasing star formation rate with redshift.   In the extreme
limit of strong redshift dependence and cosmological factors, we expect  $\E{(d_{L,*}/d_L)^2}\simeq 1$. The primary
limitation of this interpretation is the implicit assumption of a Gaussian joint likelihood, which only applies for loud
signals (e.g., comparable to GW170817).  Even after restricting our calculation to sources louder than $\rho_{c}$, the
detection rate of pertinent sources is only reduced by of order $(8/\rho_c)^3$.  For third-generation instruments with
$O(10^5)$ sources, many thousands must still make important contributions, even for the most conservative choices for
$\rho_c$.
Moreover, once in the Gaussian-likelihood regime, our calculation scales much more slowly than the naive Euclidean
factor, only as  $N \E{(d_{L,*}/d_L)^2}$.

Our expression provides an even stronger conclusion about the relative importance of
different parts of the mass distribution: the lowest mass binaries overwhelmingly dominate the mass average
$\E{m^{11/3-2p}}\simeq \E{m^{-8}}$ (more precisely = $\frac{1-2^{-(8-2/3)}}{9-2/3}$ for the average for a uniform mass distribution from $1~\unit{M_\odot}$ to $2~\unit{M_\odot}$). In retrospect, the lowest mass binaries must dominate: $\Lambda(m)$ is vastly
larger at the lowest allowed masses.  If any binaries exist at these masses their observations will dominate our
constraints on the equation of state because their large tidal effects will be easily differentiated from the noise.

 Our expression implies
the measurement accuracy $\sigma_{\Lambda_*,N}$ from combining $N$ measurements will be roughly 
\begin{align}
\sigma_{\Lambda_*,N} &\simeq 200/\sqrt{N}
\end{align}
Our simple estimate for the expected statistical uncertainty in this parameter after $N$ observations  is comparable to (but calibrated above to be slightly more optimistic than)  detailed calculations produced in other studies 
\citep{2023PhRvC.108b5811R,2020PhRvD.101l3007L,2023arXiv230705376H}; for example,  \cite{2020PhRvD.101l3007L} argues the
90\% credible interval is $\Delta
\Lambda_*\simeq 910/\sqrt{N}$ and thus that the comparable $68\%$ credible interval would be $520/\sqrt{N}$.
The aforementioned measurement accuracy on $\Lambda_*$  implies a relative measurement accuracy on $\ln R_*$ of  $\Lambda_*^{-1} |d\ln R_*/d\ln
\Lambda_*|/\sqrt{\Gamma_{**,N}} $.  To evaluate this expression and put the result in context, we'll assume for the purposes of illustration the MPA1 equation of state,
which produces for $m_*=1.4~\unit{M_\odot}$ the  fiducial pair $(\Lambda_*,R_*/\unit{km})=(513,12.5)$. 
For simplicity again following \cite{Zhao:2018nyf}, we can evaluate this change-of-variable expression by
adopting the same scaling $\Lambda = a(m/R)^{-p}$ assumed above for some constant $a\simeq 0.0085$, so $d\ln R_*/d\ln
\Lambda_*=p$.  We therefore estimate the measurement accuracy $\sigma_{R_*,N}$ in $R_*$ produced by stacking $N$
measurements to be
\begin{align}
\sigma_{R_*,N}&\simeq  R_*\frac{p\sigma_*
}{
\Lambda_* \sqrt{\E{w^2} N  \E{(d_{L,*}/d_L)^2} \E{\Big(\frac{m}{m_*}\Big)^{2/3+3-2p}}}} \\
&\simeq 5\unit{km}/\sqrt{N}
\end{align}
which is close to (but slightly wider than) the corresponding estimate presented in \citep{2020PhRvD.101l3007L}.
Our simple estimates can be easily updated (and compared to the latest best-fit value
$\Lambda_*$) as our understanding of the NS mass distribution and measurement accuracy improves.  They can likewise be
easily updated to include both GW and X-ray observations adding in quadrature.

Our analysis in this appendix suggests important trends in tension with previous preliminary investigations.  Many groups over several years have argued
that only the loudest handful of observations will dominate overall EoS inference \citep{2015PhRvD..91d3002L,PhysRevD.110.043013}.
Our analytic argument suggests a comparable and potentially even dominant contribution from more numerous fainter events.

In the discussion in this appendix, we have only provided estimates for the statistical accuracy $\sigma_{\Lambda_*,N}$ and
$\sigma_{R_*,N}$ for the tidal deformability and radius of a fiducial neutron star, respectively, after combining $N$
events.  Given proposed network sensitivity models and reference, these expressions can be immediately translated to the
operating time needed to constrain these quantities to any target precision.  For example, in terms of the proposed
detection rate of useful events (scaled to $10^5/\unit{yr}$, assuming all events contribute) we find for the simple
estimates discussed here that a target accuracy of $\Delta R_*=10~\unit{m}$ requires an observing time scaling with event
rate and fiducial neutron star radius as
\begin{align}
T_R &= R_d^{-1}(\sigma_{R_*,N}/\Delta R_*)^2  \nonumber \\
&\simeq  2.5 \unit{yr}(R_d \unit{yr}/10^5)^{-1}(R_*/12.5\unit{km})^2 \; .
\end{align}

\section{More alternate perspectives on GW170817}
\label{ap:170817}
In this appendix, we describe a few supplementary analyses of GW170817 which complement the material presented in the text.

\begin{figure}
\includegraphics[trim={0.5cm 0cm 1cm 1cm},clip,width=\columnwidth]{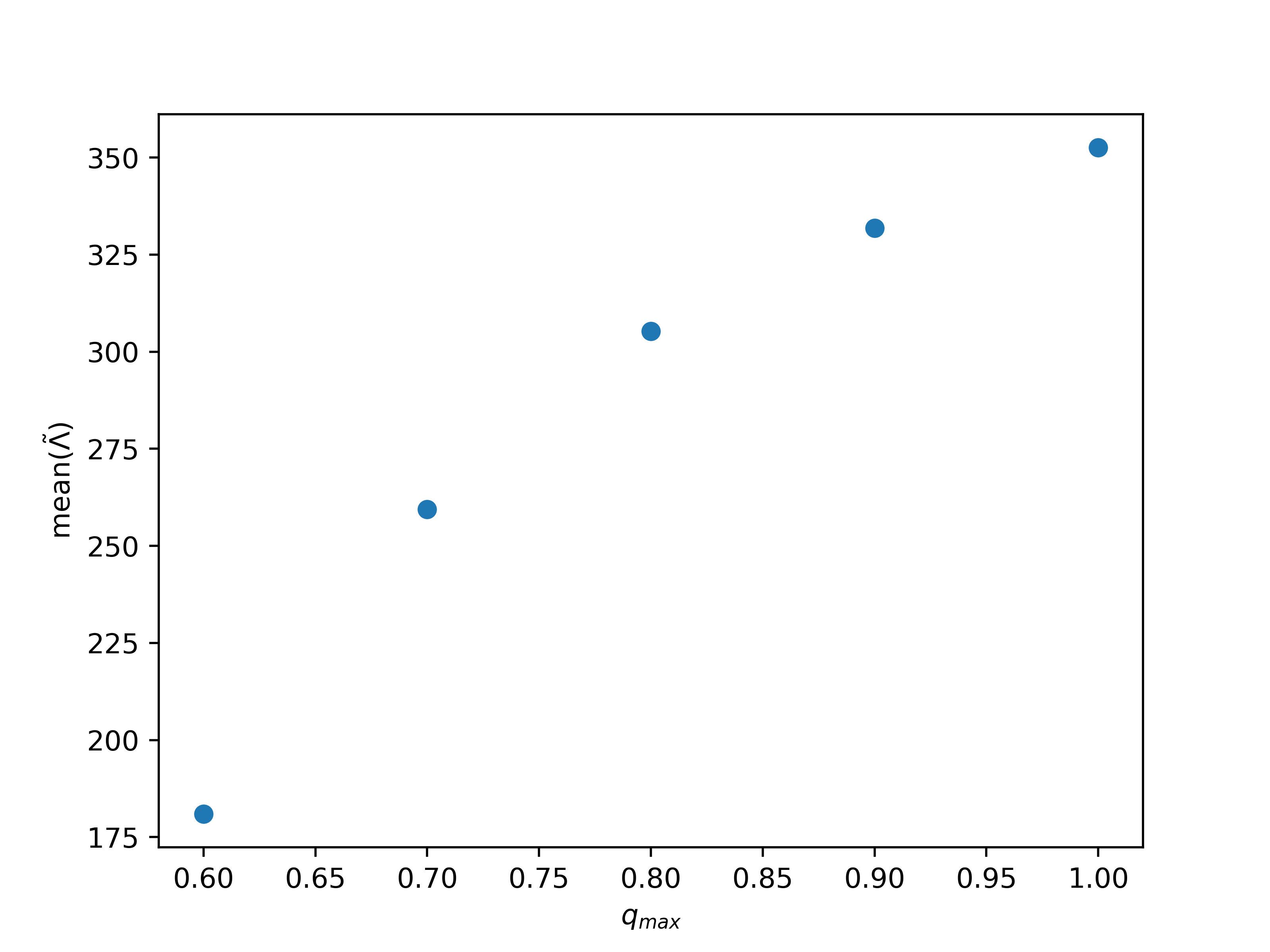}
\caption{\label{fig:ap:qmax_lambdaTilde} Median $\tilde{\Lambda}$ versus $q_{\rm max}$. 
In this figure, parameter inferences are performed with the same fiducial mass, spin, and tidal deformability priors as
adopted in Figure \ref{fig:GW_parameter_posterior}, except that the mass ratio is constrained to be less than $q_{\rm max}$.
}
\end{figure}

Figure \ref{fig:ap:qmax_lambdaTilde} shows the median $\tilde{\Lambda}$ obtained from a one-parameter family of
interpretations of GW170817, which allow for binary mass ratios $q<q_{\rm max}$ that range between the two fiducial
extremes presented in the text.  

\begin{figure}
\includegraphics[trim={0.5cm 0cm 1cm 1cm},clip,width=\columnwidth]{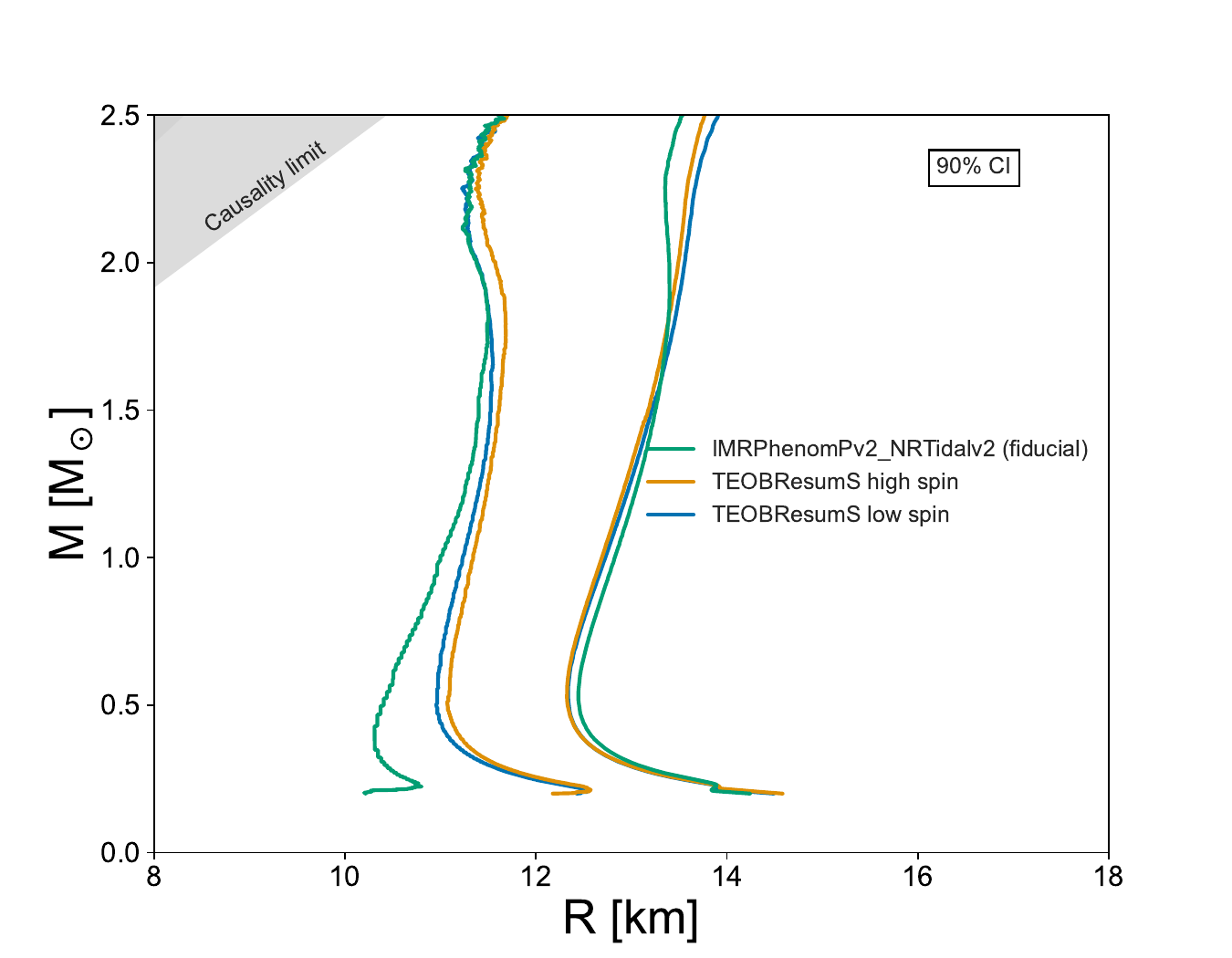}
\caption{\label{fig:ap:MR:ResumS} Mass-radius posterior distribution derived from alternative parameter inferences of
  GW170817  using a different waveform model
}
\end{figure}

Figure \ref{fig:ap:MR:ResumS} shows the results of an analysis that incorporates an alternative interpretation of
GW170817 performed by Lange et al. (in prep), obtained by comparing the TEOBResumS waveform model to the data.  The
differences in interpretation reported there propagate to  modest differences in the mass-radius posterior distribution using the baseline, broad NS population prior.



\section{Hyperpipeline examples}
\label{ap:hyperpipe}
In Section \ref{sec:sub:hyperpipe} we presented our iterative strategy for hyperparameter inference. This approach has been used for GW source parameter inference with RIFT as well as the equation of state which are presented in the Section \ref{sec:results}. Here, we provide
illustrations showing how our iterative strategy works within the context of EoS hyperparameter inference.

\begin{figure*}
\includegraphics[width=\columnwidth]{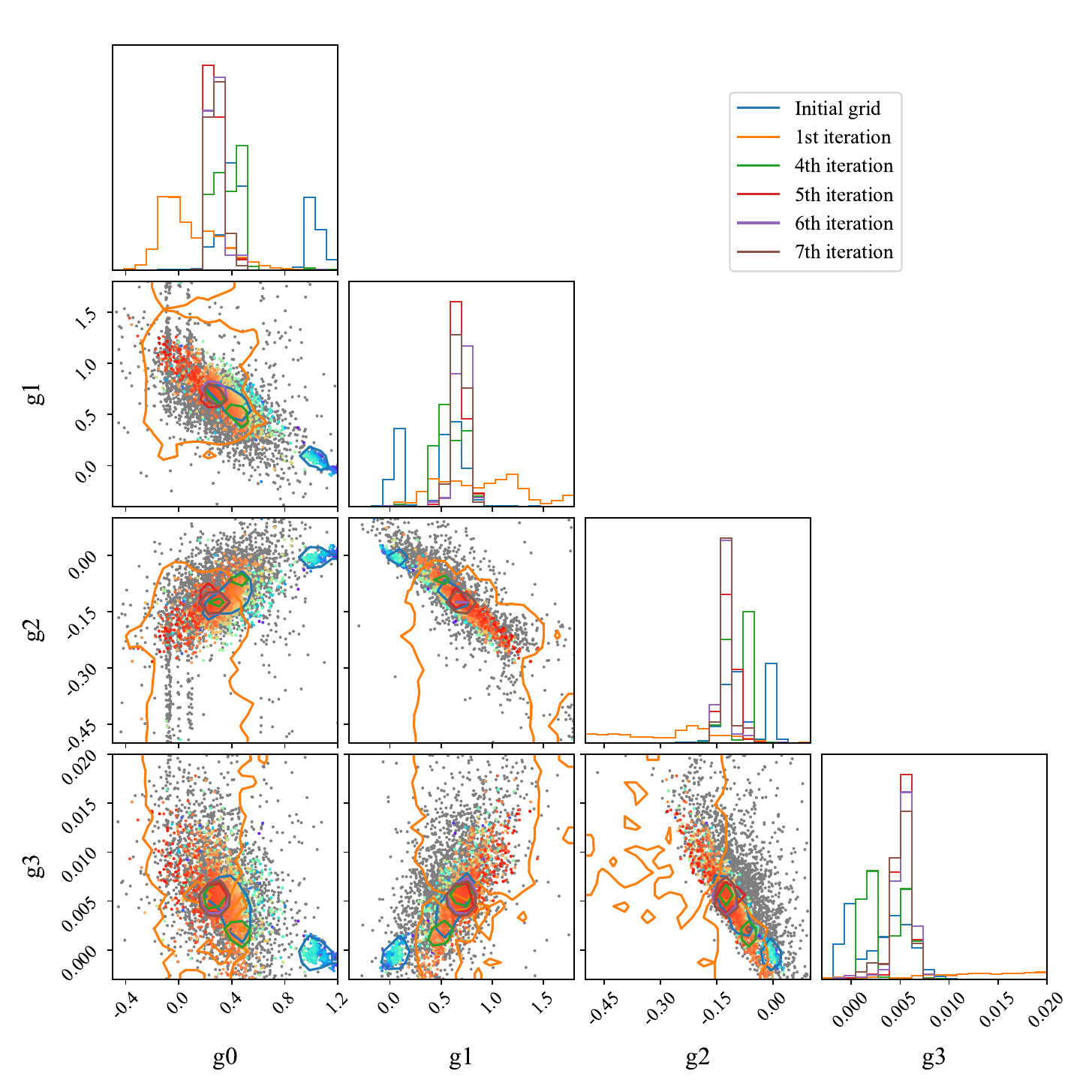}
\includegraphics[width=\columnwidth]{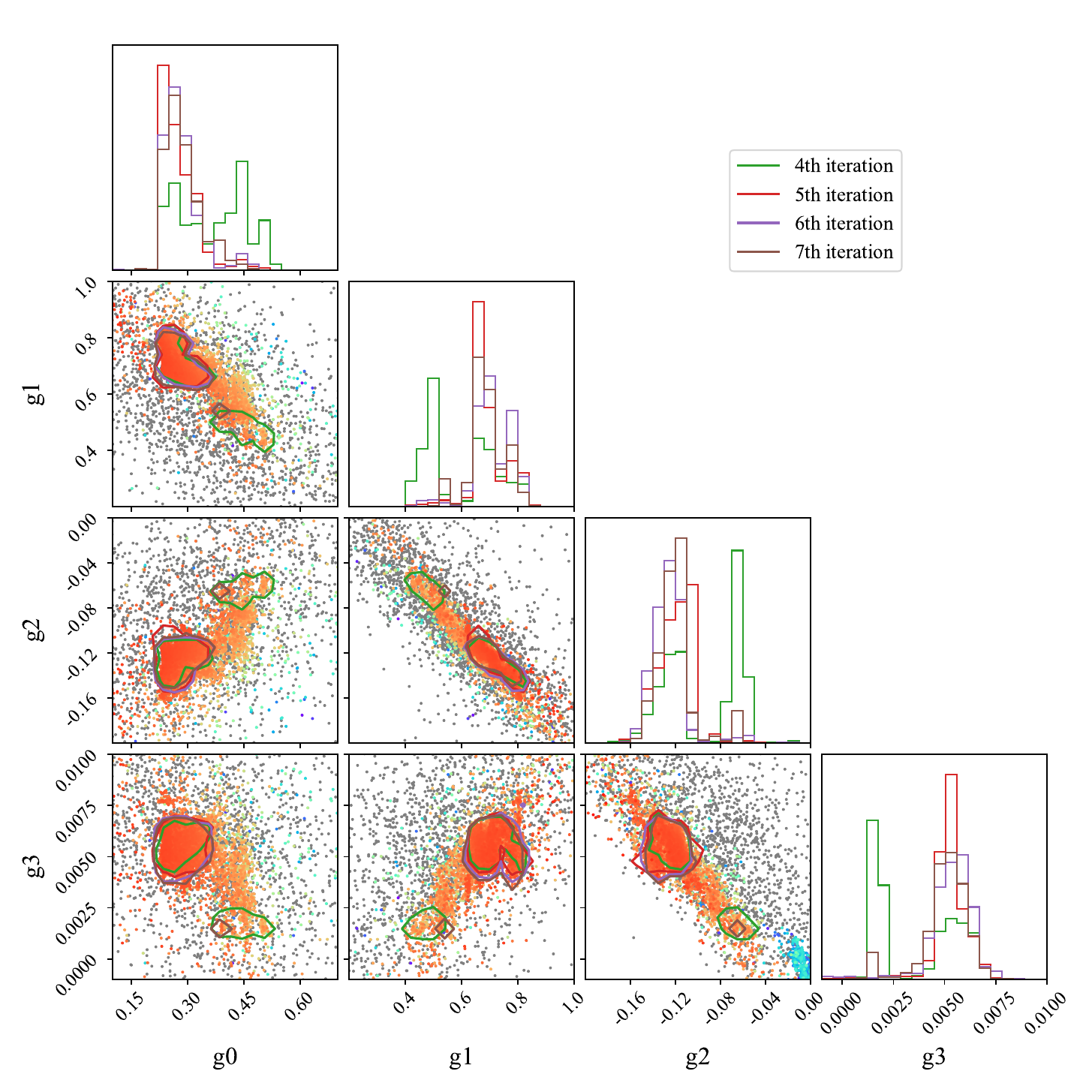}
\caption{\label{fig:joint_gw_psr:corner}Hyperparameter corner plot: Joint EoS constraint from galactic  pulsars, symmetry energy, and GW using the
  $\Gamma$-spectral EoS parameterization (the hyperparameters that lead to Fig. \ref{fig:joint_gw_psr}). The right panel zooms into the last four iterations in the first panel.}
\end{figure*}

As an example, Figure \ref{fig:joint_gw_psr:corner} shows 90\% credible intervals estimated at each stage of one
of our analyses, along with (in color) the marginal likelihoods accumulated during the analysis. Initial grid is an initial guess made during this work, however the pipeline starts exploring regions to obtain a more complete picture of the posterior. By the 5th, 6th, and 7th iteration the posterior estimation does not change significantly, leading us to a converged posterior distribution.

\section{NICER PSR-J0437}
\label{sec:sub:revised_0437}
One of the NICER collaboration groups announced the observational constraints for PSR-J0437 \citep{2024ApJ...971L..20C}. The pulsar has a strong a-priori mass constraint from radio observations \citep{2023PASA...40...49Z}. The full posterior data release has not occurred, however, here we apply this pulsar's separate 90\% credible intervals for its mass and radius and can construct an approximate Gaussian posterior to explore its effect on the EoS. The reported Radius is $R=11.36^{+0.95}_{-0.63}~\unit{km}$ and $M = 1.418 \pm 0.037 ~\unit{M_\odot}$. For this preliminary application, we employ the constraints as a Gaussian with $R=11.36 \pm 0.95~\unit{km}$ and $M = 1.418 \pm 0.037 ~\unit{M_\odot}$.

\begin{figure}
\includegraphics[trim={0.5cm 0cm 1cm 1cm},clip,width=\columnwidth]{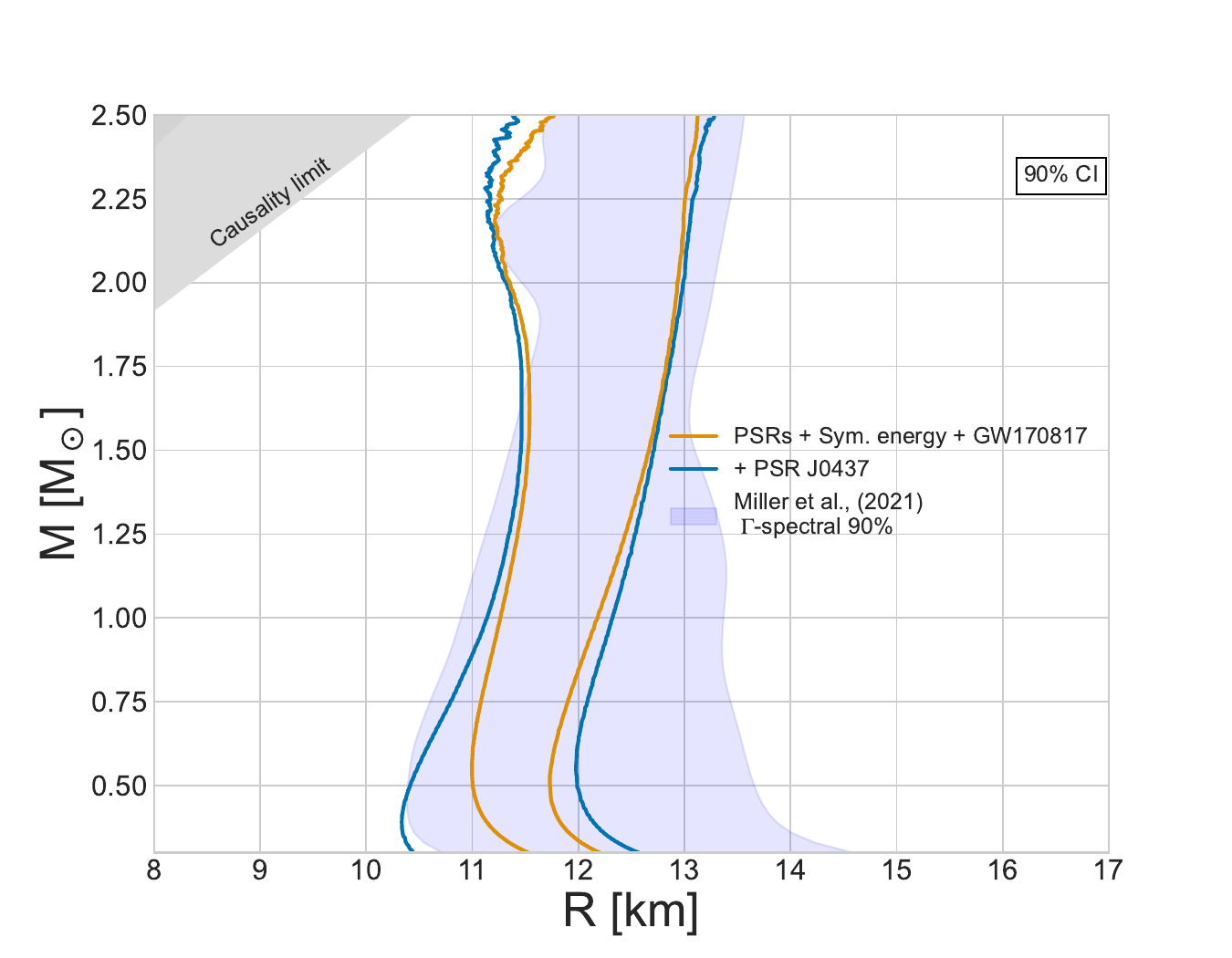}
\caption{Mass-Radius posterior comparison without (Orange) and with (Blue) the approximate PSR-J0437 posteriors in the inference.}
\label{fig:j0437}
\end{figure}

Figure \ref{fig:j0437} displays the posterior Mass-Radius 90\% credible interval shown earlier in Fig. \ref{fig:joint_gw_psr}, and a new posterior obtained upon applying the mass-radius constraint for PSR-J0437. Again, we notice on this preliminary study, the impact of this additional source on the M-R posterior is very minimal.

\end{document}